# THERMAL FLUCTUATIONS OF MAGNETIC NANOPARTICLES: FIFTY YEARS AFTER BROWN[1)]

William T. Coffey[a] and Yuri P. Kalmykov[b]

[a] *Department of Electronic and Electrical Engineering, Trinity College, Dublin 2, Ireland*

[b] *Laboratoire de Mathématiques et Physique (LAMPS), Université de Perpignan Via Domitia, 52, Avenue Paul Alduy, F-66860 Perpignan, France*

The reversal time (superparamagnetic relaxation time) of the magnetization of fine single domain ferromagnetic nanoparticles owing to thermal fluctuations plays a fundamental role in information storage, paleomagnetism, biotechnology, etc. Here a comprehensive tutorial-style review of the achievements of fifty years of development and generalizations of the seminal work of Brown [W.F. Brown, Jr., *Phys. Rev.*, **130**, 1677 (1963)] on thermal fluctuations of magnetic nanoparticles is presented. Analytical as well as numerical approaches to the estimation of the damping and temperature dependence of the reversal time based on Brown's Fokker-Planck equation for the evolution of the magnetic moment orientations on the surface of the unit sphere are critically discussed while the most promising directions for future research are emphasized.

---

# I. INTRODUCTION

**A. Thermal instability of magnetization in fine particles**

Fine single domain ferromagnetic particles exhibit unstable behavior of the magnetization due to thermal agitation resulting in superparamagnetism[1] (so-called) because each particle effectively behaves as an enormous paramagnet of magnetic moment ($\sim 10^4$–$10^5$ $\mu_B$). Now the thermal fluctuations and consequent relaxation of the magnetization of single domain particles play a major role both in information storage[2,3] and biomedical applications.[4,5] In particular, all magnetic recording media rely on the properties of such fine (~100 Å) ferromagnetic particles essentially because[1,6-9] the ferromagnetic state with a given orientation of the magnetic moment of a single domain nanoparticle has a *remanent magnetization* which led directly to the establishment of the modern magnetic recording industry. However, the apparent stable ferromagnetic state of a tape or a magnet is only one of many local minima of the free energy so that thermal agitation may cause spontaneous jumps of the magnetic moment from one metastable state to another. Thus at temperatures above a certain critical temperature called the *blocking temperature*, the magnetization may reverse its direction due to thermal agitation so that it exhibits thermal instability hence the stable magnetic behavior so characteristic of a ferromagnet is destroyed, ultimately resulting in the complete loss of the recording. It follows that the onset of thermal instability (characterized by a time dependent magnetization) in the fine magnetic particles used in magnetic recording is of profound significance as the latter are continually being reduced in size to provide both increased signal-to-noise ratio and greater storage density. The thermal instability is also of profound interest in rock magnetism[1,10] as the magnetic record keeping ability of igneous rocks depends on the fact that the fine particles preserve the direction of the earth's magnetic field from the epoch in which the temperature of the environment has fallen below (e.g., due to cooling of the primeval earth) the blocking temperature of the particles. Thus they constitute magnetic fossils and as such are indispensable in the study of paleomagnetism. Yet another consideration is that with recent progress in the development of magnetic nanotechnologies experimental studies of relaxation processes in individual particles have now become possible.[2,11]

Clearly the main parameter characterizing the thermal stability is the reversal time (superparamagnetic relaxation time) of the magnetization of the nanoparticles, which is crucially affected by thermal interactions of the particles with their surrounding heat bath resulting in fluctuations and dissipation, ultimately leading to a complete loss of the stored information. Thus it is vital for information storage purposes to determine the dependence of the reversal time on the dissipative coupling to the bath at a given temperature. Besides estimates of that time over wide ranges of temperature and damping are required in numerous other physical applications, e.g., in the determination of linear and nonlinear dynamic susceptibilities (e.g., [12-26]), the loop shape, coercive force and specific power loss in dynamic magnetic hysteresis (e.g., [27-34]), the signal-to-noise ratio in stochastic resonance (e.g., [35-40]), the switching field curves and surfaces at finite temperatures (e.g., [6,41-45]), Mössbauer spectra (e.g., [46-50]), etc.

To prepare the ground for our discussion we must first spend a little time in describing the relaxation process in a single domain particle.

A particle of ferromagnetic material[1] below a certain critical size (typically 150 Å in radius) constitutes a *single-domain particle* meaning[1] that it is in a state of *uniform magnetization* for *any applied field*. If we denote the magnetic dipole moment of such a particle by **μ** and ignore the anisotropy energy, and if we further suppose that an assembly of them has come to equilibrium at temperature $T$ under the influence of an applied magnetic field **H**, then we will have, for the mean dipole moment in the direction of the field,

$$\langle \mathbf{\mu} \cdot \mathbf{H} \rangle / H = \mu L(\xi), \tag{1}$$

where $L(\xi) = \coth \xi - \xi^{-1}$ is the Langevin function, $\xi = \mu H/(kT)$ is a dimensionless field parameter, $k$ is Boltzmann's constant, and $T$ is the temperature. The behavior is exactly analogous to that of noninteracting rigid electric dipoles in the Debye theory of the static electric susceptibility[1] or the Langevin treatment of paramagnetism; the vital difference, however, is that the moment **μ** is *not* that of a *single* atom but rather of a *single-domain particle* of volume $v$ which may be of the order of $10^4$–$10^5$ Bohr magnetons, so that *extremely large* moments and *large* susceptibilities are involved: hence the term *superparamagnetism*. The *thermal instability* of the magnetization occurs if the thermal energy $kT$ is sufficient to change the orientation of the magnetic moment **μ** of the entire particle. Then the thermal agitation causes continual changes in the orientation of **μ** and, in an ensemble of such particles, maintains a distribution of orientations characteristic of thermal equilibrium. Thus the number of particles with orientations of **μ** within solid angle $d\Omega = \sin\vartheta\, d\vartheta\, d\varphi$ is proportional to the Boltzmann distribution $e^{-vV/(kT)}d\Omega$, where $V(\vartheta, \varphi)$ is the free energy per unit volume, $v$ is the volume of the particle, and $\vartheta$ and $\varphi$ are angular coordinates which describe the orientation of the moment **μ** in the spherical polar coordinate system. In the absence of anisotropy, $vV = -\boldsymbol{\mu}\cdot\mathbf{H}$. Hence, the overall behavior is just like an assembly of paramagnetic atoms. No hysteresis exists, merely saturation behavior as predicted by Eq. (1). However, single-domain particles will in general not be *isotropic*, as is assumed in deriving Eq. (1), above but will have *anisotropic* contributions to their total energy associated with the external shape of the particle, imposed stress or the crystalline structure itself. If we consider the simplest anisotropy energy, namely the uniaxial one, then the total free energy of the particle, $vV$, will be (if the applied field **H** is assumed parallel to the polar axis)

$$vV = Kv\sin^2\vartheta - \mu H\cos\vartheta, \qquad (2)$$

($K$ is the anisotropy constant) so that the magnetization curve will no longer be the Langevin function. However, the dominant term governing the approach to saturation (as $\xi \to \infty$) will still be $1 - \xi^{-1}$.[1]

The discussion so far has been concerned with *equilibrium* behavior. We now have to consider *magnetic after-effect behavior*; i.e., under what conditions an assembly of single-domain particles can achieve thermal equilibrium in a time that is *short* compared with the time of an experiment. In 1949, Néel[7] predicted that, if a single-domain particle were *sufficiently small*, thermal fluctuations could cause its *direction of magnetization* $\mathbf{M} = \boldsymbol{\mu}/v$ to undergo a type of Brownian rotation, so that the stable magnetic behavior characteristic of a ferromagnet would ultimately be destroyed. This decay phenomenon constitutes the Néel-Brown relaxation process. An example given by Brown[9] of a tape recording is of interest: we expect that if we put this recording on a shelf that it will stay in the same magnetic state; we would be surprised if it suddenly jumped from being a recording of Beethoven to a recording of Brahms. In principle, however,[8,9] the *apparent stability* of the recording is only one of *many* local minima of the free energy: thermal agitation can cause spontaneous jumps from one such state to another. The apparent stability[1,8,9] (*ferromagnetic behavior*) arises because our tape or magnet cannot get from one magnetic state to another without surmounting an energy barrier which is very large in comparison with $kT$. Thus, the probability per unit time of a jump over such a barrier is so small that the mean time we would have to wait for it to occur *far exceeds our own lifetime*; we perceive *stable* ferromagnetic behavior even though the process is actually time dependent with a relaxation time which may be of the order of a geological epoch. However, if the barrier is neither very large nor very small in comparison with the noise strength $kT$ (our case), then the specimen neither remains in a single stable state for a *long* time nor attains thermal equilibrium in a *short* time after a change in field: it instead undergoes a change of magnetization which is not completed instantaneously but *lags behind* the field exactly analogous to the Debye relaxation process in polar dielectrics.[51,52] This is called *magnetic after-effect*, or *magnetic viscosity*, or *Néel relaxation*, and occurs only for *sufficiently fine* ferromagnetic particles. In order to illustrate the Néel mechanism,[7] consider an assembly of aligned uniaxial particles in the presence of a field **H**, whose potential energy is given by Eq. (2). Thus, the particles are fully magnetized along the polar axis, which is the axis of symmetry. A sufficiently long time after the field is switched off, the remanence will vanish as

$$M_r(t) = M_S e^{-t/\tau}, \tag{3}$$

which is the longest lived mode of the relaxation process. Here $M_S$ is the mean magnetization of a nonrelaxing particle, $t$ is the time after the removal of the field, and $\tau$ is the superparamagnetic relaxation time. Néel[7] then suggested that, from transition state theory (TST),[53] the relaxation time is given by

$$\tau \sim f_0^{-1} e^{\frac{vK}{kT}}, \tag{4}$$

where $f_0$ is the so-called attempt frequency associated with the frequency of the *gyromagnetic precession* so that, by varying the volume or the temperature of the particles, $\tau$ can be made to vary from $10^{-9}$ s to *millions of years* ($f_0^{-1}$ is often taken as small as $10^{-10}$–$10^{-11}$ s in practice).[8] The presence of the exponential factor in Eq. (4) indicates that, in order to approach the zero remanence (corresponding to thermal equilibrium), a sufficient number of particles (magnetic moments) must be reversed by thermal activation over the energy barrier $vK$. The probability of such a process occurring is proportional to $e^{-vK/(kT)}$. For example, when **H** = 0, Eq. (2) is a symmetric bistable potential with minima at $\vartheta = 0$ and $\vartheta = \pi$ and a maximum at $\vartheta = \pi/2$.

It is apparent from Eq. (4) that the superparamagnetic relaxation time $\tau$ being governed by an activation process depends *exponentially on the particle volume*; hence there is a fairly well-defined particle radius above which the magnetization will appear stable. We consider the figures given by Bean and Livingston[1] for a spherical iron particle with uniaxial anisotropy $Kv\sin^2\vartheta$. A particle of radius 115 Å will have a relaxation time of $10^{-1}$ s at 300 K, so that the moment will relax almost instantaneously. A particle of radius 150 Å, on the other hand, will have a relaxation time of $10^9$ s and so will be exceedingly stable (i.e., the moment will not reverse in this time; see the preceding example above). This situation corresponds to an energy barrier that is very large in comparison to $kT$ where, for any reasonable measurement time,[8,9] we may ignore thermal agitation and calculate the static magnetization by simply minimizing $V$ with respect to the polar and azimuthal angles ($\vartheta, \varphi$) for each value of an applied field $\mathbf{H}_0$. This is the well-known Stoner–Wohlfarth calculation;[6] it leads to hysteresis because in certain field ranges two or more minima exist and transitions between them are neglected. Here a typical potential of a particle would be[6]

$$vV(\vartheta,\varphi) = Kv\sin^2\vartheta - \mu H_0(\cos\vartheta\cos\psi + \sin\psi\sin\vartheta\cos\varphi). \qquad (5)$$

The polar axis is the easy axis of magnetization; the field $\mathbf{H}_0$ is applied in the $xz$ plane at an angle $\psi$ to the easy axis. Thus, in general, there will be only a *narrow range of particle sizes* for which the relaxation time will be of the order of experimental times, and for which measurable "magnetic viscosity" effects, manifesting themselves as an *observable* change of magnetization, *lagging behind field changes*, would be expected. Bean and Livingston[1] have given a rough measure of the size of the particle needed for transition to stable ferromagnetic behavior, taking $\tau = 10^2$ s, they find that the energy is $25kT$. The temperature at which this occurs for a given particle is called the *blocking temperature*. They obtain sizes of 40 Å for h.c.p. cobalt, 125 Å for iron, 140 Å for f.c.c. cobalt. We mention that, in an assembly consisting *solely* of single-domain particles, the remanence at a given temperature should be a measure of the amount of material with particle volume greater than the volume that is just stable at this temperature. Thus,[1] by following the increase of remanence with decreasing temperature, we can ascertain how much material lies in various ranges of volume, and so determine the particle size distribution. It is interesting to recall that Néel[7] was led to his solid-state mechanism of relaxation, that is, *rotation of the magnetic moment inside the particle due to thermal agitation*, through the study of *paleomagnetism*.

**B. Kramers escape rate theory**

The Néel theory[7] of thermal fluctuations of the magnetization $\mathbf{M}(t)$ of a single domain ferromagnetic particle was based on TST, which has its origins in the 1880s when Arrhenius[53] proposed, from an analysis of experimental data, that the rate coefficient in a chemical reaction should obey the law

$$\Gamma = \nu_0 e^{-\frac{\Delta V}{kT}}, \qquad (6)$$

where $\Delta V$ denotes the threshold energy for activation and $v_0$ is a prefactor.[53] After very many developments, summarized in the excellent review of Hänggi *et al.*,[53] this equation led to the concept of chemical reactions as being analogous to an assembly of particles situated at the bottom of a potential well. Rare members of this assembly will have enough energy to escape over the potential hill, owing to the shuttling action of thermal agitation, and will never return[53] (see Fig. 1), thus constituting a model of a chemical reaction. The escape over the potential barrier represents the breaking of a chemical bond.[53] The Arrhenius law for the escape rate $\Gamma$ (reaction velocity in the case of chemical reactions) of particles that are initially trapped in a potential well at *A*, and that may subsequently, under the influence of thermal agitation, escape over a high (>> *kT*) barrier of height $\Delta V$ at *C* and never return to *A*, may be written using TST as

$$\Gamma^{\text{TST}} = \frac{\omega_A}{2\pi} e^{-\frac{\Delta V}{kT}}. \tag{7}$$

Here the superscript TST stands for transition state theory while the attempt frequency, $\omega_A$, is the angular frequency of a particle executing oscillatory motion (i.e., libration) at the bottom of a well. The barrier arises from the potential function of some external force, which may be electrical, magnetic, gravitational, and so on. The formula represents an attempt frequency times a Boltzmann factor, which weights the escape from the well. Relaxation processes of this nature appear to have been first identified in crystalline solids by Debye.[51]

A very unsatisfactory feature of the Arrhenius formula is that it appears to predict *escape in the absence of coupling to a heat bath* contradicting the fluctuation–dissipation theorem. This defect was remedied, and reaction rate theory was firmly set in the framework of nonequilibrium statistical mechanics by Kramers.[54] His idea, motivated by the fluctuation-dissipation theorem and prompted by the fact that escape over the barrier is exponentially slow so that *quasi-stationary* conditions prevail, is to calculate a correction factor $\Lambda$ in an Arrhenius like equation for the escape rate $\Gamma$ over the potential barrier $\Delta V$ (reaction velocity in the case of chemical reactions), viz.,

$$\Gamma = \Lambda \Gamma^{\text{TST}} = \Lambda \frac{\omega_A}{2\pi} e^{-\frac{\Delta V}{kT}}. \tag{8}$$

Kramers included *nonequilibrium effects* in the barrier-crossing process (the tendency of which is always to reduce the TST rate, manifested by a frictional dependence, i.e., a coupling to the heat bath of the prefactor in the TST formula) by choosing, as a microscopic model of a chemical reaction, a classical Brownian particle of mass *m* moving in a single-well potential (see Fig. 1; for applications of Kramers' method see [53, 55-58]). Thus the dynamics are described by the Langevin equation for the random state variables coordinate and momentum $(x, p)$

$$p = m\dot{x}, \quad \dot{p}(t) + \beta p(t) + \frac{dV}{dx} = F(t), \tag{9}$$

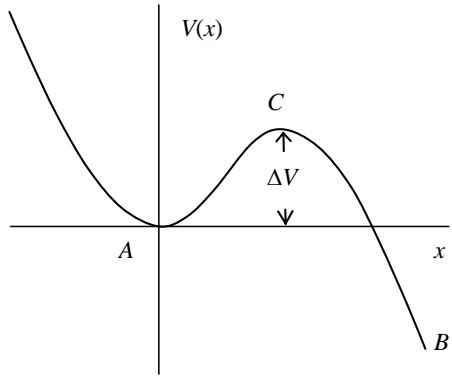

FIG. 1. Single-well potential function as the simplest example of escape over a barrier. Particles are initially trapped in the well near the point *A* by a high potential barrier at the point *C*. They are thermalized very rapidly in the well. However, very few may attain enough energy to escape over the barrier into region *B*, from which they never return (a sink of probability). The barrier *C* is assumed to be sufficiently large so that the rate of escape of particles is very small.

where $\beta p(t)$ is a systematic frictional force *slowing down* the motion of the particle and $F(t)$ is a white noise random force *maintaining the motion*, which is due to the impacts of the surrounding molecules of the liquid on the Brownian particle, (both representing the effect of the bath) and $\beta$ is a damping coefficient. The associated diffusion equation describing the evolution of the probability density function $W(x,p,t)$ in phase space $(x, p = m\dot{x})$ is the Fokker-Planck (in the present context, the Klein–Kramers) equation, viz.,[54,58]

$$\frac{\partial W}{\partial t} = -\frac{p}{m}\frac{\partial W}{\partial x} + \frac{dV}{dx}\frac{\partial W}{\partial p} + \beta\frac{\partial}{\partial p}\left(Wp + mkT\frac{\partial W}{\partial p}\right). \tag{10}$$

In this way a typical particle embedded in a heat bath is modeled by Brownian motion. In the single-particle distribution function picture, Eq. (10), Brownian motion represents (essentially through a dissipation parameter) all the remaining degrees of freedom of the system, consisting of the selected particle and the heat bath, which is in *perpetual thermal equilibrium* at temperature *T*. In Kramers' model,[54] the particle coordinate *x* represents the *reaction coordinate* (i.e., the distance between two fragments of a dissociated molecule – a concept first introduced in 1926 by Christiansen[53]). The value of this coordinate, $x_A$, at the first minimum of the potential represents the *reaction state*; the value $x_B$, significantly over the summit of the well at *B* (i.e., when the particle has crossed over the summit) represents the *product state*, and the value $x_C$, at the saddle point, represents the *transition state*. It is assumed throughout that quasi-stationarity obtains so that one may set $\dot{W}=0$ in Eq. (10) giving rise to a steady current of particles when considering the barrier crossing.

Now, in the Kramers problem originally pertaining to point Brownian particles with separable and additive Hamiltonians, three regimes of damping appear as a direct consequence of the particular asymptotic method involved in the solution of the quasi-stationary Klein-Kramers equation:

(i) Intermediate-to-high damping (IHD): the general picture here being that inside the well the distribution function is almost the equilibrium Maxwell-Boltzmann distribution obtaining in the depths of the well. However, very near the barrier the distribution function deviates from the equilibrium distribution due to the slow draining of particles across the barrier. The barrier region, in which nonequilibrium behavior prevails, is assumed to be so limited in spatial extent, however, that one may approximate the potential in this region by an inverted parabola.

(ii) Very low damping (VLD): here the damping is so small that the tacit assumption that the particles approaching the barrier region from the depths of the well have the Maxwell-Boltzmann distribution completely breaks down. Thus, the barrier region or boundary layer, where deviations from that distribution occur, now extends far beyond the interval, where the potential shape may be sensibly approximated by an inverted parabola (conceive of a particle executing large oscillations in a potential well with only a tiny dissipation to the surroundings so that the motion is almost Newtonian). Here, one proceeds using a completely different approach involving an energy-controlled diffusion equation derived by first transforming the Fokker-Planck (Klein-Kramers) equation (10) into an equation in energy $E = m\dot{x}^2/2 + V(x)$ (slow) and phase $w$ (fast) variables by supposing that the large amplitude librational motion in the well of a (barrier crossing) particle with energy equal to the barrier energy is almost conservative (the phase is defined via the constant of integration in the differential equation $\dot{x} = \pm\sqrt{2[E-V(x)]/m}$, i.e., $\int_{x(0)}^{x} \left(2[E-V(x')]/m\right)^{-1/2} dx' = t + w$).

Such an energy trajectory is called a critical energy trajectory or separatrix. The critical energy is the energy required by a particle to just escape the well and the separatrix separates the bounded motion in the well and the unbounded motion outside. Thus[55] the concept of large oscillations of a particle in a well before escape is always involved. The almost conservative assumption ensures that the Liouville terms, i.e., first two terms, in the Fokker-Planck equation (10) vanishes (unlike in IHD, where strong coupling between the Liouville and diffusive terms exists) so that only the diffusion term in the energy variable, which would not of course be present in the purely Newtonian motion, remains. The dependence on the fast phase variable is eliminated by averaging the distribution function in energy-phase variables along a *closed* trajectory of the energy since we assume a librational motion in the well.

(iii) An intermediate (turnover), i.e., a more or less critically damped region, where neither IHD nor VLD formulas apply. In contrast to the VLD case the Liouville term in the Fokker-Planck equation does not vanish meaning that one cannot simply average out the phase dependence of the distribution function which is ultimately accounted for by constructing from the quasi-stationary Fokker-Planck equation an equation for the distribution function in the barrier region with the energy and action as independent variables.

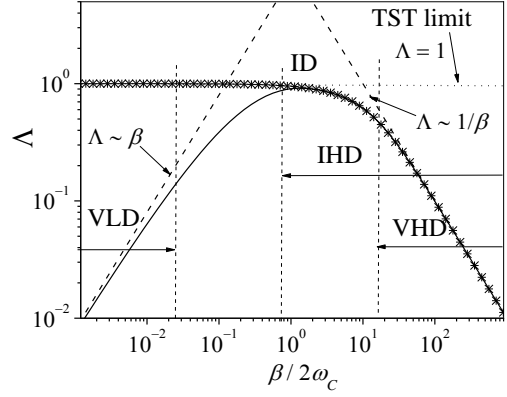

FIG. 2. Diagram of damping regions for the prefactor $\Lambda$ in Eq. (8). Asterisks: Kramers' IHD formula Eq. (11). Dashed lines: the very high damping (VHD) and VLD asymptotes, Eqs. (11) at $\beta \gg 1$ and (12). Solid line: numerical solution of the Fokker-Planck Eq. (10). Three regions exist, namely VLD, intermediate damping (ID) (TST), and VHD, and two crossovers between them [cf. Fig. 18 of Ref. 53 and Fig. 4 of Ref. 56]. Kramers' turnover refers to the underdamped region between ID and VLD. Clearly, the TST rate (dotted line) must represent the upper bound of the escape rate because all dissipation to the bath is ignored in that rate.

Kramers[54] then obtained so called IHD and VLD asymptotic formulas for the escape rate, assuming in both cases that the energy barrier is much greater than the thermal energy so that the concept of an escape rate applies (tantamount to the quasi-stationary assumption). The first is the IHD formula (see Fig. 2)

$$\Gamma^{IHD} = \left( \sqrt{1 + \frac{\beta^2}{4\omega_C^2}} - \frac{\beta}{2\omega_C} \right) \Gamma^{TST}, \qquad (11)$$

where $\omega_C$ is the characteristic frequency of the inverted oscillator approximation to the potential $V(x)$ in the vicinity of the barrier. In the IHD formula, the correction $\Lambda$ to the TST result is essentially the positive eigenvalue (characterizing the unstable barrier-crossing mode) of the Langevin equation, Eq. (9), omitting the noise, linearized about the saddle point of the potential $V(x)$. In the case considered by Kramers, this is a one-dimensional maximum. Equation (11) formally holds,[53,54] when the energy loss per cycle of the motion of a particle librating in the well with energy equal to the barrier energy $E_C = \Delta V$, is significantly greater than $kT$. The energy loss per cycle of the motion of a barrier-crossing particle is $\beta S(E_C)$, where $E_C$ is the energy contour through the saddle point of the potential, and $S$ is the action evaluated at $E = E_C$. This criterion effectively follows from the Kramers very-low-damping result (see below). The IHD asymptotic formula is derived by supposing that (i) the barrier is so high and the dissipative coupling to the bath so strong that a Maxwell–Boltzmann distribution always holds at the bottom of the well; and (ii) the Langevin equation may be linearized in the region very close to the potential barrier, meaning that all the coefficients in the corresponding quasi-stationary Klein–Kramers equation are linear in the positions and velocities.

If these simplifications can be made, then the quasi-stationary Klein–Kramers equation, although it remains an equation in the two phase variables $(x, p)$, may be integrated by introducing an independent variable which is a linear combination of $x$ and $p$ so that it becomes an *ordinary* differential equation in a *single* variable.

However, for small friction $\beta$ such that the energy loss per period $\beta S(E_C) \ll kT$, the IHD formula fails, predicting, just as with the TST formula, escape in the *absence of coupling to the bath*, because[56,58,59] the tacit assumption that the particles approaching the barrier from the depths of the well are in thermal equilibrium is violated (owing to the tiny dissipation of energy to the bath ensuing that the motion is almost purely Newtonian). Thus, as we have mentioned, the spatial region of significant departure from the Maxwell–Boltzmann distribution in the well extends far beyond the region over which the potential may sensibly be approximated by an inverted parabola. Kramers showed how his second formula, valid in the VLD or almost Newtonian case, where the energy loss per cycle $\beta S(E_C)$ of a librating particle is very much less than $kT$, may be obtained by again reducing the Klein–Kramers equation to a partial differential equation in a single spatial variable. This variable is the *energy* or, equivalently, the *action*. Here the blurred energy trajectories diffuse very slowly so that they do not differ significantly from those of the undamped librational motion in a well with energy corresponding to the saddle energy $\Delta V$ or $E_C$. The blurring effect of the noise is vividly illustrated by the calculation of the Green function pertaining to the alteration in the energy in one cycle for particles with energy infinitesimally close to the critical energy due to Mel'nikov,[55] which is a narrowly peaked Gaussian distribution rather than the delta-function associated with the purely Newtonian motion at the critical energy. Thus, the net effect of escape, which has its origin in fluctuations in the energy about the critical energy, is to produce a very slow spiraling of the closed energy trajectories towards the origin in the phase space $(x, p)$. Kramers solved the VLD problem by writing the Klein–Kramers equation in angle–action (or angle–energy) variables (the angle is the *phase* or *instantaneous state* of the system along an energy trajectory) and taking a time average of the motion along a closed energy trajectory infinitesimally close to the saddle energy trajectory. Thus, by dint of thermal fluctuations, the (noisy) trajectory may become the separatrix or the open trajectory on which the particle exits the well. Thus, once again, the time derivative of $W$ (when $W$ is written as a function of the energy using the averaging procedure above) is *exponentially small* at the saddle point. Hence, the quasi-stationary solution in the energy variable may be used. This procedure ultimately yields the Kramers' VLD formula (see Fig. 2):

$$\Gamma^{\text{VLD}} = \frac{\beta S(E_C)}{kT} \Gamma^{\text{TST}}. \tag{12}$$

This formula, which assumes that all particles that reach the separatrix exit the well, holds when $\beta S(E_C) \ll kT$; unlike the TST result it vanishes when $\beta \to 0$, so that escape is impossible without coupling to the bath.

Likewise, if the coupling to the bath is very large, the escape rate *again vanishes*. Kramers made several estimates of the range of validity of both IHD and VLD formulas and the intermediate (or moderate) damping (ID) region where the TST, Eq. (7), holds with a high degree of accuracy. He was, however, unable to give a formula in the underdamped region between IHD and VLD, as there $\beta S(E_C) \approx kT$ so that no small perturbation parameter now exists mentioning that he could not find a general method of attack for the purpose of obtaining a formula which would be valid for any damping regime. In essence, this problem, known as the *Kramers turnover*, was solved nearly 50 years later by Mel'nikov[55] and Mel'nikov and Meshkov.[60] They converted, using their Green function and the principle of superposition, their energy-action diffusion equation into an integral equation for the evolution of the energy distribution function in the vicinity of the separatrix which they solved using the Wiener–Hopf method,[61] and so obtained an escape rate formula which is valid for all values of the friction $\beta$ constituting a solution of the Kramers turnover problem for mechanical particles for the escape rate from a single well (the results also apply to rotational Brownian motion of rigid bodies), viz.,

$$\Gamma = A\left[\frac{\beta S(E_C)}{kT}\right]\Gamma^{\text{IHD}}, \tag{13}$$

where $A(\Delta)$ is called the depopulation factor given by

$$A(\Delta) = \exp\left(\frac{1}{2\pi}\int_{-\infty}^{\infty} \ln\left[1 - e^{-\Delta(\lambda^2 + 1/4)}\right]\frac{d\lambda}{\lambda^2 + 1/4}\right). \tag{14}$$

Extension of this formula for double-well and periodic potentials are given elsewhere.[53,55,60] By extending the Mel'nikov-Meshkov approach, Grabert[62] and Pollak *et al.*[63] later presented a complete solution of the Kramers turnover problem and have shown that the Mel'nikov and Meshkov formula, Eq. (13), can be obtained without *ad hoc* interpolation between the weak and strong damping regimes. We remark that the theory of Pollak *et al.*[63] is also applicable to an arbitrary memory friction and not only in the "white noise" (memoryless) limit.

As far as the verification of the turnover formula, Eq. (13), is concerned, many examples of calculations based on either the analytical and numerical solutions of the Klein-Kramers equation or on numerical simulations of the Brownian dynamics exist. They include the comparison of the turnover formula with the numerical results for the escape out of a single potential well[64,65] and that from both double-[66,67] and multi-well[68-71] potentials (see Fig. 3). Another example is the turnover treatment of the same one-dimensional problem and its generalization to diffusion on a surface which was undertaken by Pollak and collaborators in Refs. 72-74, where a comparison with numerical simulations based on the Langevin equation is given. Examples of the treatment of turnover problems for the rotational Brownian motion of a single-axis rotator in a potential are given by Coffey *et al.*[75,76] Moreover, Moro and Polimeno,[77] Pastor and Szabo,[78] and Kalmykov *et al.*[79] successfully tested the Mel'nikov-Meshkov formula for the rotational Brownian motion of a linear molecule in a uniaxial potential. The escape rate theory has also undergone experimental verifications from fields as diverse as chemical kinetics, diffusion in solids or on surfaces, diffusion processes in disordered and amorphous materials, homogeneous nucleation, electrical transport, etc.; a detailed discussion of experiments investigating the rate between metastable states is given in the review of Hänggi *et al.*[53]

Regarding the above discussion of escape rate regimes the reader should always bear in mind that the seemingly separate damping regimes encountered in the Kramers theory are merely artifacts of the asymptotic methods used. If numerical calculations of the escape rate are made via the smallest nonvanishing eigenvalue of the Fokker-Planck operator or the inverse mean first passage time,[53] all the damping regimes occur seamlessly being part of the one and same dynamical entity. However, the great merit of the Kramers calculation is that as well as yielding a clear physical picture of the processes underlying the escape rate, it also yields an analytical formula for high potential barriers in a form suitable for comparison with experiment, which is never possible via the numerical methods.

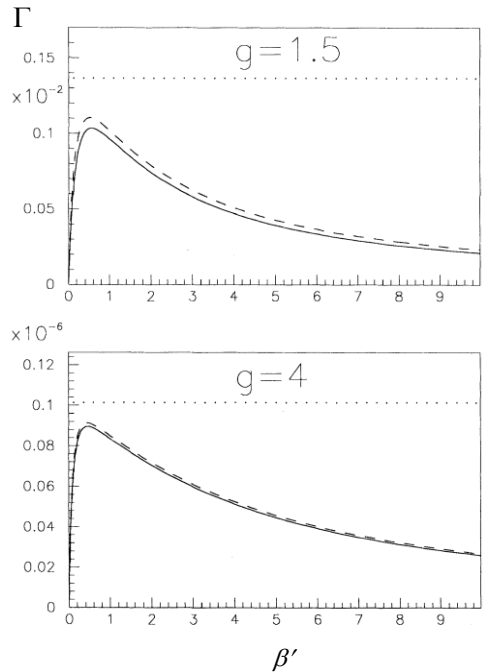

FIG. 3. The escape rate for the Brownian motion of a point particle in a cosine periodic potential $V(x) = -2gkT\cos(2\pi x/a)$ vs. the dimensionless friction parameter $\beta' = \beta a(2\pi)^{-1}\sqrt{m/(kT)}$ for two different potential barriers $g = 1.5$ (upper panel) and $g = 4$ (lower panel). Solid lines: exact numerical solution of the Klein-Kramers Eq. (10). The straight-dotted lines: the TST Eq. (7). The dashed lines: the turnover Eq. (13). Reprinted figure with permission from R. Ferrando, R. Spadacini, and G. E. Tommei, Phys. Rev. E **48**, 2437 (1993).[69] Copyright (1993) by the American Physical Society.

## C. Superparamagnetic relaxation time: Brown's approach

Returning to the magnetic problem, in order to estimate the characteristic time of reversal of the magnetic moment over the internal anisotropy potential barrier, Néel's TST calculation of the superparamagnetic relaxation time $\tau$ was set in the context of the general theory of stochastic processes by W.F. Brown[7] via the classical theory of the Brownian motion and then adapting to magnetic relaxation the ingenious method of Kramers[54] outlined above. Brown proceeded by taking as a Langevin equation, the Gilbert equation[80] for the motion of the magnetization augmented by a random field, viz.,[8,9]

$$\dot{\mathbf{u}}(t) = \mathbf{u}(t) \times \left[ \gamma \mathbf{H}(t) - \alpha \dot{\mathbf{u}}(t) + \gamma \mathbf{h}(t) \right], \tag{15}$$

where $\mathbf{u} = \mathbf{M}/M_S$ is the unit vector directed along $\mathbf{M}$, $\gamma$ is the gyromagnetic ratio, $\alpha$ is the dimensionless damping (dissipation) parameter,

$$\mathbf{H} = -\frac{\partial V}{\partial \mathbf{M}} = -\left( \mathbf{i}\frac{\partial V}{\partial M_X} + \mathbf{j}\frac{\partial V}{\partial M_Y} + \mathbf{k}\frac{\partial V}{\partial M_Z} \right), \tag{16}$$

$V$ is the Gibbs free energy density (characterizing the magnetic anisotropy and Zeeman energy density of the particle), and $\mathbf{h}(t)$ is a random Gaussian field with white noise properties (in our notation)

$$\overline{h_i(t)} = 0, \quad \overline{h_i(t_1)h_j(t_2)} = \frac{2kT\alpha}{v\gamma M_S}\delta_{ij}\delta(t_1 - t_2). \tag{17}$$

Here the indices $i, j = 1, 2, 3$ in Kronecker's delta $\delta_{ij}$ and $h_i$ correspond to the Cartesian axes $X,Y,Z$ of the laboratory coordinate system $OXYZ$, $\delta(t)$ is the Dirac delta function, and the overbar means the statistical average over an ensemble of particles which all have at time $t$ the *same* magnetization $\mathbf{M}$. The random field accounts for the thermal fluctuations of the magnetization of an individual particle without which the random orientational motion would not be sustained. Brown then derived from Eq. (15), the appropriate Fokker-Planck equation for the distribution function $W(\vartheta,\varphi,t)$ of the orientations of the magnetization vector $\mathbf{M}$ on the surface of the unit sphere[8,9] (see Sec. II.A for details)

$$\frac{\partial}{\partial t}W = \mathrm{L}_{FP}W = \frac{1}{2\tau_{\mathrm{N}}}\left\{ \frac{v}{kT}\left[ \alpha^{-1}\mathbf{u}\cdot\left( \frac{\partial V}{\partial \mathbf{u}} \times \frac{\partial W}{\partial \mathbf{u}} \right) + \frac{\partial}{\partial \mathbf{u}}\cdot\left( W\frac{\partial V}{\partial \mathbf{u}} \right) \right] + \Delta W \right\}, \tag{18}$$

where

$$\tau_{\mathrm{N}} = \frac{vM_S(1+\alpha^2)}{2\gamma\alpha kT} \tag{19}$$

is the free diffusion time of the magnetization ($\tau_{\mathrm{N}}$ is of the order of $10^{-11}$–$10^{-8}$ s), $\mathrm{L}_{FP}$ is the Fokker-Planck operator, $\Delta$ is the Laplacian on the surface of the unit sphere defined as

$$\Delta = \frac{1}{\sin\vartheta} \frac{\partial}{\partial\vartheta}\left(\sin\vartheta \frac{\partial}{\partial\vartheta}\right) + \frac{1}{\sin^2\vartheta} \frac{\partial^2}{\partial\varphi^2}, \quad (20)$$

i.e., the angular part of the Laplacian, the operator $\partial/\partial\mathbf{u}$ means the gradient on the surface of the unit sphere[9] so that, in the spherical coordinate system (Fig. 4),

$$\frac{\partial}{\partial\mathbf{u}} = \frac{\partial}{\partial\vartheta}\mathbf{e}_\vartheta + \frac{1}{\sin\vartheta}\frac{\partial}{\partial\varphi}\mathbf{e}_\varphi. \quad (21)$$

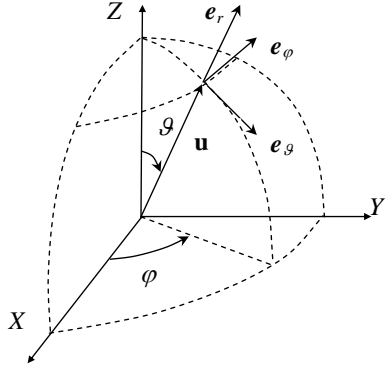

FIG. 4. Spherical polar coordinate system.

Here it is assumed that the magnetization is always uniform inside the particle and that *only* the orientation and not the magnitude of the magnetization undergoes variations. A detailed discussion of the assumptions made in the derivation of the Fokker-Planck and Gilbert equations is given elsewhere (e.g., [8,9,81]). We remark in passing that in developing his theory of the magnetization relaxation for superparamagnets (classical spins) Brown used by analogy ideas originating in the Debye theory of dielectric relaxation of polar dielectrics.[52] In Eq. (18), the term in $\alpha^{-1}$ corresponds to the *precessional* (gyromagnetic) term in Eq. (15), giving rise to ferromagnetic resonance (usually in the GHz range). When $\alpha \to \infty$ (i.e., ignoring the gyromagnetic term) Brown's equation, Eq. (18), has the same mathematical form as the *noninertial* rotational diffusion equation for a rigid body in an external potential (known as the Smoluchowski equation in configuration space).[58]

Referring to magnetic relaxation, in his earliest calculations of the reversal time of the magnetization $\tau$, *which may be defined as the inverse of the smallest nonvanishing eigenvalue $\lambda_1$ of the Fokker-Planck operator $L_{FP}$ in Eq. (18)*, Brown[8] confined himself to *axially symmetric potentials* of the magnetocrystalline anisotropy and Zeeman energy.[8] Hence *no dynamical coupling between the longitudinal and the transverse modes of motion exists* so that the longitudinal modes are governed by a single state variable, namely, the colatitude, $\vartheta$, i.e., the polar angle of **M**. The second state variable, namely, the azimuthal angle $\varphi$ of **M** gives rise only to a steady precession of that vector. Noting the decoupling between the transverse and longitudinal modes existing for axial symmetry, which results in an exact single-variable Fokker-Planck equation in the colatitude $\vartheta$, Brown demonstrated that the Kramers escape rate theory for point particles may be easily adapted to yield an expression for the escape rate for spins in axially symmetric potentials which is valid for all values of the damping parameter $\alpha$. We remark, however, that the *exact* Fokker-Planck equation in the single variable $\vartheta$ arises not from strong damping of the momentum (which in the Brownian motion of point particles or rigid bodies governed by the Klein-Kramers equation in the Euler angles and corresponding angular momenta gives rise to the *approximate* noninertial Smoluchowski equation[58]) rather it follows from the *axial symmetry* of the potential.

The magnetization reversal time problem differs fundamentally from that of point particles, because: (i) the magnetic system has two degrees of freedom, the polar $\vartheta$ and azimuthal $\varphi$ angles, (ii) the undamped equation of motion of the magnetization of a single-domain ferromagnetic particle is the gyromagnetic equation, (iii) the Hamiltonian is nonseparable, and (iv) the inertial effects play no role. The role of inertia in the mechanical system is essentially mimicked in the magnetic system for nonaxially symmetric potentials by the gyromagnetic term causing *coupling* or *entanglement* of the transverse and longitudinal modes. Thus, in order to derive escape rate formulas for superparamagnetic particles analogous to those for point particles, one has to consider in Brown's Fokker-Planck equation a nonaxially symmetric free energy density, $V(\vartheta,\varphi)$, where *explicit coupling* between the two degrees of freedom exists. Thus both regimes of damping (IHD and VLD) can occur reflecting the fact that the dynamics of the transverse response affect the dynamics of the longitudinal response and vice versa. However, this fact appears not to have been explicably recognized in the first calculations of the magnetization reversal time of superparamagnets with nonaxially symmetric anisotropy by Smith and de Rozario[82] and Brown.[9] Thus *only* IHD formulas for the escape rate $\Gamma_i^{IHD}$ (see Eq. (84) below) for nonaxially symmetric potentials were derived.[9,82] However, in 1990, Klik and Gunther[83,84] realized that the various Kramers damping regimes also applied to magnetic relaxation of single domain ferromagnetic particles and derived the corresponding VLD formula (see below Eq. (90)) which is effectively the same as the corresponding Kramers result for point particles, Eq. (12). Furthermore, they emphasized that the magnetic IHD calculations[9,56,59,82,85] are in effect a special case of Langer's general treatment of the decay of metastable states of systems with many degrees of freedom[86] which is in itself a generalization of the Becker and Döring[87] treatment of the rate of condensation of a supersaturated vapor. The conditions of applicability of these IHD and VLD solutions for superparamagnets are defined by $\alpha \gtrsim 1$ and $\alpha \ll 1$, respectively. Finally, in the turnover region, $0.01 < \alpha \lesssim 1$, Coffey *et al.*[56,88] have shown that the Mel'nikov formalism[55] for interpolating between the VLD and IHD Kramers escape rates as a function of the dissipation parameter for point particles, can be extended to include magnetization relaxation of single-domain ferromagnetic particles having nonaxially symmetric potentials of the magnetocrystalline anisotropy (see Eq. (92) below). The turnover escape rate formula for superparamagnets has been exhaustively verified by numerical calculations in many publications,[89-97] where it has been compared with that calculated numerically from either the appropriate Fokker-Planck or Langevin equations and via computer simulations in all damping ranges including VLD, IHD, and VHD limits (see Sections III and IV below).

In this review, we shall present an overview of the various theoretical approaches for the estimation of the magnetization relaxation time of superparamagnetic nanoparticles fifty years after Brown's seminal paper[8] laying particular emphasis on nanoparticles with nonaxially symmetric potential of the magnetocrystalline-Zeeman energy. During the intervening period the Brown theory for classical spins has been extensively developed and, just as the earlier Kramers' escape rate theory for point particles,[53] gives rise to distinct VLD and IHD regions as well as the turnover region between low and high damping. The latter rather complex developments, which have mainly to do with the nature of the nonaxially symmetric potentials have prompted us to write this review. We were also encouraged to write it because the community involved in nanomagnetism is ever increasing and is accompanied by a parallel increase in interest in the Néel-Brown theory of the magnetization reversal in the presence of thermal agitation, its current predictions, and its future development (for example, according to the database *Web of Science*, Brown's seminal paper[8] published in 1963 has been cited 527 times since 2007 out of a total number of 1456 citations). In large measure because now exist experimental resolutions fine enough to probe the details of the reversal time of the magnetization, in particular its damping and temperature dependence.[11]

Here, we shall consider only one mechanism of the magnetization reversal in magnetic nanoparticles, namely, the *coherent (uniform) rotation* of the magnetization, which in many cases (e.g., in almost spherical nanoparticles at relatively high temperatures) plays the most essential role. Thus we shall not consider other possible mechanisms of the magnetization reversal such as nonuniform rotation[98-100] and macroscopic quantum tunneling (MQT)[101-103] of the magnetization. The nonuniform rotation mechanism may provide an essential contribution to the magnetization relaxation process in elongated nanoparticles, nanowires and nanorods[104] while MQT may become important at low temperatures.[11] A detailed account of these mechanisms of magnetization reversal, may be found in the reviews.[2, 11, 102, 104]

The review is arranged as follows. In Section II, we present Brown's intuitive derivation of the Fokker-Planck equation for the diffusion of a classical spin and discuss the evaluation of the characteristic relaxation times from this equation. In Section III, we describe the calculation of the superparamagnetic relaxation time of a uniaxial nanoparticle, subjected to a strong uniform magnetic field $\mathbf{H}_0$ applied along the easy axis of the particle using Kramers' escape rate theory as applied to spin systems. In Section IV, we review various estimates of the reversal time of the magnetization, using the Kramers escape rate theory and its generalization to classical spin systems with nonaxially symmetric anisotropy free energy. In particular, we discuss the damping dependence of the escape rate for magnetic nanoparticles and consider in detail various approximations valid in all damping ranges including VLD, IHD, and VHD. Also in Section IV, we rigorously derive the appropriate

Fokker–Planck equations for the Gilbert, Landau–Lifshitz, and Kubo kinetic models for the Brownian motion of classical spins (the last two models are also frequently applied in the latter context). We show in particular that all three kinetic models yield, for low damping, $\alpha \ll 1$, the *same* Fokker–Planck equation, hence, the same estimate for the reversal time. However, only the Gilbert model, where the systematic and random terms in the stochastic equation for the magnetization are in the original Langevin form (i.e., a systematic slowing down of the rate of change of angular momentum due to friction superimposed on which is a rapidly fluctuating random white noise torque) can be used in all damping ranges. In contrast, neither the Kubo nor the Landau–Lifshitz models can be used for high damping, where they may predict unphysical behavior of the observables (relaxation times, escape rates, etc.). Again in Section IV, the fundamental problem of the effect of an arbitrary orientation of an external d.c. magnetic field $\mathbf{H}_0$ on the reversal time of uniaxial superparamagnets is treated in detail. Moreover, we evaluate the reversal time for cubic, biaxial, and mixed (uniaxial and cubic) anisotropy potentials respectively. We then discuss in Section V, thermal effects in switching field curves and surfaces. The results are presented in a form suitable for comparison with experiment. In order to assess the escape rate formulas used, we compare them with numerical solutions of the Gilbert-Langevin Eq. (15) and Brown's Fokker-Planck Eq. (18).

## II. BROWN'S CONTINUOUS DIFFUSION MODEL OF CLASSICAL SPINS

### A. Basic equations

The starting point of Brown's treatment of the dynamical behavior of the magnetization $\mathbf{M}$ for a single-domain particle is Gilbert's equation,[80] which neglecting thermal agitation due to the random magnetic field produced by the bath-spin interaction in Eq. (15) is

$$\dot{\mathbf{u}} = \mathbf{u} \times (\gamma \mathbf{H} - \alpha \dot{\mathbf{u}}). \tag{22}$$

In general, $\mathbf{H}$ and $-\alpha \dot{\mathbf{u}}/\gamma$ represent the *conservative* and *dissipative* parts of an "effective field", respectively. Brown now supposes that, in the presence of thermal agitation, the dissipative "effective field" $-\alpha \dot{\mathbf{u}}/\gamma$ describes only the statistical average of the rapidly fluctuating random field due to thermal agitation, and that this term must become $-\alpha \dot{\mathbf{u}}/\gamma + \mathbf{h}(t)$, where the random field $\mathbf{h}(t)$ has the white noise properties Eq. (17). Brown was then able to derive, after a long and tedious calculation using the methods of Wang and Uhlenbeck,[105] the Fokker–Planck equation for the density of magnetization orientations $W(\vartheta\,\varphi, t)$ on the sphere of radius $M_S$. This lengthy procedure may be circumvented, however, by using an alternative approach given by Brown which appears to be based on an argument originally due to Einstein[106] in order to heuristically derive the Smoluchowski equation for point particles. Einstein accomplished this by adding a diffusion current representing the effect of the heat bath on the deterministic drift current under an external force.

In order to illustrate this, we first write (cross-multiplying vectorially by **u** and using the triple vector product formula) Gilbert's equation in the absence of thermal agitation (noiseless equation) as an explicit equation for $\dot{\mathbf{u}}$; Transposing the $\alpha$ term, we have

$$\dot{\mathbf{u}} + \alpha(\mathbf{u} \times \dot{\mathbf{u}}) = \gamma(\mathbf{u} \times \mathbf{H}). \tag{23}$$

Cross-multiplying vectorially by **u** in Eq. (23), using the triple vector product formula

$$(\mathbf{u} \times \dot{\mathbf{u}}) \times \mathbf{u} = \dot{\mathbf{u}} - \mathbf{u}(\mathbf{u} \cdot \dot{\mathbf{u}}), \tag{24}$$

we obtain

$$\dot{\mathbf{u}} \times \mathbf{u} = -\alpha\dot{\mathbf{u}} + \gamma(\mathbf{u} \times \mathbf{H}) \times \mathbf{u} \tag{25}$$

because $(\mathbf{u} \cdot \dot{\mathbf{u}}) = 0$. Substituting Eq. (25) into Eq. (23) yields the explicit solution for $\dot{\mathbf{u}}$ in the Landau-Lifshitz form[9]

$$\dot{\mathbf{u}} = \alpha^{-1} h' M_S (\mathbf{u} \times \mathbf{H}) + h' M_S (\mathbf{u} \times \mathbf{H}) \times \mathbf{u}, \tag{26}$$

where $h'$ is Brown's parameter defined as $h' = \gamma / [(\alpha + \alpha^{-1}) M_S]$. With Eq. (16), Eq. (26) becomes

$$\dot{\mathbf{u}} = -\frac{h'}{\alpha} \mathbf{u} \times \frac{\partial V}{\partial \mathbf{u}} + h' \mathbf{u} \times \left( \mathbf{u} \times \frac{\partial V}{\partial \mathbf{u}} \right). \tag{27}$$

Now the instantaneous orientation $(\vartheta, \varphi)$ of the magnetization **M** of a single-domain particle may be represented by a point on the unit sphere (1, $\vartheta$, $\varphi$). As the magnetization changes its direction the representative point moves over the surface of the sphere. Following,[9] consider now a statistical ensemble of identical particles and let $W(\vartheta, \varphi, t) d\Omega$ be the probability that **M** has orientation $(\vartheta, \varphi)$ to within solid angle $d\Omega = \sin\vartheta \, d\vartheta \, d\varphi$. The time derivative of $W(\vartheta, \varphi, t)$ is then related to the probability current **J** of such representative points swarming over the surface $S$ of the sphere by the continuity equation

$$\dot{W} + \text{div}\mathbf{J} = 0. \tag{28}$$

Equation (28) states that the swarming representative points are neither created nor destroyed, merely moving to new positions on the surface of the sphere.[9] Now in the absence of thermal agitation, we have $\mathbf{J} = W\dot{\mathbf{u}}$, where $\dot{\mathbf{u}}$ is given by Eq. (27). Next add to this **J** a diffusion term $-k' \partial_{\mathbf{u}} W$ ($k'$ is a proportionality constant to be determined later), which represents the effect of thermal agitation; its tendency is to smooth out the distribution, i.e., to make it more uniform. Recall the Langevin picture of a systematic retarding torque tending to slow down the spin superimposed on a rapidly fluctuating random torque maintaining the motion. This intuitive procedure essentially due to Einstein gives for the components of **J** (on evaluating $\mathbf{u} \times \partial_{\mathbf{u}} V$, etc. in spherical polar coordinates)

$$\mathbf{J}_\vartheta = -h' \left[ \left( \frac{\partial V}{\partial \vartheta} - \frac{1}{\alpha \sin\vartheta} \frac{\partial V}{\partial \varphi} \right) W + \frac{k'}{h'} \frac{\partial W}{\partial \vartheta} \right], \tag{29}$$

$$\mathbf{J}_\varphi = -h' \left[ \left( \frac{1}{\alpha} \frac{\partial V}{\partial \vartheta} + \frac{1}{\sin \vartheta} \frac{\partial V}{\partial \varphi} \right) W + \frac{k'}{h' \sin \vartheta} \frac{\partial W}{\partial \varphi} \right]. \tag{30}$$

Equations (29) and (30), when substituted into the continuity Eq. (28), now yield Brown's Fokker–Planck equation for the surface density of magnetic moment orientations on the unit sphere, which may be written in the compact vector form of Eq. (18) noting that if the gyromagnetic term is neglected, the equation is a replica of that occurring in the theory of dielectric relaxation of nematic liquid crystals ignoring inertial effects.[58] The constant $k'$ is evaluated by requiring that the Boltzmann distribution $W_0(\vartheta,\varphi) = Ae^{-vV(\vartheta,\varphi)/(kT)}$ of orientations ($A$ is a normalizing constant) should be the stationary (equilibrium) solution of Eq. (18). The imposition of the Boltzmann distribution of orientations yields

$$k' = kTh'/v = (2\tau_N)^{-1} \tag{31}$$

Here we have given Brown's intuitive derivation of his Fokker–Planck equation, Eq. (18). A rigorous derivation of that equation from the Gilbert-Langevin equation (15) is given in Section IV.B below.

Now Brown's Fokker–Planck equation (18) for the probability density function (PDF) $W(\vartheta,\varphi,t)$ of orientations of the unit vector $\mathbf{u}$ in configuration space $(\vartheta,\varphi)$, can be solved by separation of the variables. This gives rise to a Sturm–Liouville problem so that $W(\vartheta,\varphi,t)$ can be written as[107,108]

$$W(\vartheta,\varphi,t) = W_0(\vartheta,\varphi) + \sum_{k=1}^\infty \Phi_k(\vartheta,\varphi) e^{-\lambda_k t}, \tag{32}$$

where $\Phi_k(\vartheta,\varphi)$ and $\lambda_k$ are the eigenfunctions and eigenvalues of the Fokker–Planck operator $\mathrm{L}_{\mathrm{FP}}$ and $W_0(\vartheta,\varphi)$ is the stationary solution of that equation, i.e., $\mathrm{L}_{\mathrm{FP}} W_0 = 0$, corresponding to Boltzmann equilibrium. Then, the reversal time of the magnetization $\tau$ can be estimated[8,9,58] as the inverse of the smallest nonvanishing eigenvalue $\lambda_1$ of the operator $\mathrm{L}_{FP}$ in Eq. (18), i.e.,

$$\tau = 1/\lambda_1. \tag{33}$$

An alternative method involving the observables directly is to expand $W(\vartheta,\varphi,t)$ as a Fourier series of appropriate orthogonal functions forming an orthonormal basis related to them; here these are the spherical harmonics $Y_{l,m}(\vartheta,\varphi)$,[109] viz.,[58]

$$W(\vartheta,\varphi,t) = \sum_{l=0}^\infty \sum_{m=-l}^l c_{l,m}(t) Y_{l,m}^*(\vartheta,\varphi), \tag{34}$$

where $Y_{l,m}(\vartheta,\varphi)$ are defined by

$$Y_{l,m}(\vartheta,\varphi) = \sqrt{\frac{(2l+1)(l-m)!}{4\pi(l+m)!}} e^{im\varphi} P_l^m(\cos\vartheta), \quad Y_{l,-m} = (-1)^m Y_{l,m}^*,$$

$P_l^m(x)$ are the associated Legendre functions defined as[109]

$$P_l^m(\cos\vartheta) = \frac{(-1)^m}{2^l l!}(\sin\vartheta)^m \frac{d^{l+m}}{(d\cos\vartheta)^{l+m}}(\cos^2\vartheta - 1)^l,$$

and the asterisk denotes the complex conjugate while the orthogonality property of the spherical harmonics may be written as[109]

$$\int_0^{2\pi}\int_0^\pi Y_{l_1,m_1}^*(\vartheta,\varphi)Y_{l_2,m_2}(\vartheta,\varphi)\sin\vartheta\, d\vartheta\, d\varphi = \delta_{l_1 l_2}\delta_{m_1 m_2}.$$

Moreover, for *arbitrary* magnetocrystalline anisotropy, which can be expressed in terms of spherical harmonics as

$$\frac{vV}{kT} = \sum_{R=1}^\infty \sum_{S=-R}^R A_{R,S} Y_{R,S}, \tag{35}$$

we have (by assuming a solution in the form of the Fourier expansion Eq. (34) for the Fokker-Planck equation (18)) an infinite hierarchy of differential-recurrence equations for the statistical moments, viz., (details are in Refs. 58 and 110 and in Appendix A)

$$\tau_N \frac{d}{dt}\langle Y_{l,m}\rangle(t) = \sum_{s,r} e_{l,m,l+r,m+s}\langle Y_{l+r,m+s}\rangle(t), \tag{36}$$

where by orthogonality the expectation values of the spherical harmonics are given by

$$\langle Y_{l,m}\rangle(t) = \int_0^{2\pi}\int_0^\pi W(\vartheta,\varphi,t)Y_{l,m}(\vartheta,\varphi)\sin\vartheta\, d\vartheta\, d\varphi. \tag{37}$$

In Eq. (36), $e_{l,m,l',m\pm s}$ are the matrix elements of the Fokker-Planck operator expressed as

$$e_{l,m,l',m\pm s} = -\frac{l(l+1)}{2}\delta_{ll'}\delta_{s0} + (-1)^m \frac{1}{4}\sqrt{\frac{(2l+1)(2l'+1)}{\pi}}$$

$$\times \sum_{r=s}^\infty A_{r,\pm s}\left\{\frac{[l'(l'+1)-r(r+1)-l(l+1)]}{2\sqrt{2r+1}}C_{l,0,l',0}^{r,0}C_{l,m,l',-m\mp s}^{r,\mp s} + \frac{i}{\alpha}\sqrt{\frac{(2r+1)(r-s)!}{(r+s)!}}\right.$$

$$\left.\times \sum_{\substack{L=s-\varepsilon_{r,s},\\ \Delta L=2}}^{r-1}\sqrt{\frac{(L+s)!}{(L-s)!}}C_{l,0,l',0}^{L,0}\left[mC_{l,m,l',-m\mp s}^{L,\mp s} \pm s\sqrt{\frac{(l\mp m)(l\pm m+1)}{(L+s)(L-s+1)}}C_{l,m\pm 1,l',-m\mp s}^{L,\mp s\pm 1}\right]\right\}, \tag{38}$$

where $s \geq 0$ and $C_{l,m,l',m'}^{r,s}$ are the Clebsch–Gordan coefficients various definitions of which are available, e.g., in Ref. 109. (The built-in function Clebsch-Gordan[{a, α},{b, β},{c, γ}] of *Mathematica*® facilitates the calculation of these coefficients.) We remark that Eq. (38) determines the coefficients of the linear combination $e_{l,m,l',m'}$ for *arbitrary* magnetocrystalline anisotropy and Zeeman energy densities.

The Gilbert-Langevin equation, Eq. (15), can also be reduced to the moment system for $\langle Y_{l,m}\rangle(t)$, Eq. (36), by an appropriate transformation of variables and by direct averaging (without recourse to the Fokker–Planck equation) of the stochastic equation thereby obtained[58, 109] (see Appendix A). Examples of explicit calculations of $e_{l,m,l',m'}$ for particular magnetocrystalline anisotropies, are available in Refs. 20, 58, 90, and 110.

## B. Evaluation of the reversal time of the magnetization and other observables

By solving Eq. (36), we can calculate observables such as the reversal time of the magnetization, the dynamic susceptibility, etc.[58] Hence we can compare theoretical predictions with experimental data on superparamagnetic relaxation. Furthermore, the numerical calculation of the statistical moments from Eq. (36) renders benchmark solutions for comparison with results predicted by complementary methods, such as Brownian (Langevin) dynamics simulations[111–115] and the previously mentioned generalization of the Kramers escape-rate theory to magnetic systems.[8,9,56,58,59,82-84,88] In particular, by solving the differential-recurrence equations (36), we have the Cartesian components of the magnetization $\langle M_i\rangle(t)$, $i = X, Y, Z,$ which may be expressed in terms of the averaged spherical harmonics as[58]

$$\langle M_X\rangle(t) = M_S\sqrt{2\pi/3}\left[\langle Y_{1,-1}\rangle(t) - \langle Y_{1,1}\rangle(t)\right],$$

$$\langle M_Y\rangle(t) = iM_S\sqrt{2\pi/3}\left[\langle Y_{1,-1}\rangle(t) + \langle Y_{1,1}\rangle(t)\right],$$

$$\langle M_Z\rangle(t) = M_S\sqrt{4\pi/3}\langle Y_{1,0}\rangle(t). \qquad (39)$$

Furthermore we can evaluate the characteristic times of the magnetization and the equilibrium correlation functions of the longitudinal and transverse components of the magnetization,

$$C_i(t) = \frac{\langle M_i(0)M_i(t)\rangle_0 - \langle M_i(0)\rangle_0^2}{\langle M_i^2(0)\rangle_0 - \langle M_i(0)\rangle_0^2} \qquad (40)$$

and so on. Here the angular brackets designate the equilibrium ensemble average of a dynamical variable $A$ defined as

$$\langle A\rangle_0 = \int_0^{2\pi}\int_0^\pi A(\vartheta,\varphi)W_0(\vartheta,\varphi)\sin\vartheta\,d\vartheta\,d\varphi, \qquad (41)$$

Now, to characterize the overall time behavior of $C_i(t)$, we may formally introduce (see Ref. 58) the integral relaxation time $\tau_{int}^i$, viz.,

$$\tau_{int}^i = \int_0^\infty C_i(t)dt, \qquad (42)$$

which is the area under $C_i(t)$. The time $\tau_{int}^i$ may equivalently be defined using the eigenvalues ($\lambda_k$) of the Fokker–Planck operator from Eq. (18) because (Ref. 58, Chapter 2) $C_i(t)$ may formally be written as

$$C_i(t) = \sum_k c_k^i e^{-\lambda_k t}, \qquad (43)$$

so that, from Eqs. (42) and (43),

$$\tau_{int}^i = \sum_k c_k^i / \lambda_k. \qquad (44)$$

The integral relaxation time $\tau_{int}^i$ contains contributions from *all* the eigenvalues $\lambda_k$. In general, in order to evaluate both $C_i(t)$, and $\tau_{int}^i$ numerically, a knowledge of each $\lambda_k$ and $c_k^i$ is required. However, in the low temperature (high barrier) limit, for the *longitudinal* relaxation of the magnetization, $\lambda_1 \ll |\lambda_k|$ and $c_1^Z \approx 1 \gg c_k^Z$ ($k \neq 1$) provided the wells of the potential remain equivalent or nearly equivalent, the approximation $\tau_{int}^Z \approx 1/\lambda_1$ is valid.[58] In other words, the inverse of the smallest nonvanishing eigenvalue $\lambda_1$ closely approximates the longitudinal correlation time $\tau_{int}^Z$ in the low temperature limit for zero or very weak external fields. Furthermore, in the *longitudinal* relaxation of the magnetization, the smallest nonvanishing eigenvalue(s) $\lambda_1$ of the Fokker–Planck operator characterizes the long-time behavior of

$$\langle M_Z \rangle(t) - \langle M_Z \rangle_0 \sim C_Z(t) \sim e^{-\lambda_1 t} = e^{-t/\tau}, \; t \gg \tau, \qquad (45)$$

and may be associated with the longest relaxation (reversal) time of the magnetization.

In order to evaluate the reversal time $\tau$ numerically, the recurrence Eq. (36) may always be written in matrix form as

$$\dot{\mathbf{X}}(t) = \mathbf{A}\mathbf{X}(t), \qquad (46)$$

where $\mathbf{A}$ is the system matrix and $\mathbf{X}(t)$ is an infinite column vector formed from $c_{l,m}(t) = \langle Y_{l,m} \rangle(t)$. Thus the reversal time of the magnetization $\tau$ may be then determined as the smallest nonvanishing root of the characteristic equation

$$\det(\lambda \mathbf{I} - \mathbf{A}) = 0 \qquad (47)$$

by selecting a sufficiently large number of equations. The general solution of Eq. (46) is determined by successively increasing the size of $\mathbf{A}$ until convergence is attained. Alternatively, we can always transform the moment systems, Eqs. (36), governing the magnetization relaxation into the tri-diagonal vector differential-recurrence equation

$$\tau_N \dot{\mathbf{C}}_n(t) = \mathbf{Q}_n^- \mathbf{C}_{n-1}(t) + \mathbf{Q}_n \mathbf{C}_n(t) + \mathbf{Q}_n^+ \mathbf{C}_{n+1}(t), \qquad (48)$$

where $\mathbf{C}_n(t)$ are the column vectors arranged in an appropriate way from $c_{l,m}(t)$, and $\mathbf{Q}_n^\pm, \mathbf{Q}_n$ are matrices with elements $e_{l',m',l,m}$. As shown in Ref. 116 (see also Ref. 58, Chapter 2), the *exact matrix continued fraction solution* of Eq. (48) for the Laplace transform of $\mathbf{C}_1(t)$ is given by

$$\tilde{\mathbf{C}}_1(s) = \tau_N \boldsymbol{\Delta}_1(s) \left\{ \mathbf{C}_1(0) + \sum_{n=2}^{\infty} \left[ \prod_{k=2}^{n} \mathbf{Q}_{k-1}^+ \boldsymbol{\Delta}_k(s) \right] \mathbf{C}_n(0) \right\}, \tag{49}$$

where $\tilde{\mathbf{C}}_1(s) = \int_0^\infty \mathbf{C}_1(t) e^{-st} dt$, $\boldsymbol{\Delta}_n(s)$ is the matrix continued fraction defined by the recurrence equation

$$\boldsymbol{\Delta}_n(s) = \left[ \tau_N s \mathbf{I} - \mathbf{Q}_n - \mathbf{Q}_n^+ \boldsymbol{\Delta}_{n+1}(s) \mathbf{Q}_{n+1}^- \right]^{-1}, \tag{50}$$

and $\mathbf{I}$ is the unit matrix. Having determined $\tilde{\mathbf{C}}_1(s)$, one may evaluate all the relevant observables. In a similar way, one can also calculate the smallest nonvanishing eigenvalue(s) (yielding the reversal time of the magnetization) from the secular or characteristic equation[117]

$$\det(\lambda \tau_N \mathbf{I} - \mathbf{S}) = 0, \tag{51}$$

where the matrix $\mathbf{S}$ is defined via the matrix continued fractions

$$\mathbf{S} = -\left[ \mathbf{Q}_1 + \mathbf{Q}_1^+ \boldsymbol{\Delta}_2(0) \mathbf{Q}_2^- \right] \left[ \mathbf{I} - \mathbf{Q}_1^+ \boldsymbol{\Delta}_2'(0) \mathbf{Q}_2^- \right]^{-1} \tag{52}$$

and the prime designates the derivative of $\boldsymbol{\Delta}_2(s)$ with respect to $\tau_N s$ (see Ref. 58, Chapter 2, Section 2.11.2). Thus $\lambda_1$ is the smallest nonvanishing eigenvalue of $\mathbf{S}$. The integral relaxation times $\tau_{\text{int}}^i$ can also be calculated using matrix continued fractions via the one-sided Fourier transform of the appropriate correlation function $\tilde{C}_i(-i\omega) = \int_0^\infty C_i(t) e^{i\omega t} dt$ as $\tau_{\text{int}}^i = \tilde{C}_i(0)$. In practical applications, such as to magnetization reversal, matrix continued fractions due to their rapid convergence are much better suited to numerical calculations than standard direct matrix inversion based on the matrix representation, Eq. (46), of the infinite system of linear differential-recurrence relations for the averaged spherical harmonics. We remark that for some cases, e.g., for particles with cubic anisotropy, the long-time overbarrier relaxation processes are due to the *two* slowest relaxation modes with two distinct eigenvalues $\lambda_1$ and $\lambda_2$, which are of the same order of magnitude.[92] Here we may evaluate the reversal time $\tau$ via the one-sided Fourier transform of the longitudinal correlation function $\tilde{C}_Z(-i\omega) = \int_0^\infty C_Z(t) e^{i\omega t} dt$ as follows. First consider the long-time behavior of $C_Z(t)$ which can be approximated at long times by an exponential

$$C_Z(t) \approx C_0 e^{-t/\tau}. \tag{53}$$

It follows that the longest relaxation time $\tau$ can then be extracted from $\tilde{C}_Z(-i\omega)$ (by eliminating $C_0$) as[79]

$$\tau = \lim_{\omega \to 0} \frac{\tilde{C}_Z(0) - \tilde{C}_Z(-i\omega)}{i\omega \tilde{C}_Z(-i\omega)}. \tag{54}$$

Examples of applications of matrix continued fractions to magnetization relaxation in superparamagnets are given in, e.g., Refs. 19, 20, 22–24, 89–92, 95, 119, and 120. Some of these results will be summarized in Section IV. However, for very low damping $\alpha < 0.001$, and/or very high potential barriers, the continued fraction method (as well as the standard matrix method based on Eq. (46)) may not be applicable because the matrices involved become ill-conditioned, meaning that numerical inversions are no longer possible. In such cases, alternative methods (e.g., escape-rate theory) should be used.

## III. REVERSAL TIME IN SUPERPARAMAGNETS WITH AXIALLY SYMMETRIC MAGNETOCRYSTALLINE ANISOTROPY

**A. Formulation of the problem**

At the time Brown was writing (1963), the lack of advanced computing facilities, without which the reversal time $\tau$ cannot be calculated from Eq. (46), compelled him to seek simple analytic formulas for $\tau$ in the high-energy barrier approximation. This was accomplished by utilizing the Kramers escape-rate theory,[54] suitably modified for rotation in space and for a nonseparable Hamiltonian, in the same manner as the Kramers theory had originally been formulated for translational Brownian motion of point particles. In order to estimate the reversal time of the magnetic moment over the internal anisotropy potential barrier, which is the inverse of the smallest nonvanishing eigenvalue of the Fokker–Planck operator in Eq. (18), Brown[8,9] adapted the Kramers method[54] to classical spins. Now as briefly outlined, Kramers' idea, motivated by the fluctuation–dissipation theorem, is to calculate the prefactor $\Lambda$ in the Arrhenius-like equation (8) for the escape rate $\Gamma$ over the potential barrier $\Delta V$ (reaction velocity in the case of chemical reactions). Referring to magnetic relaxation, in his first calculations (see Section III.B below) of approximate expressions for $\Gamma$, Brown confined himself to *axially symmetric potentials* of the magnetocrystalline anisotropy and Zeeman energy.[8] Hence, *no dynamical coupling* between the longitudinal and the transverse modes of motion exists, so that the longitudinal modes are governed by a single state variable, namely the colatitude $\vartheta$, i.e., the polar angle of the magnetization vector **M**. This is so because in an axially symmetric potential $V(\vartheta)$, the Fokker–Planck equation (18) for the distribution function $W(\vartheta,t)$ is effectively a one-space-variable equation

$$2\tau_N \frac{\partial W}{\partial t} = \frac{1}{\sin\vartheta} \frac{\partial}{\partial \vartheta} \left[ \sin\vartheta \left( \frac{\partial W}{\partial \vartheta} + W \frac{\partial V}{\partial \vartheta} \right) \right] \tag{55}$$

(from now on the abbreviation $vV(\vartheta)/(kT) \to V(\vartheta)$ will be used). The second state variable, namely the azimuthal angle $\varphi$ of **M**, gives rise only to a steady precession of that vector. Noting the decoupling between the transverse and longitudinal modes existing for axial symmetry, yielding an *exact* single-variable Fokker–Planck equation in the colatitude $\vartheta$, Brown demonstrated that the Kramers escape-rate theory for point particles may be easily adapted to yield an expression for the escape rate of spins in *axially symmetric* potentials, which is valid for all values of $\alpha$. However, in magnetic applications the Fokker–Planck equation in the single state variable $\vartheta$ does not now arise from strong damping of the angular momentum (which, in the Brownian motion of point particles or rigid inertial rotators gives rise to the *approximate* Fokker–Planck equation in configuration space, known as the Smoluchowski equation); rather, it follows from the *axial symmetry* of the potential.

Before we proceed to the more sophisticated treatment of Brown[8,9] based on the Langevin equation, we shall briefly describe the *discrete orientation model*[52] for the calculation of the Néel relaxation time (this model is described in detail in Appendix D). We shall suppose throughout that the energy barriers are so large in comparison with $kT$ that the magnetization lies always along only one of the directions ($\vartheta_i$, $\varphi_i$) of easy magnetization; nevertheless, the barriers are not so high as to preclude changes of orientation altogether. Thus, in orientation $i$, there is a probability $v_{ij}$ per unit time of a jump to orientation $j$. This probability $v_{ij}$ depends on $K$, $H$, and $kT$. Let us now suppose that we have only two orientations as for a uniaxial anisotropy given by Eq. (2). Let 1 and 2 refer to the positive and negative orientations, respectively. If we have a large number $n$ of identical noninteracting particles, the number of particles $n_i$ in orientation $i$ then changes with time in accordance with the rate equations

$$\dot{n}_1 = -\dot{n}_2 = v_{21}n_2 - v_{12}n_1. \tag{56}$$

Hence, we have the evolution equation[8,9]

$$\frac{d}{dt}(n_2 - n_1) = -(v_{21} + v_{12})(n_2 - n_1) + (v_{12} - v_{21})n,$$

so that $n_1$ and $n_2$ approach their final values when $v_{12}$ and $v_{21}$ are constant according to the factor $e^{-(v_{12}+v_{21})t}$, that is, with time constant

$$\tau = (v_{12} + v_{21})^{-1}. \tag{57}$$

If $v_{ij}^0$ is the frequency of oscillation of a particle in a potential well (called in TST the attempt frequency), the probability per second for the flip of a particle from orientation $i$ to orientation $j$ is given by

$$v_{ij} = v_{ij}^0 e^{-(V_0 - V_i)}, \quad (i = 1, j = 2 \text{ or } i = 2, j = 1), \tag{58}$$

where $V_i$ is the free energy density in orientation $i$, and $V_0$ is the free energy density at the top of the barrier between the orientations $i$ and $j$; $v$ is, as usual, the particle volume. The frequencies $v_{ij}^0$, if they vary with temperature, are assumed to change so slowly in comparison with the exponential factor, and are often taken to be constant, although Néel[7] has calculated them explicitly (see Ref. 121). We reiterate that, regardless of the precise form of $v_{ij}^0$, if the ratio $v/T$ changes by a factor of less than three in a certain critical part of its range, the time constant, Eq. (57), changes from $10^{-1}$ to $10^9$ s. We emphasize that the discrete orientation model of overbarrier relaxation was originally proposed for dielectric relaxation in polar dielectrics.[51,52]

## B. Estimation of the reversal time via Kramers' theory

Now it is instructive to first give the solution for the particular case of axially symmetric potentials, as this transparently illustrates the application of Kramers' theory to the magnetic problem. Here, the escape rate has the interesting particular property that it is valid for all values of the damping parameter $\alpha$. In Kramers' mechanical problem, on the other hand, the governing equation, namely the Klein–Kramers equation, is always an equation in a two-dimensional state space, and can only be converted to a one-dimensional equation in the limiting cases (VLD and IHD). *Thus in magnetic relaxation, the three friction regimes of Kramers' problem, namely VLD, the crossover region, and IHD, will only appear when nonaxially symmetric potentials are involved.*

For an axially symmetric potential $V(\vartheta)$ Eq. (2) with two wells at $\vartheta_1 = 0$ and $\vartheta_2 = \pi$ separated by a potential barrier at $\vartheta_m$, we have $\partial J_\varphi / \partial \varphi = 0$ since $W = W(\vartheta)$. Hence referring to Eq. (29), and recalling that, in the quasi-stationary case $\dot{W} = 0$, the total current over the barrier $J = 2\pi J_\vartheta \sin\vartheta$ is constant. Thus, with Eq. (29) we obtain

$$\frac{\partial W}{\partial \vartheta} + \frac{\partial V}{\partial \vartheta}W = e^{-V}\frac{\partial}{\partial \vartheta}\left(e^V W\right) = -\frac{\tau_N J}{\pi \sin \vartheta},$$

and so

$$e^{V(\vartheta)}W(\vartheta) = \frac{\tau_N J}{\pi}\int_\vartheta^{\vartheta_0} \frac{e^{V(\vartheta')}}{\sin \vartheta'}d\vartheta'. \qquad (59)$$

Suppose now that $W$ vanishes at the barrier angle $\vartheta = \vartheta_0$ (i.e., particles which arrive at this boundary are no longer counted), so that $W(\vartheta_0) = 0$, i.e., all the particles are absorbed. Then

$$W(\vartheta) = \frac{\tau_N J}{\pi}e^{-V(\vartheta)}\int_\vartheta^{\vartheta_0}\frac{e^{V(\vartheta')}}{\sin \vartheta'}d\vartheta' \qquad (60)$$

and the number of particles $N_i$ in the well $i$ is

$$N_i = 2\pi \int_{\vartheta_i}^{\vartheta_0} W \sin\vartheta\, d\vartheta = 2\tau_N J \int_{\vartheta_i}^{\vartheta_0} e^{-V(\vartheta)} \sin\vartheta \int_{\vartheta}^{\vartheta_0} \frac{e^{V(\vartheta')}}{\sin\vartheta'} d\vartheta' d\vartheta. \tag{61}$$

Thus, the characteristic escape (mean first-passage) time $\tau(\vartheta_i)$ from the well $i$ is, via the flux-over-population method,[53,122]

$$\tau(\vartheta_i) = \frac{N_i}{J} = 2\tau_N \int_{\vartheta_i}^{\vartheta_0} e^{-V(\vartheta)} \sin\vartheta \int_{\vartheta}^{\vartheta_0} \frac{e^{V(\vartheta')}}{\sin\vartheta'} d\vartheta' d\vartheta.$$

On integrating by parts, we obtain

$$\tau(\vartheta_i) = 2\tau_N \int_{\vartheta_i}^{\vartheta_0} \frac{e^{V(\vartheta')}}{\sin\vartheta'} \int_{\vartheta_i}^{\vartheta'} e^{-V(\vartheta)} \sin\vartheta\, d\vartheta\, d\vartheta'. \tag{62}$$

This is the time to reach the top of the barrier, provided that all particles reaching the top are absorbed there, which is the boundary condition that $W$ vanishes at $\vartheta = \vartheta_0$. Equation (62) can also be derived using the mean first-passage time (MFPT) approach [53,122] by solving the equation

$$L_{FP}^\dagger \tau(\vartheta) = \frac{1}{2\tau_N \sin\vartheta} \frac{\partial}{\partial\vartheta}\left[e^V \sin\vartheta \frac{\partial}{\partial\vartheta} e^{-V} \tau(\vartheta)\right] = -1$$

for $\tau(\vartheta)$ with appropriate boundary conditions; here $L_{FP}^\dagger$ is the adjoint Fokker–Planck operator.[53,56]

In practice, a particle has a 50:50 chance of crossing the barrier top, which means that the corresponding Kramers escape rate $\Gamma_i$ from the well $i$ is given by

$$\Gamma_i \approx [2\tau(\vartheta_i)]^{-1}. \tag{63}$$

In the limit of very high potential barriers, the integrals in Eq. (62) may be approximately evaluated using steepest descents[58,122] as follows. We have for the exact time to go from the well at $\vartheta_1 = 0$ to the top of the barrier at $\vartheta = \vartheta_0$

$$\tau(0) = 2\tau_N \int_0^{\vartheta_0} \frac{e^{V(\vartheta')}}{\sin\vartheta'} \int_0^{\vartheta'} e^{-V(\vartheta)} \sin\vartheta\, d\vartheta\, d\vartheta'. \tag{64}$$

Since almost all the particles (i.e., the population) are situated near the minimum at $\vartheta_1 = 0$, then $\vartheta$ is a very small angle. The well (inner) integral in Eq. (64) may then be evaluated using steepest descents, yielding the well population as

$$\int_{\text{near } \vartheta_i} e^{-V(\vartheta)} \sin\vartheta\, d\vartheta \approx \int_0^\infty \vartheta e^{-[V(0)+V_{\vartheta\vartheta}''(0)\vartheta^2/2]} d\vartheta \sim \frac{e^{-V(0)}}{V_{\vartheta\vartheta}''(0)}. \tag{65}$$

The integral may be extended to infinity without significant error, since the particles are almost all concentrated at the bottom of the well. Likewise, near the barrier $\vartheta_0$, the Taylor series in $V(\vartheta)$ can be approximated by its first two nonvanishing terms

$$V(\vartheta) \approx V(\vartheta_0) - |V_{\vartheta\vartheta}''(\vartheta_0)|(\vartheta - \vartheta_0)^2/2.$$

Hence, we have for the outer integral

$$\int_{\text{near }\vartheta_0} \frac{e^{V(\vartheta')}}{\sin \vartheta'} d\vartheta' \approx \frac{e^{V(\vartheta_0)}}{\sin \vartheta_0} \int_{-\infty}^{\vartheta_0} e^{-|V''_{\vartheta\vartheta}(\vartheta_0)|[\vartheta'-\vartheta_0]^2/2} d\vartheta' \sim \frac{\sqrt{\pi} e^{-V(\vartheta_0)}}{\sin \vartheta_0 \sqrt{2|V''_{\vartheta\vartheta}(\vartheta_0)|}} \qquad (66)$$

(here the range of integration in Eq. (66) may be extended to $-\infty$ since the integral has its main contribution from values near to $\vartheta_0$ and almost no contribution lying outside these values). Hence, in the *high-barrier limit*, the mean first-passage time $\tau(0)$ for transitions from the minimum at $\vartheta = 0$ is

$$\tau(0) \sim \frac{\tau_N}{V''_{\vartheta\vartheta}(0)} \frac{\sqrt{2\pi}}{\sqrt{|V''_{\vartheta\vartheta}(\vartheta_0)|}} \frac{e^{V(\vartheta_0)-V(0)}}{\sin \vartheta_0}. \qquad (67)$$

Likewise, the time to go from the minimum at $\vartheta_2 = \pi$ to $\vartheta_0$ is

$$\tau(\pi) = 2\tau_N \int_{\vartheta_0}^{\pi} \frac{e^{V(\vartheta')}}{\sin \vartheta'} d\vartheta' \int_{\vartheta'}^{\pi} e^{-V} \sin \vartheta d\vartheta, \qquad (68)$$

which can be estimated in the high-barrier approximation as

$$\tau(\pi) \sim \frac{\tau_N}{V''_{\vartheta\vartheta}(\pi)} \frac{\sqrt{2\pi}}{\sqrt{|V''_{\vartheta\vartheta}(\vartheta_0)|}} \frac{e^{V(\vartheta_0)-V(\pi)}}{\sin \vartheta_0}. \qquad (69)$$

These are the times to reach the barrier from the depths of the well. According to Eq. (57), the corresponding reversal time of the magnetization $\tau$ is, in the high-barrier limit, given by

$$\tau \approx \frac{1}{\Gamma_1 + \Gamma_2} = \left(\frac{1}{2\tau(0)} + \frac{1}{2\tau(\pi)}\right)^{-1} = \frac{2\tau(0)\tau(\pi)}{\tau(0) + \tau(\pi)}. \qquad (70)$$

The above method may be used for an arbitrary axially symmetric potential. Furthermore, the results we have just given may also be derived from a variational principle,[8] namely, the method of approximate minimization from the definition of longest relaxation time as the inverse of the smallest nonvanishing eigenvalue of the Fokker-Planck operator in Eq. (55) when posed as a Sturm-Liouville problem. To paraphrase Brown, the minimization method has the advantage that it justifies on the basis of a purely mathematical approximation simplifications, which have to be injected arbitrarily into the Kramers calculation.

**C. Uniaxial superparamagnet subjected to a d.c. bias field parallel to the easy axis**

The particular axially symmetric potential relevant in superparamagnetism is when a d.c. bias field $\mathbf{H}_0$ is superimposed on the uniaxial anisotropy potential, Eq. (71) below. In general, the field can be applied at *some angle* to the easy axis of the magnetization (this case will be treated in detail in Section IV.C). However, in order to preserve axial symmetry and the consequent mathematical simplifications, it is often assumed that the field is applied along the easy axis, so that the potential $V$ becomes (Fig. 5)

$$V(\vartheta) = \sigma \sin^2 \vartheta - \xi \cos \vartheta = \sigma(\sin^2 \vartheta - 2h\cos \vartheta), \quad (71)$$

where $h = \xi/(2\sigma) = M_S H_0/(2K)$. The potential in Eq. (71) was used by Néel,[7] who gave an expression for the reversal time $\tau$ of the magnetization using the discrete orientation approximation (see Appendix D). The potential was further studied by Brown[8,9] who obtained approximate expressions for $\tau$ in the limit of large and small $\sigma$ using the Kramers escape rate method and perturbation theory, respectively. Later, $\tau$ was calculated numerically by Aharoni.[118] Coffey et al.[123] have studied the effect of a dc magnetic field on the reversal time $\tau$, the integral relaxation time $\tau_{\text{int}}^{\parallel}$, and the dynamic susceptibility. Klik and Yao[124] presented a detailed study of the eigenvalue spectrum of Brown's Fokker–Planck equation. Other aspects of the magnetization kinetics of uniaxial particles in the presence of an external d.c. field have been treated, e.g., in Refs. 13, 58, 125, and 126.

For the axially symmetric potential Eq. (71), the mean first-passage time, Eq. (70), yields for arbitrary barrier heights and $h < 1$ [58,122]

$$\tau = \frac{2\tau(0)\tau(\pi)}{\tau(0) + \tau(\pi)}, \quad (72)$$

where

$$\tau(0) = 2\tau_N \int_{-h}^{1} \frac{e^{-\sigma(z^2+2hz)}}{1-z^2} \int_{z}^{1} e^{\sigma(z'^2+2hz')} dz' dz = \tau_N \sqrt{\frac{\pi}{\sigma}} \int_{-h}^{1} \frac{e^{-\sigma(z+h)^2} \left\{ \text{erfi}\left[\sqrt{\sigma}(1+h)\right] - \text{erfi}\left[\sqrt{\sigma}(z+h)\right] \right\}}{1-z^2} dz,$$

$$\tau(\pi) = 2\tau_N \int_{-1}^{-h} \frac{e^{-\sigma(z^2+2hz)}}{1-z^2} \int_{-1}^{z} e^{\sigma(z'^2+2hz')} dz' dz = \tau_N \sqrt{\frac{\pi}{\sigma}} \int_{-1}^{-h} \frac{e^{-\sigma(z+h)^2} \left\{ \text{erfi}\left[\sqrt{\sigma}(1-h)\right] + \text{erfi}\left[\sqrt{\sigma}(z+h)\right] \right\}}{1-z^2} dz,$$

and

$$\text{erfi}(z) = \frac{2}{\sqrt{\pi}} \int_0^z e^{t^2} dt$$

is the error function of imaginary argument. Furthermore, we have in the high barrier approximation, $\sigma(1-h)^2 \gg 1$, from Eqs. (67)–(70) (details in Refs. 58 and 122),

$$\tau \sim \frac{\tau_N \sqrt{\pi}}{\sigma^{3/2}(1-h^2)} \left[(1+h)e^{-\sigma(1+h)^2} + (1-h)e^{-\sigma(1-h)^2}\right]^{-1}, \quad (73)$$

which in the limit $h \to 0$, reduces to

$$\tau = \frac{\tau_N \sqrt{\pi} e^{\sigma}}{2\sigma^{3/2}}. \quad (74)$$

Equation (74) is Brown's well known asymptotic formula[8] for the reversal time of the magnetization for uniaxial superparamagnets. Figure 6 indicates that Eq. (72) provides a good approximation for the reversal time $\tau \approx \lambda_1^{-1}$ for any barrier height, while Eq. (73) allows one to estimate $\tau$ for $\sigma \geq 3$.

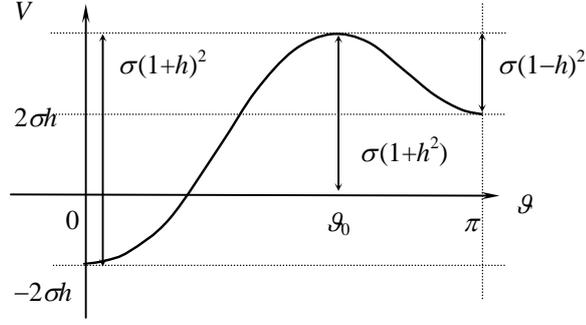

FIG. 5. The profile of the uniaxial potential $V = \sigma(\sin^2\vartheta - 2h\cos\vartheta)$, showing a maximum at $\vartheta_0 = \arccos(-h)$ and minima at $\vartheta = 0$ and $\pi$. Particles in the shallower well are inhibited from crossing into the deeper well by the potential barrier of height $\sigma(1-h)^2$. However, the particles populating the deeper of the two wells must possess much greater thermal energy to be able to cross into the shallower well, owing to the elevated potential barrier height $\sigma(1+h)^2$.

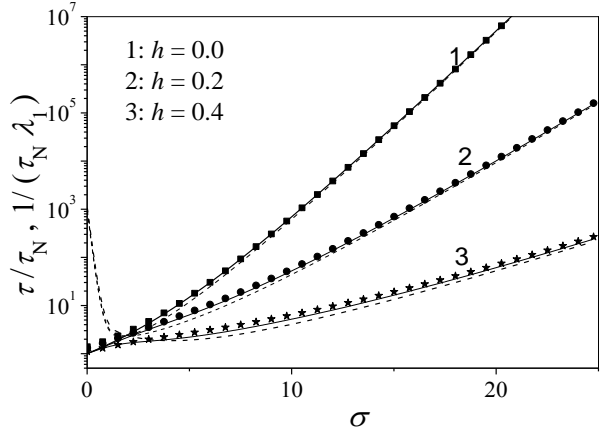

FIG. 6. Reversal time of the magnetization $\tau$ vs. $\sigma$ (inverse temperature parameter) for $h = 0, 0.2$, and $0.4$. Solid lines: numerical calculation of the inverse of the smallest nonvanishing eigenvalue of the Fokker–Planck operator $\lambda_1$.[58] Dashed lines: Brown's asymptotic equation, Eq. (73). Symbols: the MFPT equation, Eq. (72).

For the uniaxial potential $V$ given by Eq. (71), the dynamics of the system are described by the *single-variable* Fokker–Planck equation, Eq. (55), so that the integral relaxation time $\tau_{\text{int}}^{\parallel}$ may also be expressed in closed-integral form as[126]

$$\tau_{\text{int}}^{\parallel} = \frac{2\tau_N / Z}{\langle \cos^2\vartheta \rangle_0 - \langle \cos\vartheta \rangle_0^2} \int_{-1}^{1} \left[ \int_{-1}^{z} (z' - \langle \cos\vartheta \rangle_0) e^{\sigma z'^2 + \xi z'} dz' \right]^2 \frac{e^{-\sigma z^2 - \xi z}}{1 - z^2} dz, \qquad (75)$$

where

$$\langle \cos\vartheta \rangle_0 = \frac{1}{Z} \int_{-1}^{1} x e^{\sigma x^2 + \xi x} dx = \frac{e^\sigma \sinh(2\sigma h)}{\sigma Z} - h,$$

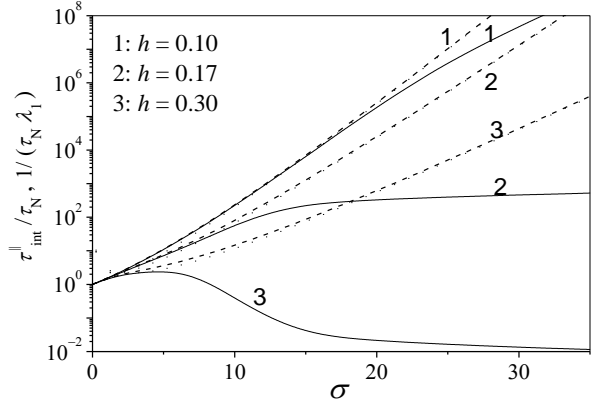

FIG. 7. $\tau_{int}^{\|}$ (solid lines), and $\lambda_1^{-1}$ (dashed lines) vs. $\sigma$ for various values of $h$.

$$\langle \cos^2 \vartheta \rangle_0 = \frac{1}{Z} \int_{-1}^{1} x^2 e^{\sigma x^2 + \xi x} dx = \frac{e^\sigma [\cosh(2\sigma h) - h \sinh(2\sigma h)]}{\sigma Z} + h^2 - \frac{1}{2\sigma},$$

$$Z = \int_{-1}^{1} e^{\sigma x^2 + \xi x} dx = \frac{1}{2} \sqrt{\frac{\pi}{\sigma}} e^{-\sigma h^2} \left\{ \mathrm{erfi}[(1+h)\sqrt{\sigma}] + \mathrm{erfi}[(1-h)\sqrt{\sigma}] \right\}.$$

We compare the two time constants $\lambda_1^{-1}$ and $\tau_{int}^{\|}$, in Fig. 7 for different $h$. Here, the most interesting effect is the behavior of the integral relaxation time $\tau_{int}^{\|}$ as a function of the barrier-height parameter $\sigma$ for sufficiently large bias parameter $h$. When $h$ exceeds a certain critical value $h_c \approx 1/6$, $\tau_{int}^{\|}$ no longer has an activation character at large $\sigma$ (solid curve 2 in Fig. 7). At this critical value of $h$, the relaxation switches from being dominated by the slowest overbarrier mode to being dominated by the fast intrawell modes. Thus, $\tau_{int}^{\|}$ decreases as the height of the potential barrier increases. This effect was discovered numerically by Coffey et al.[123] and later explained in analytic fashion by Garanin.[125]

In applications to superparamagnets, the uniaxial potential

$$V(\vartheta) = \sigma \sin^2 \vartheta. \tag{76}$$

i.e., the particular case $h = 0$ of Eq. (71), is the most frequently used approximation. Various aspects of the magnetization relaxation in uniaxial superparamagnets have been treated using this simple symmetric double-well potential; for example, see Refs. 8, 9, 14, 15, 58, and 127. In particular, Coffey et al.[127] have evaluated the reversal time $\tau$, the integral relaxation time $\tau_{int}^{\|}$, and the dynamic susceptibility for uniaxial superparamagnets. The uniaxial potential, Eq.(76), has also been used to analyze nonlinear magnetic susceptibilities,[17, 18, 21, 25] stochastic resonance,[37–39] dynamic hysteresis,[29, 32, 128, 129] Mössbauer spectra,[46–50] and other related parameters of fine particle systems. Here we briefly summarize the principal findings.

For $h=0$, Eq. (75) for $\tau_{int}^{\|}$ can be considerably simplified and is given by[127]

$$\tau_{\text{int}}^{\parallel} = \frac{3\tau_{N}e^{\sigma}}{\sigma^{2}M(3/2,5/2,\sigma)} \int_{0}^{1} \frac{\cosh[\sigma(1-z^{2})]-1}{1-z^{2}} dz, \quad (77)$$

In the high-barrier limit, $\sigma \gg 1$, $\tau_{\text{int}}^{\parallel}$ has asymptotic behavior[58]

$$\tau_{\text{int}}^{\parallel} \approx \tau_{N} \frac{\sqrt{\pi}}{2} e^{\sigma} \sigma^{-3/2} \left(1 + \frac{1}{\sigma} + \frac{3}{2\sigma^{2}} + \cdots\right), \quad (78)$$

The inverse of $\lambda_{1}$, yielding the reversal time of the magnetization $\tau$, is given by (in the low-temperature limit),[8,127] viz.,

$$\tau = \frac{1}{\lambda_{1}} \approx \tau_{N} \frac{\sqrt{\pi}}{2} e^{\sigma} \sigma^{-3/2} \left(1 + \frac{1}{\sigma} + \frac{7}{4\sigma^{2}} + \cdots\right). \quad (79)$$

In addition, for practical calculations of both $\tau_{\text{int}}^{\parallel}$ and $\lambda_{1}$, one may use the empirical equation[130]

$$\lambda_{1} \approx \frac{1}{\tau_{\text{int}}^{\parallel}} \approx \frac{\sigma}{\tau_{N}(e^{\sigma}-1)} \left(2^{-\sigma} + \frac{2\sigma^{3/2}}{\sqrt{\pi}(1+\sigma)}\right) \quad (80)$$

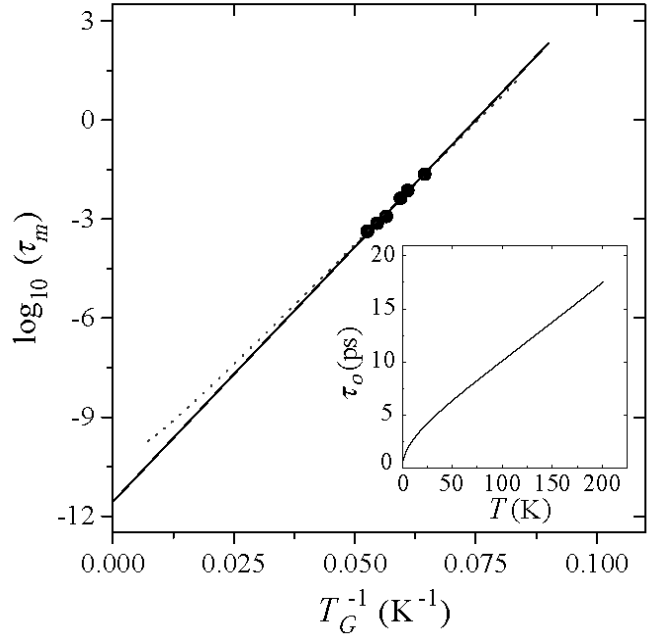

FIG. 8. The reversal time of the magnetization of noninteracting cobalt nanoparticles dispersed in a polymer $\tau_{m} = \tau_{o}e^{E_{B}/(kT)}$ vs. the inverse of the blocking temperature $T_{G}$. The solid line is the adjustment of the experimental results considering a Néel-Brown relaxation process with a constant $\tau_{o}$ value equal to $4 \cdot 10^{-12}$ s. The dashed line presents the theoretical variation of the relaxation time considering the thermal dependence of $\tau_{o}(T)$ from Eq. (79). The inset shows the theoretical thermal dependence of the pre-exponential relaxation time $\tau_{o}(T)$ calculated from Eq. (79). Reprinted from M. Respaud, M. Goiran, J.M. Broto, F. Lionti, L. Thomas, B. Barbara, T. Ould Ely, C. Amiens; and B. Chaudret, Dynamical properties of noninteracting Co nanoparticles, Europhys. Lett. **47**, 122 (1999).[21] Copyright (1999) by EDP Sciences.

which is valid for all $\sigma$. A comparison of Brown's asymptotic formulas for uniaxial supermagnets with the experimental reversal time of the magnetization of noninteracting cobalt nanoparticles dispersed in a polymer is illustrated by Fig. 8. Other examples of the use of these formulas are available in, e.g., Refs. 17, 18, 21, and 25.

We now consider nonaxially symmetric problems, so that the various cases (IHD, VLD, etc.) of the Kramers calculation will appear.

## IV  REVERSAL TIME OF THE MAGNETIZATION IN SUPERPARAMAGNETS WITH NONAXIALLY SYMMETRIC ANISOTROPY

### A. Escape rate equations

We recall that Kramers[54] obtained his IHD and VLD formulas for the escape rate for *point* particles. Moreover, he stated that he could not find a general method of attack in order to obtain a formula which would be valid for any damping regime. Much later this Kramers turnover problem was solved by Mel'nikov and Meshkov,[55, 60] Grabert,[62] and Pollak *et al.*[63] They obtained the escape rate for point particles, which of course also applies to rigid mechanical rotators,[75-79] in the whole damping range by expressing the energy distribution function in the separatrix region in the underdamped regime (extending from zero to intermediate damping) at a given action as an integral equation, which may be posed as a Wiener–Hopf equation. Effectively, the solution of this equation as obtained by the Wiener-Hopf method, *simply* multiplied by the IHD escape rate, yields an integral formula for the relaxation time which is valid for all values of the damping, and constitutes a solution of the Kramers turnover problem for point particles.

We also recall that the analogous magnetic-spin problem, as formulated by Brown,[8, 9] differs fundamentally from that of mechanical particles with separable and additive Hamiltonians. First, the magnetic system has two degrees of freedom, namely the polar $\vartheta$ and azimuthal $\varphi$ angles; second, the undamped equation of motion of the magnetization of a single-domain ferromagnetic particle is the gyromagnetic equation; third, the Hamiltonian is not separable; fourth, inertial effects play no role. Nevertheless, the role of inertia in the mechanical system is essentially mimicked in the magnetic system for nonaxially symmetric potentials by the gyromagnetic term, which causes *coupling* or *entanglement* of the transverse (which give rise to ferromagnetic resonance) and longitudinal modes. Thus, in order to derive escape-rate formulas for superparamagnetic particles equivalent to those for mechanical particles, one must introduce in Brown's Fokker–Planck equation a nonaxially symmetric free energy density $V(\vartheta,\varphi)$, where explicit coupling between the two degrees of freedom now exists. Thus both regimes of damping (IHD and VLD) can occur, reflecting the fact that the dynamics of the transverse response affect the dynamics of the longitudinal response, and vice versa.

As we saw, IHD formulas for nonaxially symmetric potentials were first derived by Smith and de Rozario[82] and Brown.[9] In evaluating the escape rate in the IHD range, it is supposed that the free energy per unit volume $V(\mathbf{M})$ of a single-domain particle has a multistable structure with minima at $\mathbf{n}_i$ and $\mathbf{n}_j$ separated by a potential barrier with a saddle point at $\mathbf{n}_0$. If $\mathbf{M}$ is close to a stationary point $\mathbf{n}_p$ ($p = 0, i, j$) and $(u_1^{(p)}, u_2^{(p)}, u_3^{(p)})$ denote the direction cosines of $\mathbf{M}$, then $V(\mathbf{M})$ can be approximated to second order in $u_1^{(p)}$ and $u_2^{(p)}$ as

$$V = V_p + \frac{1}{2}\left[c_1^{(p)}\left(u_1^{(p)}\right)^2 + c_2^{(p)}\left(u_2^{(p)}\right)^2\right]. \tag{81}$$

To determine $c_1^{(p)}$, $c_2^{(p)}$, and $V_p$, we recall that the transformation matrix $\mathbf{R}^{(p)}$ relating the basic polar coordinate system $P$ and a new polar coordinate system $P'$ with the origin at the stationary point $\mathbf{n}_p$, is defined as[59]

$$\mathbf{R}^{(p)} = \begin{pmatrix} \cos\varphi_p \cos\vartheta_p & \sin\varphi_p \cos\vartheta_p & -\sin\vartheta_p \\ -\sin\varphi_p & \cos\varphi_p & 0 \\ \cos\varphi_p \sin\vartheta_p & \sin\varphi_p \sin\vartheta_p & \cos\vartheta_p \end{pmatrix},$$

so that the relationship between the direction cosines $u_n^{(p)}$ and $u_m'^{(p)}$ in systems $P$ and $P'$ is given by

$$u_n^{(p)} = R_{1n}^{(p)} u_1'^{(p)} + R_{2n}^{(p)} u_2'^{(p)} + R_{3n}^{(p)} u_3'^{(p)} \tag{82}$$

($n = 1, 2, 3$). Because

$$u_3'^{(p)} = (1 - u_1'^{(p)2} - u_2'^{(p)2})^{1/2} \approx 1 - (u_1'^{(p)2} + u_2'^{(p)2})/2,$$

$c_1^{(p)}$, $c_2^{(p)}$, and $V_p$ can be evaluated from Eqs. (81) and (82) as

$$V_p = V_p(u_1^{(p)}, u_2^{(p)})\Big|_{u_1'^{(p)}, u_2'^{(p)}=0}, \quad c_1^{(p)} = \frac{\partial^2 V}{\partial u_1'^{(p)2}}\Big|_{u_1'^{(p)}, u_2'^{(p)}=0}, \quad c_2^{(p)} = \frac{\partial^2 V}{\partial u_2'^{(p)2}}\Big|_{u_1'^{(p)}, u_2'^{(p)}=0}. \tag{83}$$

Substituting Eq. (81) into the Fokker–Planck Eq. (18), the latter may be solved in the saddle point region (which has the shape of a hyperbolic paraboloid, while the well has the shape of an elliptic paraboloid) yielding the escape rate $\Gamma_i^{\text{IHD}}$ from the well $i$ over a saddle point 0 as[9,59]

$$\Gamma_i^{\text{IHD}} = \Gamma_i^{\text{TST}} \frac{\Omega_0(\alpha)}{\omega_0}, \tag{84}$$

with $\Gamma_i^{\text{TST}}$ as the escape rate from the well $i$ for TST as applied to classical spins,[7] namely

$$\Gamma_i^{\text{TST}} = \frac{\omega_i}{2\pi} e^{-\Delta V_i}, \tag{85}$$

where $\Delta V_i = V_0 - V_i$ is the dimensionless barrier height,

$$\omega_i = \frac{\gamma kT}{vM_S}\sqrt{c_1^{(i)} c_2^{(i)}} \quad \text{and} \quad \omega_0 = \frac{\gamma kT}{vM_S}\sqrt{-c_1^{(0)} c_2^{(0)}} \tag{86}$$

are the well and saddle angular frequencies, respectively,

$$\Omega_0(\alpha) = \frac{1}{4\tau_0(\alpha+\alpha^{-1})}\left[-c_1^{(0)} - c_2^{(0)} + \sqrt{(c_2^{(0)} - c_1^{(0)})^2 - 4\alpha^{-2}c_1^{(0)}c_2^{(0)}}\right] \quad (87)$$

is the damped saddle angular frequency, and $\tau_0$ is the normalizing damping-independent time defined as

$$\tau_0 = \frac{\tau_N}{\alpha+\alpha^{-1}} = \frac{vM_S}{2kT\gamma}. \quad (88)$$

Equation (84) as recognized in Ref. 83 is simply a special case of Langer's extension[86] (Appendixes B and C) of the IHD Kramers escape rate to many degrees of freedom and nonseparable Hamiltonians. Using Langer's method,[86] $\Gamma_i^{IHD}$ can be estimated from his expression, viz., (Appendixes C)

$$\Gamma_i^{IHD} = \frac{\Omega_0}{2\pi}\frac{Z_0}{Z_i}, \quad (89)$$

where $Z_i \approx \int\int_{well} e^{-V(\vartheta,\varphi)}\sin\vartheta\,d\vartheta\,d\varphi$ and $Z_0 \approx \int\int_{saddle} e^{-V(\vartheta,\varphi)}\sin\vartheta\,d\vartheta\,d\varphi$ are the well and saddle partition functions, respectively.

Just as with the Brown IHD equation, Eq. (84), the IHD escape rate, Eq. (89), is only valid in the IHD region and so it *cannot be used* to estimate the reversal time for low damping. Indeed for vanishing damping, $\alpha \to 0$, the IHD escape rate $\Gamma_i^{IHD}$ from Eq. (84) reduces to the TST escape rate $\Gamma_i^{TST}$, Eq. (85), which is obviously independent of $\alpha$ and just as with point particles yields the upper bound for the escape rate or lower bound for the relaxation time. However, by analogy with the almost Newtonian VLD dynamics of point particles, this is not the *true* VLD limit or *energy-controlled diffusion*,[2] where the energy loss per cycle of the almost-periodic noisy motion of the magnetization on the saddle-point energy (escape) trajectory is much less than the thermal energy, as noted by Klik and Gunther.[83] Instead, it comprises the *intermediate damping (ID) limit* corresponding to the TST result. Recognizing this, Klik and Gunther derived the *correct* magnetization escape rate[83] in the VLD range, where the dynamics are almost determined by the pure gyromagnetic (Larmor) equation, viz.,

$$\Gamma_i^{VLD} = \alpha S_i \Gamma_i^{TST}, \quad (90)$$

where $S_i$ is the dimensionless action at the saddle-point energy given by

$$S_i = \frac{v}{kT}\oint_{V=V_0}\left(1-p^2\right)\frac{\partial V}{\partial p}d\varphi - \frac{1}{1-p^2}\frac{\partial V}{\partial \varphi}dp, \quad (91)$$

---

[2] A detailed treatment of the VLD limit via the *energy-controlled diffusion equation* for classical spins is given in W.T. Coffey *et al.*, Phys. Rev. B **89**, 054408 (2014).

and $p = \cos\vartheta$. The contour integral in Eq. (91) is taken along the critical energy trajectory, or separatrix, on which the magnetization may reverse by passing through the saddle point(s) of the energy $V_0$. The conditions of applicability of these IHD and VLD solutions for classical spin systems are defined by $\alpha S_i \gtrsim 1$ and $\alpha S_i \ll 1$, respectively. However, experimental values of $\alpha$ usually lie in the turnover region characterized by $\alpha S_i \sim 1$ prompting Coffey et al.[56, 88] to derive a turnover formula for classical spin systems. Thus, they obtained[88] for the escape rate $\Gamma_i$ from a single well

$$\Gamma_i = A(\alpha S_i)\Gamma_i^{IHD} = A(\alpha S_i)\frac{\Omega_0(\alpha)}{\omega_0}\Gamma_i^{TST}, \qquad (92)$$

where $A$ is the magnetization depopulation factor given by Eq. (14). Equation (92) may be deemed universal, insofar as it accurately describes the magnetization escape rate for all damping $\alpha$. The asymptotic behavior of $A(\alpha S_i)$ from Eq. (92) as a function of $\alpha$, namely,

$$A(\alpha S_i) \to 1 \text{ as } \alpha S_i \to \infty \text{ and } A(\alpha S_i) \to \alpha S_i \text{ as } \alpha S_i \to 0, \qquad (93)$$

ensures that the IHD and VLD limits of the magnetization escape rate, i.e., Eqs. (84) and (90), respectively, are reproduced correctly.

The range of validity as a function of damping $\alpha$ of presently available asymptotic formulas for magnetization escape rates is summarized in Table 1 and Fig. 9. In practical applications, the conditions of validity of these formulas, namely, that they only hold in the low-temperature (high-barrier) limit and for the elliptic and hyperbolic paraboloid approximations to the free energy in the vicinity of the stationary points, must be taken into account.

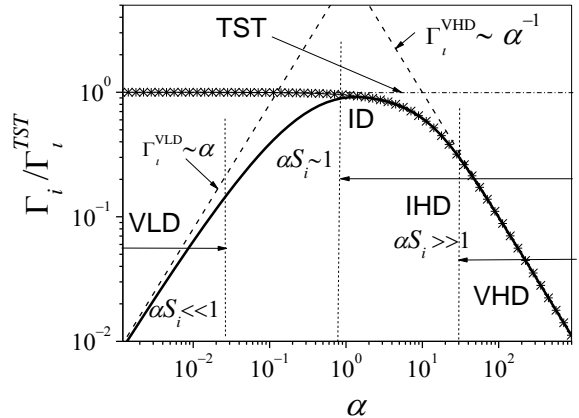

FIG. 9. Qualitative behavior of the normalized escape rate $\Gamma_i / \Gamma_i^{TST}$ vs. damping $\alpha$ (solid line) for classical spins . Three regions exist just as the Kramers theory for particles (cf. Fig. 2), namely, VLD, intermediate damping (ID) (TST), and VHD, and two crossovers between them . Dashed lines: the VHD and VLD asymptotes, Eqs. (89) at $\alpha \gg 1$ and (90). Dashed-dotted line: TST. Asterisks: Brown's IHD formula Eq. (89). Solid line: numerical solution of the Fokker-Planck Eq. (18) and the turnover Eq. (92).

**Table 1.** Range of validity of the asymptotic formulas for the escape rate $\Gamma_i$.

| Escape rate | $\Gamma_i^{\text{TST}}$, Eq. (85) | $\Gamma_i^{\text{IHD}}$, Eq. (84) | $\Gamma_i^{\text{VLD}}$, Eq. (90) | $\Gamma_i$, Eq. (92) |
|---|---|---|---|---|
| Range of validity | $\alpha S_i \sim 1$ | $\alpha S_i \gtrsim 1$ | $\alpha S_i \ll 1$ | all $\alpha$ |
| References | 7 | 9, 82 | 83 | 88 |

We emphasize that throughout the derivation of the above formulas, it is assumed that the potential is *nonaxially symmetric*. If the departures from axial symmetry become small, the nonaxially symmetric asymptotic formulas for the escape rate may be smoothly connected to the axially symmetric results via suitable interpolating integrals. This procedure is described in Refs. 56 and 131 for the particular case of a uniform field transversally applied to the easy axis of the magnetization, for a particle with uniaxial anisotropy. Yet another method of treating uniaxial crossover is to evaluate[132] the integrals with respect to the azimuthal angle $\varphi$ at the saddle point, analytically in terms of the error function, without using steepest descents. This method also smoothly connects axially symmetric and nonaxially symmetric IHD asymptotes for the relaxation time.

We also emphasize the difference between the (overall) reversal time of the magnetization $\tau$ and the individual inverse escape rates $\Gamma_i$, which may differ considerably from each other. In general, both depend on the energy-scape as well as the damping regime. In the IHD damping range, the relation between $\tau$ and $\Gamma_i$ can be found using the discrete orientation model (see Appendix D). For example, for IHD (i) for a potential with two equivalent wells 1 and 2 and one saddle point, $\tau \approx (2\Gamma_1^{\text{IHD}})^{-1}$ (here the factor 2 occurs because *two* equivalent wells are involved in the relaxation process); (ii) for a potential with two strongly nonequivalent wells ($\Gamma_1 \gg \Gamma_2$) and one saddle point, $\tau \approx (\Gamma_1^{\text{IHD}})^{-1}$, where $\Gamma_1^{\text{IHD}}$ is the escape rate from the shallow well 1, (iii) for a potential with two equivalent wells with two saddle points, e.g., for biaxial anisotropy (see Section IV.E), we have $\tau \approx (4\Gamma_1^{\text{IHD}})^{-1}$ (here the factor 4 occurs because *two* escape routes from the well over the saddle points exist and *two* equivalent wells are involved in the relaxation process). Other examples are given below.

All the results we have presented in this section for the reversal time pertain to the memoryless (white-noise) limit of the magnetization reversal. If long-time memory is included, all the basic equations, e.g., Gilbert's equation, escape rate equations, etc., must be generalized in an appropriate manner (see, for example, Ref. 84, where a generalized Gilbert equation taking into account memory effects as well as generalized escape rate formulas are given). For a generalized (non-Markovian)

Langevin description of the dynamics of stochastic systems, the escape rates may differ from the predictions of the Kramers theory (see, for example, Refs. 53 and 63). Nevertheless, as conjectured in Ref. 84, if the memory effects are included, the high-barrier asymptotes for the reversal time should hold with an *effective* (decreased) damping constant. Furthermore, the description of the relaxation processes using classical escape-rate theory neglects quantum effects such as macroscopic quantum tunneling (a mechanism of magnetization reversal suggested in Ref. [1]). The subject of tunneling in the context of the Kramers escape rate is of particular relevance in superparamagnetism (see, e.g., Refs. 11,101–103), because the magnetization of such particles is a *macroscopic* object: $10^4$–$10^5$ spins are collectively involved. A very important question first posed by Bean and Livingston[1] is: does reversal of magnetization by tunneling occur in such particles? If it occurs, then one would have an important example of *macroscopic quantum tunneling*. It follows, therefore, that an accurate analytical formula for the Kramers escape rate incorporating tunneling effects is vital to the investigation of magnetization reversal mechanisms in superparamagnets with a relatively small magnetic moment (~10–100 $\mu_B$), and to the question of the existence of macroscopic quantum tunneling in such systems.

## B. Comparison of the Gilbert, Landau–Lifshitz, and Kubo kinetic models of the Brownian rotation of classical spins

Hitherto we have used Gilbert's form of the Langevin equation, namely, Eq. (15) and its accompanying Fokker–Planck equation, Eq. (18). Equations (15) and (18) often occur in stochastic magnetization dynamics. Brown[133] justified using Gilbert's equation because all the terms in it can be derived from a Lagrangian function and a Rayleigh dissipation function. Moreover, Gilbert's equation fits naturally into the Kramers escape rate theory in all damping ranges if the damping torque is regarded as the time average of a fluctuating torque, whose instantaneous value contains also a random term with statistical properties. However, in the literature, alternative forms of the Langevin equations governing the magnetization $\mathbf{M}(t)$ have also been proposed. Two other frequently used Langevin equations for stochastic spin dynamics are the Landau–Lifshitz (e.g., [134,112]) and Kubo[135] forms, respectively,

$$\dot{\mathbf{u}}(t) = \gamma \mathbf{u}(t) \times [\mathbf{H}(t)+\mathbf{h}(t)] - \gamma\alpha \mathbf{u}(t) \times [\mathbf{u}(t) \times [\mathbf{H}(t)+\mathbf{h}(t)]] \qquad (94)$$

and

$$\dot{\mathbf{u}}(t) = \gamma \mathbf{u}(t) \times [\mathbf{H}(t)+\mathbf{h}(t)] - \gamma\alpha \mathbf{u}(t) \times [\mathbf{u}(t) \times \mathbf{H}(t)] \qquad (95)$$

($\mathbf{u} = \mathbf{M}/M_S$). The difference between these two models is that in the Kubo Eq. (95) the random field $\mathbf{h}(t)$ appears only in the gyromagnetic term.

In general, the explicit form of the infinite hierarchy of differential-recurrence equations for the statistical moments depends on the Langevin equation. Furthermore, the corresponding Fokker–Planck equation is also determined by that equation. Nevertheless, all the Langevin equations, Eqs. (15), (94), and (95), yield very similar hierarchies and Fokker–Planck equations, the only difference being in the definition of the free diffusion time. To illustrate this, we give a detailed derivation of the Fokker–Planck equations corresponding to Eqs. (15), (94), and (95), and show that each of the three Langevin equations gives rise to the *same* mathematical form for the corresponding Fokker–Planck equation. The only difference lies in the characteristic time constants. This difference becomes negligible at low damping, which is the case of most interest in superparamagnetism, however, for *high damping*, the models may predict a very different behavior.

We start from Gilbert's equation, Eq. (15), written in the Landau–Lifshitz form, viz.,[9,58]

$$\dot{\mathbf{u}}(t) = bM_S \alpha^{-1} [\mathbf{u}(t) \times (\mathbf{H}(t) + \mathbf{h}(t))] - bM_S [\mathbf{u}(t) \times [\mathbf{u}(t) \times (\mathbf{H}(t) + \mathbf{h}(t))]], \quad (96)$$

where $b = v/(2kT\tau_N)$ with $\tau_N$ defined by Eq. (19). Following Ref. 110, we use a spherical coordinate system[109] as shown in Fig. 4 above. In the basis $(\mathbf{e}_r, \mathbf{e}_\vartheta, \mathbf{e}_\varphi)$

$$\mathbf{u} = \begin{pmatrix} 1 \\ 0 \\ 0 \end{pmatrix}, \quad \dot{\mathbf{u}} = \begin{pmatrix} 0 \\ \dot{\vartheta} \\ \dot{\varphi} \sin\vartheta \end{pmatrix}, \quad \mathbf{H} = -\frac{1}{M_S} \frac{\partial V}{\partial \mathbf{u}} = -\frac{1}{M_S} \begin{pmatrix} 0 \\ \partial_\vartheta V \\ \csc\vartheta \, \partial_\varphi V \end{pmatrix}. \quad (97)$$

Thus Eq. (96) is equivalent to two stochastic equations for $\vartheta$ and $\varphi$:

$$\dot{\vartheta}(t) = bM_S [h_\vartheta(t) - \alpha^{-1} h_\varphi(t)] - b\left(\frac{\partial V}{\partial \vartheta}(t) - \frac{1}{\alpha \sin\vartheta(t)} \frac{\partial V}{\partial \varphi}(t)\right), \quad (98)$$

$$\dot{\varphi}(t) = \frac{bM_S}{\sin\vartheta(t)} [\alpha^{-1} h_\vartheta(t) + h_\varphi(t)] - \frac{b}{\sin\vartheta(t)} \left(\frac{1}{\sin\vartheta(t)} \frac{\partial V}{\partial \varphi}(t) + \frac{1}{\alpha} \frac{\partial V}{\partial \vartheta}(t)\right), \quad (99)$$

where $V(t) = V[\vartheta(t), \varphi(t), t]$ and the components $h_\vartheta(t)$ and $h_\varphi(t)$, of the random noise field $\mathbf{h}(t)$ in the spherical coordinate system are expressed in terms of the components $h_X(t)$, $h_Y(t)$, and $h_Z(t)$ in the Cartesian coordinate system as[8,109]

$$h_\vartheta(t) = h_X(t)\cos\vartheta(t)\cos\varphi(t) + h_Y(t)\cos\vartheta(t)\sin\varphi(t) - h_Z(t)\sin\vartheta, \quad (100)$$

$$h_\varphi(t) = -h_X(t)\sin\varphi(t) + h_Y(t)\cos\varphi(t), \quad (101)$$

and the components $h_X(t), h_Y(t), h_Z(t)$ in the Cartesian basis satisfy Eq. (17). We use here the Stratonovich definition[58,107,136] of the stochastic differential equations, Eqs. (98) and (99), since this definition always constitutes the mathematical idealization of the magnetic relaxation of superparamagnetic particles. Here one can again apply the usual rules of calculus.[107]

In order to derive the Fokker–Planck equation corresponding to Gilbert's equation (96)[8]

$$\frac{\partial P}{\partial t} = -\sum_{i=1}^{2} \frac{\partial}{\partial x_i}(D_i P) + \sum_{i,j=1}^{2} \frac{\partial^2}{\partial x_i \partial x_j}(D_{ij} P). \quad (102)$$

($D_i$ and $D_{ij}$ are drift and diffusion coefficients, respectively) for the PDF $P(\vartheta,\varphi,t) = W(\vartheta,\varphi,t)\sin\vartheta$, we recall that, the stochastic differential equations (98) and (99) involving two random variables $\{x_1(t) = \vartheta(t), x_2(t) = \varphi(t)\}$ may be written formally as

$$\dot{x}_i(t) = H_i[x_1(t), x_2(t)] + \sum_{j=1}^{3} G_{ij}[x_1(t), x_2(t)]h_j(t), \tag{103}$$

where

$$H_1 = -b\left(\frac{\partial V}{\partial \vartheta} - \frac{1}{\alpha \sin\vartheta}\frac{\partial V}{\partial \varphi}\right), \tag{104}$$

$$H_2 = -\frac{b}{\sin\vartheta}\left(\frac{1}{\sin\vartheta}\frac{\partial V}{\partial \varphi} + \frac{1}{\alpha}\frac{\partial V}{\partial \vartheta}\right) \tag{105}$$

$$\begin{aligned} G_{11} &= bM_S\left(\cos\vartheta\cos\varphi + \alpha^{-1}\sin\varphi\right), \\ G_{12} &= bM_S\left(\cos\vartheta\sin\varphi - \alpha^{-1}\cos\varphi\right), \\ G_{13} &= -bM_S\sin\vartheta, \end{aligned} \tag{106}$$

$$\begin{aligned} G_{21} &= bM_S\left(\alpha^{-1}\cot\vartheta\cos\varphi - \frac{\sin\varphi}{\sin\vartheta}\right), \\ G_{22} &= bM_S\left(\alpha^{-1}\cot\vartheta\sin\varphi + \frac{\cos\varphi}{\sin\vartheta}\right), \\ G_{23} &= -\alpha^{-1}bM_S. \end{aligned} \tag{107}$$

Now from the general theory of the Brownian motion,[58, 106-108] the drift $D_i$ and the diffusion $D_{ij}$ coefficients in the Fokker–Planck equation Eq. (102) are related to the coefficients $H_i$ and $G_{ij}$ in Eq. (103) via[8]

$$D_i = H_i + \frac{\alpha kT}{v\gamma M_S}\sum_{k,j=1}^{3} G_{kj}\frac{\partial}{\partial x_k}G_{ij}, \tag{108}$$

$$D_{ij} = \frac{\alpha kT}{v\gamma M_S}\sum_{k=1}^{3} G_{ik}G_{jk}. \tag{109}$$

The coefficients $D_i$ and $D_{ij}$ can then be evaluated from Eqs. (104)–(109), yielding

$$D_1 = \frac{1}{2\tau_N}\left[\cot\vartheta - \frac{v}{kT}\left(\frac{\partial V}{\partial \vartheta} - \frac{1}{\alpha\sin\vartheta}\frac{\partial V}{\partial \varphi}\right)\right], \tag{110}$$

$$D_2 = -\frac{v/(kT)}{2\tau_N \sin\vartheta}\left(\frac{1}{\sin\vartheta}\frac{\partial V}{\partial \varphi} + \frac{1}{\alpha}\frac{\partial V}{\partial \vartheta}\right), \tag{111}$$

$$D_{11} = (2\tau_N)^{-1}, \quad D_{22} = (2\tau_N \sin^2\vartheta)^{-1}, \quad D_{12} = D_{21} = 0. \tag{112}$$

The general Fokker–Planck equation, Eq. (102) with $D_i$ and $D_{ij}$ given by Eqs. (110)–(112) ultimately reduces to the Fokker–Planck equation Eq. (18) for the PDF $W(\vartheta,\varphi,t)$.

For the Kubo model, Eq. (95), the Langevin equations for the random variables $\vartheta$ and $\varphi$ read, in the Stratonovich interpretation,

$$\dot{\vartheta}(t) = -b_K \left\{ \frac{\partial V}{\partial \vartheta}(t) - \frac{1}{\alpha \sin \vartheta(t)} \frac{\partial V}{\partial \varphi}(t) + \frac{M_S}{\alpha}\left[-h_X(t)\sin\varphi(t) + h_Y(t)\cos\varphi(t)\right]\right\}, \quad (113)$$

$$\dot{\varphi}(t) = -\frac{b_K}{\sin\vartheta(t)}\left(\frac{1}{\sin\vartheta(t)}\frac{\partial V}{\partial \varphi}(t) + \frac{1}{\alpha}\frac{\partial V}{\partial \vartheta}(t)\right) \\ + \frac{b_K M_S}{\alpha}\left[h_X(t)\cot\vartheta(t)\cos\varphi(t) + h_Y(t)\cot\vartheta(t)\sin\varphi(t) - h_Z(t)\right], \quad (114)$$

where $b_K = v/(2kT\tau_K)$ and $\tau_K = \tau_N(1+\alpha^2)^{-1} = vM_S/(2kT\gamma\alpha)$ is the characteristic time constant. Now, the drift $D_i$ and the diffusion $D_{ij}$ coefficients can be evaluated as before, and are given by

$$D_1 = \frac{1}{2\tau_K}\left[\cot\vartheta - \frac{v}{kT}\left(\frac{\partial V}{\partial \vartheta} - \frac{1}{\alpha\sin\vartheta}\frac{\partial V}{\partial \varphi}\right)\right], \quad (115)$$

$$D_2 = -\frac{v/(kT)}{2\tau_K \sin\vartheta}\left(\frac{1}{\sin\vartheta}\frac{\partial V}{\partial \varphi} + \frac{1}{\alpha}\frac{\partial V}{\partial \vartheta}\right), \quad (116)$$

$$D_{11} = (2\tau_K)^{-1},\ D_{22} = (2\tau_K \sin^2\vartheta)^{-1},\ D_{12} = D_{21} = 0. \quad (117)$$

Clearly, by comparing Eqs. (115)-(117) with Eqs. (110)-(112), the drift and diffusion coefficients $D_i$ and $D_{ij}$ from Eqs. (115)-(117) coincide with those of the Gilbert equation, the only difference being that the time $\tau_N$ is replaced by $\tau_K$. Hence, one may conclude that the Fokker–Planck equation for Kubo's model also coincides with the Fokker–Planck equation, Eq. (18). This implies, in particular, that the differential-recurrence equation for the statistical moments becomes, for the Kubo model,

$$\tau_K \frac{d}{dt}\langle Y_{l,m}\rangle(t) = \sum_{s,r} e_{l,m,l+r,m+s}\langle Y_{l+r,m+s}\rangle(t), \quad (118)$$

i.e., Eq. (36), where $\tau_N$ is replaced by $\tau_K$.

Now, because the Gilbert Langevin equation, Eq. (15), can be reduced to the Landau–Lifshitz form, Eq. (96), which is Eq. (94), with an effective gyromagnetic constant, namely, $\gamma \to \gamma/(1+\alpha^2)$, one can conclude that the Fokker–Planck equation for the Landau–Lifshitz form, Eq. (94), coincides with that of Gilbert, Eq. (18). The only difference is that the free diffusion time $\tau_N$ in Eqs. (110)–(112) is replaced by the free-diffusion time $\tau_K = \tau_N(1+\alpha^2)^{-1}$, which is the same as in the Kubo model. Hence the Kubo and Landau–Lifshitz models, despite the different forms of the Langevin equations Eqs. (94) and (95), yield *identical* mathematical forms for the corresponding Fokker–Planck equations.

Thus the Gilbert, Kubo, and Landau–Lifshitz models for Brownian motion of classical spins irrespective of the Langevin equations, yield the *same* form of the corresponding Fokker–Planck equations, as well as the infinite hierarchy of differential-recurrence equations for the statistical moments, the only difference being in the free-diffusion time constants $\tau_N$ and $\tau_K$, a difference that is negligible at low damping (the most interesting damping range from an experimental point of view). However, only the Gilbert model, where the systematic and random terms in the stochastic equation, Eq. (15), viewed as the kinematic relation $\dot{\mathbf{u}} = \boldsymbol{\omega} \times \mathbf{u}$, are in the original Langevin form, i.e., with the rate of change of the angular momentum systematically slowed down superimposed on which is a rapidly fluctuating white noise random torque, can be used in all damping ranges. In contrast, neither the Kubo nor the Landau–Lifshitz models can be used at high damping, where they may predict unphysical behavior of the observables (relaxation times, escape rates, etc.). For example, for high damping, $\alpha > 1$, the Kubo and the Landau–Lifshitz models both predict an escape rate $\geq \Gamma^{TST}$, which is obviously an unphysical result (we recall that the upper bound for the escape rate is $\Gamma^{TST}$ [53]). A simple physical explanation of the behavior of the IHD escape rates for the Kubo and the Landau–Lifshitz models relates to the qualitative behavior of the overbarrier current density of representative points $J_s$ at the saddle point for $\alpha \geq 1$. In both of these equations, $J_s^{KLL} \sim \alpha$ *increases with increasing* $\alpha$, i.e., *damping tends to enhance the overbarrier current density*. Thus, the *escape rate* $\sim J_s^{KLL}$ *also increases*. In contrast, in Gilbert's equation, $J_s^G \sim \alpha/(1+\alpha^2)$ *decreases with increasing* $\alpha$, i.e., *damping tends to retard the current density* over the saddle point so that the *escape rate* $\sim J_s^G$ *also decreases*. Here $\Gamma^{IHD}/\Gamma^{TST} \ll 1$ for $\alpha \gg 1$ (see Fig. 9). In contrast in the low-damping range, $\alpha < 1$, the escape rates calculated from the Kubo, Landau–Lifshitz and Gilbert equations are *the same* as they should be. Thus, in order to avoid unphysical behavior of the escape rate for classical spins in the IHD range, $\alpha > 1$, the Langer IHD formula should only be used in conjunction with Gilbert's equation and not with the Landau-Lifshitz and/or Kubo equations, *which strictly apply only in the underdamped limit*, $\alpha < 1$, where energy controlled diffusion dominates. Unfortunately, some authors (see, e.g., Refs. 137 and 138)[*] have ignored this property of the Landau-Lifshitz equation and, in consequence, have used this *intrinsically underdamped* equation in conjunction with the *intrinsically IHD* Langer formula for the calculation of the escape rate in all damping ranges. Thus the ensuing escape rate formulas[137, 138] are misleading and not valid for experimental comparison both at low damping, where they coincide with the TST rate, and also in the IHD range, where they predict unphysical behavior of the rate, namely, a rate in excess of the TST one.

---

[*] The authors of Refs. 137 and 138 have uploaded to arXiv.org a comment (**arXiv:1210.2436**) on the present review; our reply to this comment is given below.[169]

In the following sections, we estimate reversal times of the magnetization for classical superparamagnets with various anisotropy potentials. These times are universal in the sense that they are valid for all values of damping, including the VLD and IHD regions. In order to assess the range of applicability of analytic formulas so obtained, we compare them with the results of numerical solutions of the Fokker-Planck equation, Eq. (18).

**C. Uniaxial superparamagnets in a uniform d.c. magnetic field of an arbitrary orientation**

As discussed above, Brown[8,9] estimated the reversal time $\tau$ for a uniaxial superparamagnet when $\mathbf{H}_0$ is applied *along* the easy axis of the magnetization. Brown's uniaxial asymptote, Eq. (73) is valid for all values of $\alpha$ due to the axial symmetry of the potential $V$. However, by applying an uniform magnetic field $\mathbf{H}_0$ at an angle $\psi$ with respect to the easy axis, one can break the symmetry of $V$, which will then also depend on the azimuthal angle $\varphi$. For axially symmetric anisotropy with $\mathbf{H}_0$ parallel to the easy axis, Eq. (71), the energy-landscape is a uniform equatorial ridge (zone) separating two (essentially) polar minima and has no saddle points; in contrast, the external field $\mathbf{H}_0$ generates azimuthally nonuniform energy distributions with a saddle point. The nonaxially symmetric energy-landscape leads to a new effect, viz., *intrinsic* dependence of magnetic characteristics (such as the dynamic susceptibility and relaxation times) on $\alpha$, arising from *coupling or entanglement of the longitudinal and transverse relaxation modes* as identified by Raikher and Shliomis.[139] A detailed treatment of the oblique-field problem has been given by Coffey *et al.*,[59,140] Kennedy,[141] Kalmykov *et al.*,[19,24,89] and Fukushima *et al.*[142] In particular, they evaluated $\tau$ as a function of the field strength, temperature, damping, and angle $\psi$, and showed that the analytical calculations based on the escape rate theory are in agreement with their numerical results. These analytical calculations also agree with computer simulations[93,96,97] and with experiments,[143] emphasizing the vital importance of an accurate determination of the damping dependence of $\tau$.

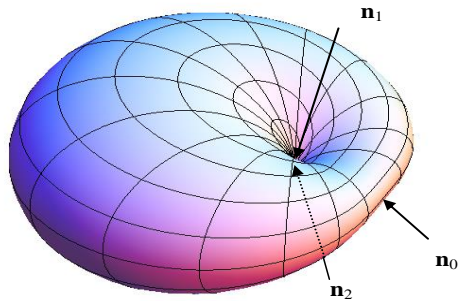

FIG. 10. 3D plot of the oblique field potential, Eq. (5) for the field parameter $h = \mu H_0 / (2vK) = 0.5$ and $\psi = \pi/2$.

The normalized free energy of a uniaxial superparamagnet in a d.c. magnetic field applied at an angle $\psi$ to the easy axis is given by Eq. (5). It has a bistable structure with minima at $\mathbf{n}_1$ and $\mathbf{n}_2$ separated by a potential barrier with a saddle point at $\mathbf{n}_0$. The saddle point lies generally in the equatorial region, while $\mathbf{n}_1$ and $\mathbf{n}_2$ lie in the north and south polar regions, respectively (see Fig. 10). In general, the potential Eq. (5) retains its *asymmetric bistable* form for $0 < h < h_c$ and $\psi \neq \pi/2$, i.e., the function $V(\vartheta, 0)$ has two minima separated by a maximum, which corresponds to a saddle point of the potential $V(\vartheta, \varphi)$. The two minima usually have different energies, so that the energy barriers $\Delta V_1 = V(\vartheta_0, 0) - V(\vartheta_1, 0)$ and $\Delta V_2 = V(\vartheta_0, 0) - V(\vartheta_2, 0)$ are not equal. For $h = h_c$, the second minimum becomes a point of inflexion; if $h > h_c$, the two-well structure of the potential disappears. For $\psi \neq 0$, it is not easy to evaluate the barrier heights as a function of $h$ and $\psi$ (see below). However, for particular values of $\psi$, e.g., $\psi = 0$, we find that $\Delta V_{1,2} = \sigma(1 \pm h)^2$ (see Fig. 5). For $\psi = \pi/2$, the barriers $\Delta V_1$ and $\Delta V_2$ coincide, so that $\Delta V_{1,2} = \sigma(1 - h)^2$ and the potential assumes a *symmetric* bistable form.

The stationary points of the potential energy occur for $\varphi = 0$ and $\varphi = \pi$. The stationary point for $\varphi = \pi$ corresponds to a maximum of $V(\vartheta, \varphi)$ in Eq.(5), and so is of no interest to us. The stationary points at $\varphi = 0$, however, correspond to a *saddle point* of $V(\vartheta, \varphi)$ at $\vartheta_0$, and two minima at $\vartheta_1$ and $\vartheta_2$ for $h < h_c$, where $h_c$ is some critical (nucleation) value of $h$ at which the potential Eq. (5) loses its bistable character (to be determined below). As already mentioned, the saddle point is generally in the equatorial region. The two equilibrium directions of the magnetization (in polar regions) and their associated polar angles $\vartheta_1$ and $\vartheta_2$ lie in the x–z plane ($\varphi = 0$) and are determined by the conditions for a minimum of $V(\vartheta, 0)$, namely $\partial_\vartheta V = 0$, $\partial^2_{\vartheta\vartheta} V > 0$. The position of the saddle point follows from the conditions for a maximum of $V(\vartheta, 0)$, namely $\partial_\vartheta V = 0$, $\partial^2_{\vartheta\vartheta} V < 0$. The critical value $h_c$ follows from the condition for a point of inflexion of $V(\vartheta, 0)$, namely $\partial_\vartheta V = \partial^2_{\vartheta\vartheta} V = 0$, yielding $\sin 2\vartheta = -2h_c \sin(\vartheta - \psi)$ and $\cos 2\vartheta = -h_c \cos(\vartheta - \psi)$, i.e.,

$$\tan 2\vartheta = 2\tan(\vartheta - \psi). \tag{119}$$

The only real root of Eq. (119) in the range $(0, \pi)$ is $\tan \vartheta = -(\tan \psi)^{1/3}$, so that

$$h_c = (\cos^{2/3} \psi + \sin^{2/3} \psi)^{-3/2}. \tag{120}$$

In the calculations of Stoner and Wohlfarth,[6] the external field axis is taken as the polar axis. Thus in their treatment the quantities $\vartheta - \psi$ and $\vartheta$ in Eq. (119) are interchanged.

The corresponding turnover formula for $\tau$ for the oblique field potential, Eq. (5), with two nonequivalent wells, is formally given by[89]

$$\tau = \left(\Gamma_1^{\text{IHD}} + \Gamma_2^{\text{IHD}}\right)^{-1} \frac{A(\alpha S_1 + \alpha S_2)}{A(\alpha S_1)A(\alpha S_2)}, \tag{121}$$

where

$$\Gamma_i^{\text{IHD}} = \frac{\omega_i\, e^{-\Delta V_i}}{8\pi\, \omega_0 \tau_0 (\alpha + \alpha^{-1})} \left[ -c_1^{(0)} - c_2^{(0)} + \sqrt{(c_2^{(0)} - c_1^{(0)})^2 - 4\alpha^{-2} c_1^{(0)} c_2^{(0)}} \right], \tag{122}$$

$\omega_i / \omega_0 = \sqrt{-c_1^{(i)} c_2^{(i)} / (c_1^{(0)} c_2^{(0)})}$ ($i = 1, 2$), and $\tau_0$ is defined by Eq. (88). In the VLD limit, Eq. (121) yields the VLD asymptote $\tau_{\text{VLD}}$, viz.,

$$\tau_{\text{VLD}} = \frac{4\pi \tau_0 (S_1^{-1} + S_2^{-1})}{\alpha \left[ \sqrt{c_1^{(1)} c_2^{(1)}}\, e^{-\Delta V_1} + \sqrt{c_1^{(2)} c_2^{(2)}}\, e^{-\Delta V_2} \right]}. \tag{123}$$

Equations (83) for $c_1^{(p)}$, $c_2^{(p)}$, and $V_p$ ($p = 0, 1, 2$) become[59]

$$c_1^{(p)} = 2\sigma \left[ \cos 2\vartheta_p + h \cos(\vartheta_p - \psi) \right], \quad c_2^{(p)} = 2\sigma \left[ \cos^2 \vartheta_p + h \cos(\vartheta_p - \psi) \right], \tag{124}$$

$$V_p = \sigma \left[ \sin^2 \vartheta_p - 2h \cos(\vartheta_p - \psi) \right], \tag{125}$$

where $\vartheta_p$ are the solutions of the transcendental equation

$$\sin 2\vartheta = 2h \sin(\psi - \vartheta). \tag{126}$$

So far our solution is purely formal. In order to derive an explicit analytic formula for $\tau$, we require all the parameters in the escape-rate equations, Eqs. (122) and (121), viz., $\Delta V_{1,2}$, $\omega_{1,2}$, $\omega_0$, $c_1^{(0)}$, and $c_2^{(0)}$. This may be accomplished as follows. Equation (126) may be rewritten as the quartic equation[59]

$$(x + h\cos\psi)^2 (1 - x^2) = (xh\sin\psi)^2$$

($x = \cos\vartheta$). The roots of this quartic equation $-1 \leq x_1 = \cos\vartheta_2 < x_2 = \cos\vartheta_0 < x_3 < x_4 = \cos\vartheta_1 \leq 1$ have complicated algebraic forms (see Ref. 59 for details). However, they can be written as a converging Taylor series to any desired order of $h$ ($h < h_c(\psi) \leq 1$), viz.,[89]

$$\cos\vartheta_0 = -h\cos\psi - \frac{h^2}{2}\sin 2\psi - \frac{h^3}{2}\sin\psi \sin 2\psi - \frac{h^4}{8}(3 - \cos 2\psi)\sin 2\psi$$
$$- \frac{h^5}{4}(3 + \cos 2\psi)\sin\psi \sin 2\psi - \frac{h^6}{128}(73 - 20\cos 2\psi - 29\cos 4\psi)\sin 2\psi + \ldots. \tag{127}$$

$$\cos\vartheta_{1,2} = \pm 1 \mp \frac{h^2}{2}\sin^2\psi + h^3 \sin^2\psi \cos\psi \mp \frac{h^4}{16}(13 + 11\cos 2\psi)\sin^2\psi$$
$$+ \frac{h^5}{2}(3 + \cos 2\psi)\sin^2\psi \cos\psi \mp \frac{h^6}{64}(183 + 156\cos 2\psi - 19\cos 4\psi)\sin^2\psi + \ldots \tag{128}$$

The corresponding Taylor series for $\Delta V_1$, $\Delta V_2$, $\omega_1$, $\omega_2$, $\omega_0$, $c_1^{(0)}$, and $c_2^{(0)}$ can be obtained from Eqs. (125), (127), and (128). We have[89]

$$\Delta V_{1,2} = \sigma \left[ 1 - 2h(\sin\psi \mp \cos\psi) + h^2 + \frac{h^3}{2}\sin 2\psi (\cos\psi \mp \sin\psi) + \frac{h^4}{2}\sin^2 2\psi \right.$$
$$\left. + \frac{h^5}{32}\sin 2\psi (7\cos\psi - 3\cos 3\psi \mp 7\sin\psi \mp 3\sin 3\psi) + \frac{h^6}{2}\sin^2 2\psi + ... \right], \quad (129)$$

$$\omega_{1,2} = \frac{2\gamma K}{M_S}\left[ 1 \pm h\cos\psi - \frac{h^2}{2}\sin^2\psi \pm \frac{3}{2}h^3\cos\psi\sin^2\psi - \frac{h^4}{16}(21 + 19\cos 2\psi)\sin^2\psi \right.$$
$$\left. \pm \frac{h^5}{2}\cos\psi\sin^2\psi(5 + 2\cos 2\psi) - \frac{h^6}{128}\sin^2\psi(321 + 284\cos 2\psi - 29\cos 4\psi) + \cdots \right], \quad (130)$$

$$\omega_0 = \frac{2K\gamma\sqrt{h\sin\psi}}{M_S}\left[ 1 - \frac{h}{2}\sin\psi - \frac{h^2}{16}(3 + \cos 2\psi) - \frac{h^3}{32}(19 + 17\cos 2\psi)\sin\psi \right.$$
$$\left. - \frac{h^4}{1024}(351 + 28\cos 2\psi - 283\cos 4\psi) - \frac{h^5}{2048}(2021 + 1332\cos 2\psi - 633\cos 4\psi)\sin\psi + ... \right], \quad (131)$$

$$c_1^{(0)} = 2\sigma\left[ -1 + h\sin\psi + h^2\cos^2\psi + \frac{5}{4}h^3\cos\psi\sin 2\psi + h^4\sin^2 2\psi \right.$$
$$\left. + \frac{11}{32}h^5\cos^2\psi(13\sin\psi - 3\sin 3\psi) + \frac{7}{4}h^6\sin^2 2\psi + ... \right] \quad (132)$$

$$c_2^{(0)} = 2\sigma\left[ h\sin\psi + \frac{h^3}{4}\cos\psi\sin 2\psi + \frac{h^4}{4}\sin^2 2\psi \right.$$
$$\left. + \frac{3}{32}h^5(13\sin\psi - 3\sin 3\psi)\cos^2\psi + \frac{h^6}{2}\sin^2 2\psi + ... \right]. \quad (133)$$

Now, in order to evaluate the actions $S_1$ and $S_2$ defined by the line integral, Eq. (91), it is necessary to determine explicit equations for the critical trajectories (separatrixes), which are solutions of the equation

$$\sin^2\vartheta - 2h(\cos\psi\cos\vartheta + \sin\psi\sin\vartheta\cos\varphi) = V_0/\sigma.$$

These trajectories are

$$\cos\vartheta_{1,2}(\varphi)\big|_{V=V_0} = -h\cos\psi\left(1 + h\sin\psi\cos\varphi + h^2\sin^2\psi\cos\varphi + ...\right)$$
$$\mp \sin(\varphi/2)\sqrt{h\sin\psi}\left[2 + h\sin\psi\cos\varphi + \cdots\right]\},$$

so that one can analytically evaluate $S_1$ and $S_2$ from Eq. (91) as[89]

$$S_{1,2} = \sigma\sqrt{h\sin\psi}\left[16 - \frac{104}{3}h\sin\psi + h^2(1 - 21\cos 2\psi) + \frac{h^3}{2}\sin\psi(45 + 51\cos 2\psi) + ...\right]$$
$$\pm 2\pi\sigma h^2\sin 2\psi\left(4 - 3h\sin\psi - 2h^2\sin^2\psi + \cdots\right). \quad (134)$$

Thus, by using Eqs. (129)–(134), one can estimate $\tau$ in analytic fashion from the turnover Eq. (121).

The analytic Eq. (121) allows one to easily estimate the reversal time $\tau$ in wide ranges of model parameters (see Fig. 11). In this figure, $\tau$ from the turnover, Eq. (121), VLD, Eq. (123), and TST is compared with the results of numerical calculations using matrix continued fractions.[19,117] The TST characteristic time $\tau_{TST}$ is independent of damping $\alpha$, and so may only be used to estimate $\tau$ for uniaxial superparamagnets in a very narrow range of $\alpha$. However, in the VLD and VHD limits, the deviation between $\tau$ from Eq. (121) and $\tau_{TST}$ is of several orders of magnitude; see Fig. 11. It can be seen that, in the high-barrier limit, Eq. (121) provides a good approximation to the reversal time for all values of the damping parameter $\alpha$. We emphasize that Eq. (121) is not valid for very small values of $h\sin\psi$, viz., $h\sin\psi \leq 0.03$ meaning that the departures from axial symmetry are small. The limiting value $h\sin\psi = 0$ corresponds to uniaxial anisotropy. Here $\tau$ is accurately given by Brown's asymptotic equation, Eq. (73).

For $\psi = \pi/2$ (transverse field), when the potential, Eq. (5), has *two equivalent wells*, the above equations can be simplified. We have[90]

$$\tau = \frac{2\tau_0(\alpha + \alpha^{-1})\pi\sqrt{h}e^{\sigma(1-h)^2}A(2\alpha S_i)}{\sigma\sqrt{1+h}[1-2h+\sqrt{1+4h(1-h)\alpha^{-2}}]A^2(\alpha S_i)}, \qquad (135)$$

where

$$S_i = \sigma\sqrt{h}\left(16 - \frac{104}{3}h + 22h^2 - 3h^3 + \frac{7}{24}h^4 + \frac{1}{16}h^5 + \cdots\right). \qquad (136)$$

We stress that Eq. (135) is not applicable when calculating $\tau$ for low fields, $4\sigma h < 1$ and $\sigma \gg 1$.[131] In such cases, the dependence of $V$ on the azimuthal angle $\varphi$ is weak (i.e., the potential, Eq. (5), is almost axially symmetric), and *all* escape paths from the potential wells are thus *roughly equivalent*. To estimate $\tau$ in this case, we can use the following relation, which was first obtained in Ref. 131 using perturbation theory (details in Ref. 56):

$$\tau \cong \tau_B\left\{1 + h^2\sigma^2\left[1 + 2\left(2\sigma\alpha^2 e\right)^{1/(2\sigma\alpha^2)}\gamma\left(1 + \frac{1}{2\sigma\alpha^2}, \frac{1}{2\sigma\alpha^2}\right)\right]\right\}^{-1}, \qquad (137)$$

where $\tau_B$ is the reversal time for a uniaxial (axially symmetric) potential [8,9] given by Eq.(79), and

$$\gamma(a,z) = \int_0^z t^{a-1}e^{-t}dt \qquad (138)$$

is the incomplete gamma function. It can be shown[131] that the expression in square brackets in Eq. (137) is about 1 for $\alpha \gg 1$ and about $\approx \alpha^{-1}\sqrt{\pi/\sigma}$ for $\alpha \ll 1$. The range of validity of Eq. (137) is determined by the inequalities

$$h^2\sigma^2 \ll 1, \quad \alpha > 4h^2\sigma^{3/2}, \text{ and } \sigma \gtrsim 4.$$

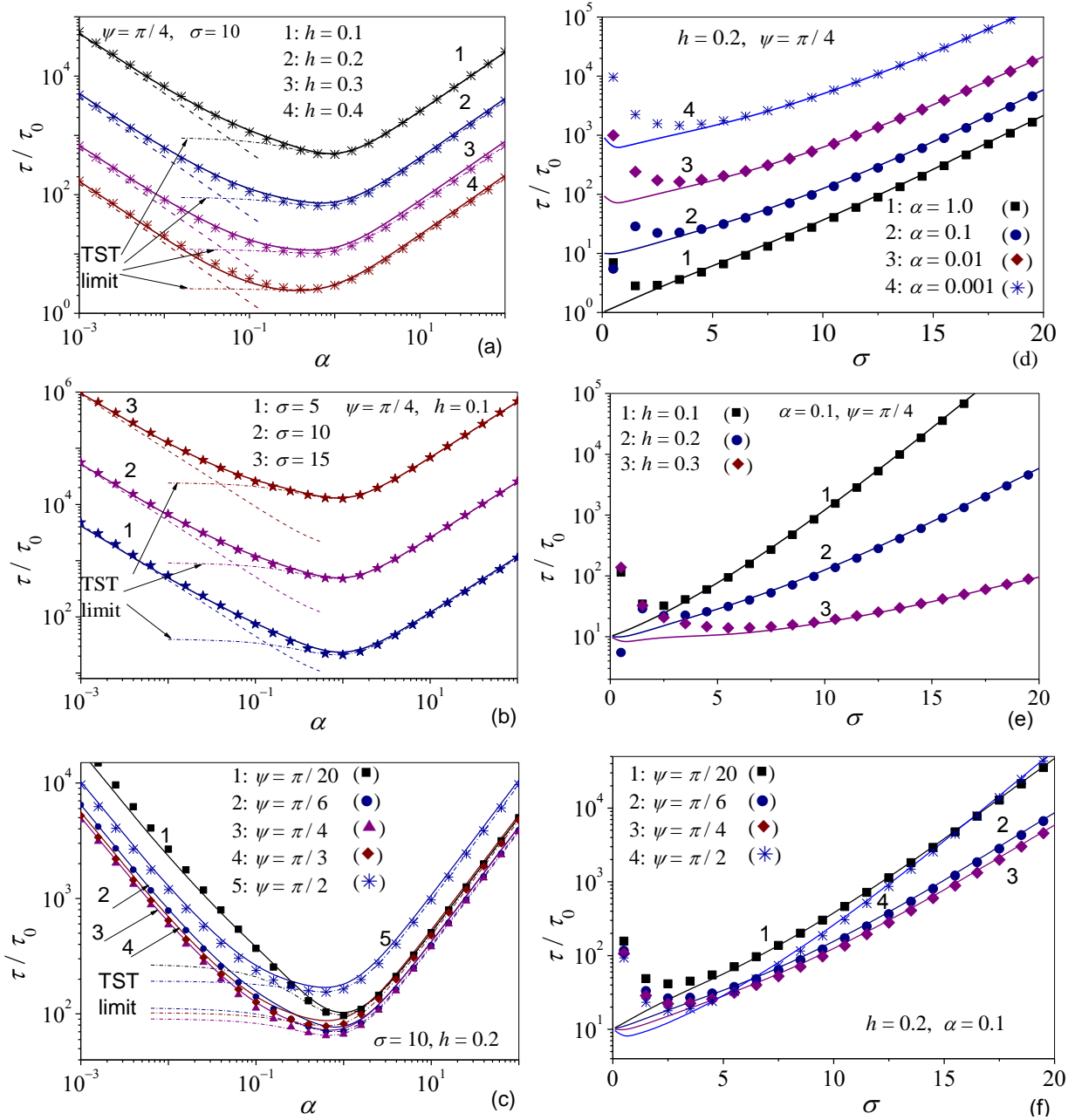

FIG. 11. (Left) $\tau/\tau_0$ vs. $\alpha$ (a) for $\sigma = 10$, $\psi = \pi/4$, and various values of $h$, (b) for $h = 0.1$, $\psi = \pi/4$, and various $\sigma$, and (c) for $h = 0.2$, $\sigma = 10$, and various $\psi$. Solid lines: matrix continued fraction solution;[117] dashed lines: the VLD asymptotes; dashed-dotted lines: the IHD, Eq. (122); symbols: Eq. (121). (Right) $\tau/\tau_0$ vs. $\sigma$ (d) for $h = 0.2$, $\psi = \pi/4$, and various $\alpha$, (e) for $\alpha = 0.1$, $\psi = \pi/4$, and various $h$, and (f) for $\alpha = 0.1$, $h = 0.2$, and various $\psi$. Solid lines: matrix continued fraction solution;[58, 117] symbols: Eq. (121).

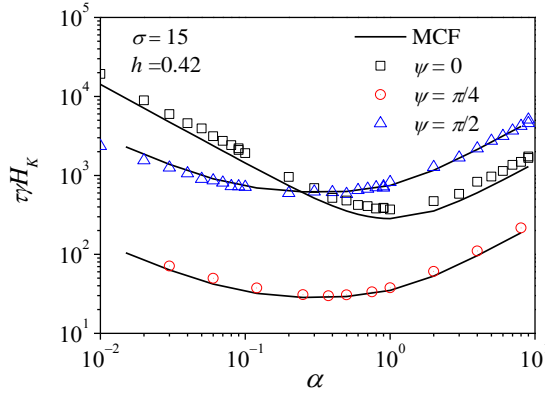

FIG. 12. (Color online) The normalized reversal time of a uniaxial superparamagnet vs. damping parameter $\alpha$ for $\sigma =15$, $h = 0.42$, and various values of the oblique angle $\psi = 0$, $\psi/4$, and $\psi/2$ ($H_K = 2K/M_S$). Solid lines: matrix continued fraction solution of Brown's Fokker-Planck equation. Symbols: Langevin dynamics simulation results. Reprinted figure with permission from Y. P. Kalmykov, W. T. Coffey, U. Atxitia, O. Chubykalo-Fesenko, P. M. Déjardin, and R. W. Chantrell, Phys. Rev. B **82**, 024412 (2010).[97] Copyright (2010) by the American Physical Society.

Equation (121) for $\tau$ can be used to deduce experimental values of the damping parameter $\alpha$[143] and to reproduce the angular variation of the switching field of individual nanoparticles at finite temperatures.[45] In particular, it can be used for comparison with temperature-dependent switching curves measured on a single (Co) nanoparticle.[11,43,44] Here it should be possible to determine $\alpha$ by fitting the theory to the experimental switching field curves (see Section V). The above results concerning the damping dependence of the escape rate are of the upmost importance in both Monte Carlo and Langevin dynamics simulations of the reversal time of the magnetization of fine particles (see, e.g., Refs. 97, 114, and 115). In analyzing the results of such simulations, the value of the analytical solutions of the Néel-Brown theory for $\lambda_1^{-1}$ is that they provide rigorous benchmark solutions with which the simulations must comply. In Fig. 12, we present the switching time obtained by the matrix continued fraction solution of Brown's Fokker-Planck equation and Langevin dynamics simulations.[45] As seen, the results of both methods are in perfect agreement.

### D. Reversal time for cubic anisotropy

The IHD formulas for the escape rates for cubic crystals were derived by Smith and de Rozario[82] and by Brown.[9] The several low-order eigenvalues of cubic crystals were first calculated numerically in the IHD limit by Aharoni[144] and Eisenstein and Aharoni.[145] Klik and Gunther[84] have also derived corresponding formulas for the VLD escape rate. The reversal time of the magnetization was also estimated in the IHD, turnover, and VLD ranges in Refs. 20 and 90. Furthermore, the effect of a d.c. bias field on the reversal time was studied in Refs. 22 and 92.

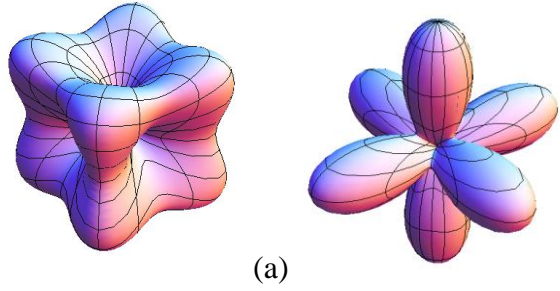

(a) (b)

FIG. 13. Cubic anisotropy potential for (a) $K > 0$ and (b) $K < 0$.

We now rewrite the cubic anisotropy potential in the form

$$V(\vartheta,\varphi) = \sigma(\sin^4\vartheta\sin^2 2\varphi + \sin^2 2\vartheta), \tag{139}$$

where $\sigma = vK/(4kT)$ is the dimensionless barrier height parameter, and $K$ is the anisotropy constant, which may have either positive or negative values.

For $K > 0$, the potential in Eq. (139) has six minima (wells), eight maxima and twelve saddle points (see Fig. 13a). Then the turnover formula is[90]

$$\tau \sim \frac{\tau_{\text{IHD}}}{A(\alpha S_i)}. \tag{140}$$

As shown in Appendix D, for the discrete-orientation model, the mean magnetization decays with time constant $1/(4\Gamma_i^{\text{IHD}})$, where $\Gamma_i^{\text{IHD}}$ is given by Eq. (84) *et seq.*, where $c_1^{(p)}$, $c_2^{(p)}$, and $V_p$ from Eq. (83) are now[90]

$$V_p = \sigma\sin^2 2\vartheta_p, \quad c_1^{(p)} = 8\sigma\cos 4\vartheta_p, \quad c_2^{(p)} = 2\sigma(3+\cos 4\vartheta_p). \tag{141}$$

Here the $\vartheta_p$ are the solutions of the trigonometric equation $\partial_\vartheta V|_{\varphi=0} = 0$, viz., $\vartheta_1 = 0$, $\vartheta_2 = \pi/2$, $\vartheta_3 = \pi$ (wells) and $\vartheta_0 = \pi/4$, $\vartheta_0 = 3\pi/4$ (saddle points). Thus, we obtain[90]

$$\Delta V_i = V_0 - V_i = \sigma, \quad c_1^{(i)} = c_2^{(i)} = 8\sigma, \quad c_1^{(0)} = 4\sigma, \quad c_2^{(0)} = -8\sigma, \tag{142}$$

and

$$\tau_{\text{IHD}} = \frac{1}{4\Gamma_i^{\text{IHD}}} = \frac{\tau_0(\alpha+\alpha^{-1})\pi e^\sigma}{2\sqrt{2}\sigma(\sqrt{9+8/\alpha^2}+1)}. \tag{143}$$

Now in order to evaluate $S_i$ in the turnover equation, Eq. (140), we need an explicit solution for the critical trajectory (separatrix). For the well with a minimum at $\vartheta = 0$, this critical trajectory, passing between the two saddle points at $\varphi = 0$, $\vartheta = \arccos 2^{-1/2}$ and $\varphi = \pi/2$, $\vartheta = \arccos 2^{-1/2}$, is determined from the trigonometric equation $\sin^4\vartheta\sin^2 2\varphi + \sin^2 2\vartheta = 1$, where the physically meaningful solution is

$$\vartheta(\varphi)|_{V=V_0} = \arccos\sqrt{\frac{1+\sin 2\varphi}{2+\sin 2\varphi}}. \tag{144}$$

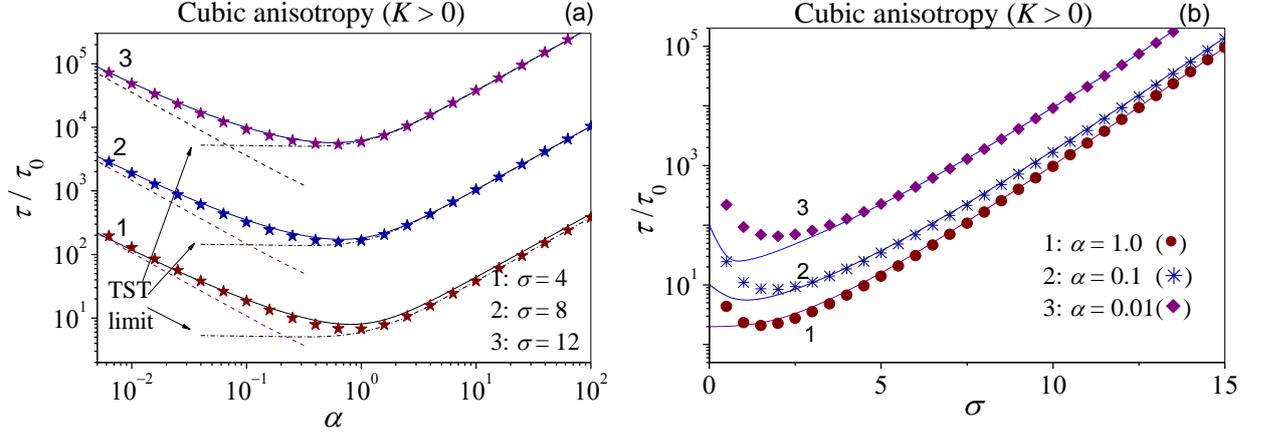

FIG. 14. (a) $\tau/\tau_0$ vs. $\alpha$ for various $\sigma$ and (b) $\tau/\tau_0$ vs. $\sigma$ for various $\alpha$. Solid lines: matrix continued fraction solution;[20] dashed lines: the VLD equation, Eq. (147); dashed-dotted lines: the IHD equation, Eq.(143); stars: the turnover equation, Eq. (146).

Hence, we have from Eqs. (91) and (144)[90]

$$S_i = 12\sigma \int_0^{\pi/2} \frac{\sin 2\varphi (1+\sin 2\varphi)^{1/2}}{(2+\sin 2\varphi)^{5/2}} d\varphi = \frac{8\sqrt{2}\sigma}{9}. \tag{145}$$

Thus, using Eqs. (140), (143), and (145), we have the turnover formula[90]

$$\tau = \frac{\tau_0(\alpha + \alpha^{-1})\pi e^\sigma}{2\sqrt{2}\sigma(\sqrt{9+8/\alpha^2}+1)A(8\sqrt{2}\sigma\alpha/9)} \tag{146}$$

In the VLD limit, Eq. (146) yields the asymptote, viz.,

$$\tau_{VLD} = \frac{\tau_0 9\pi e^\sigma}{64\alpha\sqrt{2}\sigma^2}. \tag{147}$$

By formally setting $\alpha = 0$ in Eq. (143), we obtain the TST reversal time $\tau_{TST}$ predicted by the Néel theory, namely,

$$\tau_{TST} = \frac{\tau_0 \pi e^\sigma}{8\sigma}. \tag{148}$$

The analytic equation (146) so obtained allows one to easily estimate $\tau$ for arbitrary damping $\alpha$. In Fig. 14, $\tau$ evaluated from Eqs. (146), (147), and (148) is compared with numerical calculations using the matrix continued fraction technique.[20,58] The TST (Néel) time $\tau_{TST}$ from Eq. (148) constituting the lower bound for $\tau$ is of course independent of damping, and may be used for superparamagnets with positive cubic anisotropy *only in a very narrow range of damping,* $\alpha S_i \sim 1$. In the VLD and VHD limits, deviations between $\tau$ from Eq. (146) and $\tau_{TST}$ from Eq. (148) can again be of *several orders of magnitude* (see Fig. 14).

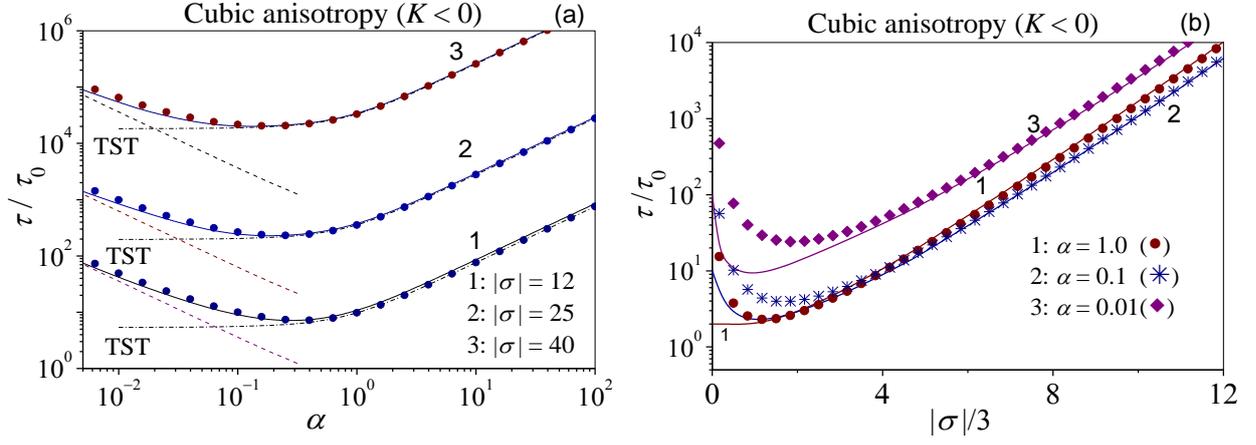

FIG. 15. (a) $\tau/\tau_0$ vs. $\alpha$ for various $|\sigma|$. Solid lines: matrix continued fraction solution;[20] dashed lines: the VLD equation, Eq. (151); dashed-dotted lines: the IHD equation, Eq. (149); filled circles: the turnover equation Eq. (150). (b) $\tau/\tau_0$ vs. $|\sigma|/3$ for various $\alpha$. Solid lines: matrix continued fraction solution;[20] symbols: Eq. (150).

For $K < 0$, the maxima and minima are interchanged, and the barrier-height parameter is now given by $v|K|/(12kT)$ (see Fig. 13b) From the discrete-orientation model, the mean magnetization decays with time constant $1/(2\Gamma_i^{IHD})$ (see Appendix D). Because[90]

$$c_1^{(p)} = -|\sigma|(8\cos 2\vartheta_p - 3\sin^2 2\vartheta_p),\ c_2^{(p)} = -2|\sigma|(\cos 2\vartheta_p + 3\cos 4\vartheta_p),\ V_p = -|\sigma|(\sin^4 \vartheta_p + \sin^2 2\vartheta_p),$$

where $p = 0,1,2$, $\vartheta_1 = \arctan\sqrt{2}$, $\vartheta_2 = \pi - \arctan\sqrt{2}$ (wells), and $\vartheta_0 = \pi/2$ (saddle point) are the solutions of the trigonometric equation $\partial_\vartheta V|_{\varphi=\pi/4} = 0$, so that

$$\Delta V_i = V_0 - V_i = |\sigma|/3,\ c_1^{(i)} = c_2^{(i)} = 16|\sigma|/3,\ c_1^{(0)} = 8|\sigma|,\ c_2^{(0)} = -4|\sigma|,$$

we then have[9,82,91]

$$\tau_{IHD} = \frac{1}{2\Gamma_i^{IHD}} = \frac{3\tau_0(\alpha+\alpha^{-1})\pi e^{|\sigma|/3}}{2\sqrt{2}|\sigma|(\sqrt{9+8/\alpha^2}-1)}. \tag{149}$$

The turnover formula for $\tau$ is then[90]

$$\tau = \frac{\tau_{IHD}}{A(\alpha S_i)} = \frac{3\tau_0(\alpha+\alpha^{-1})\pi e^{|\sigma|/3}}{2\sqrt{2}|\sigma|(\sqrt{9+8/\alpha^2}-1)A(\alpha|\sigma|8\sqrt{2}/9)}, \tag{150}$$

where $S_i = 8\sqrt{2}|\sigma|/9$ just as $K > 0$ while the VLD asymptote is

$$\tau_{VLD} = \frac{\tau_0 27\pi e^{|\sigma|/3}}{64\sqrt{2}\sigma^2\alpha}. \tag{151}$$

By setting $\alpha = 0$ in Eq. (149), we obtain

$$\tau_{TST} = \frac{\tau_0 3\pi e^{|\sigma|/3}}{8|\sigma|}. \qquad (152)$$

In Fig. 15, $\tau$ from Eqs. (150), (151), and (152) is compared with numerical calculations using matrix continued fractions.[20,58] The TST relaxation time $\tau_{TST}$ from Eq. (152) is independent of damping, and may be used to estimate $\tau$ for superparamagnets with negative cubic anisotropy, again only in the very narrow range of damping, $\alpha S_i \sim 1$. In the VLD and VHD limits, deviations between $\tau$ from Eq. (150) and $\tau_{TST}$ from Eq. (152) can once again be of *several orders of magnitude* (see Fig. 15).

**E. Biaxial superparamagnet in a uniform d.c. magnetic field applied along the easy axis**

Now we consider the effects of an external field on the relaxation dynamics of the magnetization of single-domain particles with orthorhombic anisotropy when the field $\mathbf{H}_0$ is applied along the easy axis of the magnetization. For zero d.c. bias field, $\tau$ for orthorhombic crystals has been given by Smith and de Rozario[82] in the IHD limit, and by Kalmykov and Ouari[91] for all $\alpha$. The appropriate formula for the IHD reversal time for the similar problem of the magnetization dynamics in elongated *biaxial* particles subjected to a strong d.c. magnetic field has been given by Braun.[85] Moreover, Ouari and Kalmykov[146] have evaluated $\tau$ in the presence of a d.c. bias field for all $\alpha$.

The anisotropy potential is given by[82,91]

$$V = \sigma \sin^2 \vartheta + \Delta \sin^2 \vartheta \cos^2 \varphi - 2\sigma h \cos \vartheta, \qquad (153)$$

where $\Delta$ and $\sigma$ are the biaxiality and barrier-height parameters, respectively, and $h = \xi/(2\sigma)$. For $0 < h < 1$, the potential, Eq. (153), has two *nonequivalent wells* and two *equivalent saddle points* (see Fig. 16). The biaxial anisotropy may yield a noticeable contribution to the free energy density of magnetic nanoparticles.[11] In particular, Eq. (153) describes the magnetic anisotropy energy of a spheroidal single-domain particle, with the axis of symmetry inclined at a certain angle to the easy anisotropy axis of the particle[147] as well that of elongated particles, where easy- and hard-axis anisotropy terms are present.[85]

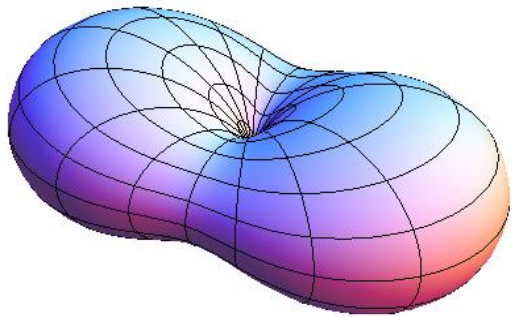

FIG. 16. Biaxial anisotropy for $h = 0$ and $\Delta/\sigma = 1$.

The reversal time is formally given by the turnover formula[146]

$$\tau = \tau_{IHD} \frac{A(\alpha S_1 + \alpha S_2)}{A(\alpha S_1)A(\alpha S_2)}. \tag{154}$$

Here the discrete-state orientation model (see Appendix D) indicates that the mean magnetization decays according to

$$\tau_{IHD} = [2(\Gamma_1^{IHD} + \Gamma_2^{IHD})]^{-1}, \tag{155}$$

where $\Gamma_1^{IHD}$ is the escape rate from (deeper) well 1 to well 2, and $\Gamma_{21}^{IHD}$ is the escape rate from well 2 to well 1 over one saddle point. The factor 2 occurs in Eq. (155) because there are *two* magnetization escape routes from the wells over the two saddle points. The escape rates $\Gamma_i^{IHD}$ are defined by Eq. (84), where $c_1^{(p)}$, $c_2^{(p)}$, and $V_p$ are now given by[146]

$$c_1^{(p)} = 2\sigma(\cos 2\vartheta_p + h\cos\vartheta_p), \quad c_2^{(p)} = 2\sigma(\delta + \cos^2\vartheta_p + h\cos\vartheta_p), \quad V_p = \sigma(\sin^2\vartheta_p - 2h\cos\vartheta_p), \tag{156}$$

where $\delta = \Delta/\sigma$ and the $\vartheta_p$ are the solutions of the equation $\partial_\vartheta V|_{\varphi=\pi/2} = 0$. These are $\vartheta_1 = 0$, $\vartheta_2 = \pi$, and $\cos\vartheta_0 = -h$. Thus

$$\Delta V_{1,2} = \sigma(1\pm h)^2, \quad c_1^{(i)} = 2\sigma(1\pm h), \quad c_2^{(i)} = 2\sigma(1+\delta\pm h), \quad c_1^{(0)} = -2\sigma(1-h^2), \quad c_2^{(0)} = 2\sigma\delta.$$

The escape rates $\Gamma_2^{IHD}$ and $\Gamma_1^{IHD}$ are now

$$\Gamma_2^{IHD}(h) = \frac{\sigma e^{-(1-h)^2\sigma}}{2\pi\tau_0(\alpha+\alpha^{-1})}\sqrt{\frac{1-h+\delta}{\delta(1+h)}}\left[1-h^2-\delta+\sqrt{(1-h^2+\delta)^2+4\delta(1-h^2)\alpha^{-2}}\right] \tag{157}$$

and $\Gamma_1(h) = \Gamma_2(-h)$. The dimensionless actions $S_1$ and $S_2$ are given by the contour integral, Eq. (91), taken along the separatrixes $p_{1,2}(\varphi)$ which are determined by the equation $V(\vartheta,\varphi) = V_0$, where $V_0$ is the value of $V$ at the saddle points. From Eq. (153) and $V_0 = \sigma(1+h^2)$, this equation is

$$p^2(1+\delta\cos^2\varphi) \pm 2hp + h^2 - \delta\cos^2\varphi = 0.$$

$$p_{1,2}(\varphi)|_{V=V_0} = \left[\mp h + \cos\varphi\sqrt{\delta(1-h^2) + \delta^2\cos^2\varphi}\right](1+\delta\cos^2\varphi)^{-1}.$$

The contour integrals $S_1$ and $S_2$ can be evaluated analytically as[146]

$$S_{1,2} = 2\sigma(1-h^2+\delta)\int_{-\pi/2}^{\pi/2}\left\{\frac{1-h^2+\delta(1+h^2)\cos^2\varphi}{\sqrt{1-h^2+\delta\cos^2\varphi}} \pm 2h\sqrt{\delta}\cos\varphi\right\}\frac{\cos\varphi d\varphi}{(1+\delta\cos^2\varphi)^2}$$

$$= \frac{4\delta\sigma(1-h^2+\delta)}{(1+\delta)^{3/2}}\left\{\left[\left(1+\frac{1}{\delta}\right)(1-h^2)\right]^{1/2} + h\arctan\left[h\left[(1-h^2)\left(1+\frac{1}{\delta}\right)\right]^{-1/2}\right] \pm \frac{h\pi}{2}\right\}. \tag{158}$$

Equation (154) with $h = 0$ also yields the known equation for biaxial anisotropy[91]

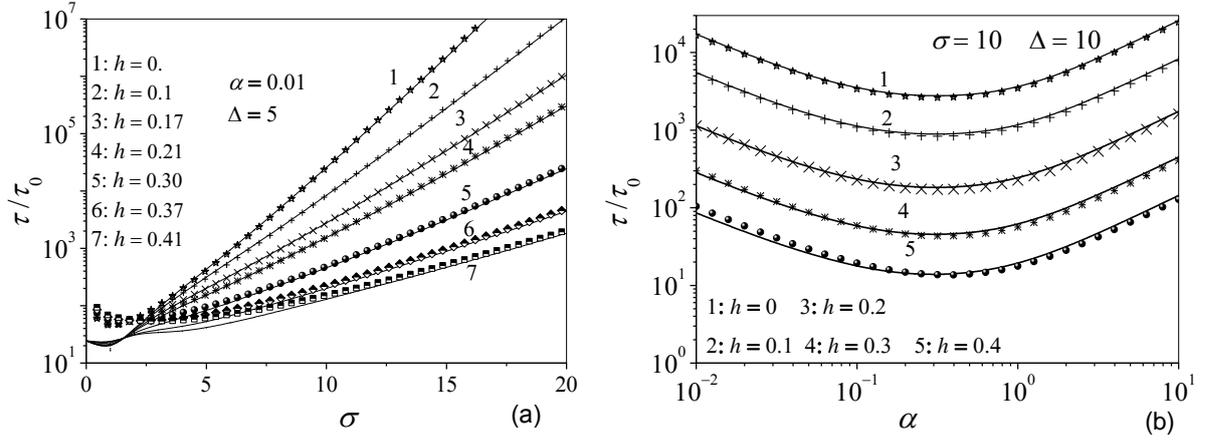

FIG. 17. (a) Reversal time $\tau/\tau_0$ vs. the barrier-height parameter $\sigma$ for various values of the field parameter $h$ and $\alpha = 0.01$ and $\Delta = 5$. Solid lines 1–7: matrix continued fraction solution.[146] Symbols: the turnover Eq. (154). (b) $\tau/\tau_0$ vs. the damping parameter $\alpha$ for various values of $h$ for $\Delta = 10$ and $\sigma = 10$. Solid lines 1–5: matrix continued fraction solution.[146] Symbols: the turnover Eq. (154).

Appropriate solutions of the above equation yield two separatrixes

$$\tau = \frac{\tau_0(\alpha+\alpha^{-1})\pi e^\sigma A(8\alpha\sigma\sqrt{\delta})}{\sigma\sqrt{1+1/\delta}\left[1-\delta+\sqrt{(1+\delta)^2+4\delta/\alpha^2}\right]A^2(4\alpha\sigma\sqrt{\delta})}. \qquad (159)$$

Equations (157)–(158) again allow us to evaluate $\tau$ from Eq. (154) *for all* $\alpha$. The time $\tau$ as predicted by the turnover equation, Eq. (154), and $\tau$ calculated numerically by the matrix continued fraction method for a biaxial potential,[146] are shown in Fig. 17. Once again, in the high-barrier limit, $\sigma \gg 1$, Eq. (154) provides a good approximation to $\tau$ for all $\alpha$. We again emphasize that Eq. (154) is not valid for $\delta = 0$ corresponding to uniaxial anisotropy; here $\tau$ is given by Brown's uniaxial asymptote equation, Eq. (74). We remark that the matrix continued fraction method developed in Ref. 146 also allows one to calculate the reversal time for an arbitrary orientation of the d.c. bias field $\mathbf{H}_0$.

### F. Mixed anisotropy: breakdown of the paraboloid approximation

The salient feature of the IHD equation, Eq. (84), is the *elliptic* and *hyperbolic paraboloid* approximation, Eq. (81), for the free energy density $V(\mathbf{M})$ near the relevant stationary points. However, there are certain situations where either the *well* or the *damped* saddle frequencies or indeed both are zero, so that Eq. (84) now predicts zero escape rate. This obviously incorrect result may occur (i) when $V(\mathbf{M})$ is approximately axially symmetric, leading to the *uniaxial crossover phenomenon*, where the saddle points become simple maxima, or (ii) if the paraboloid approximation, Eq.(81), fails. The breakdown of the hyperbolic/elliptic paraboloid approximation (ii) is encountered, for example, for $V(\mathbf{M})$ with either *flat* saddles or *flat* well bottoms, or both. A particular example is *mixed* uniaxial and cubic anisotropy

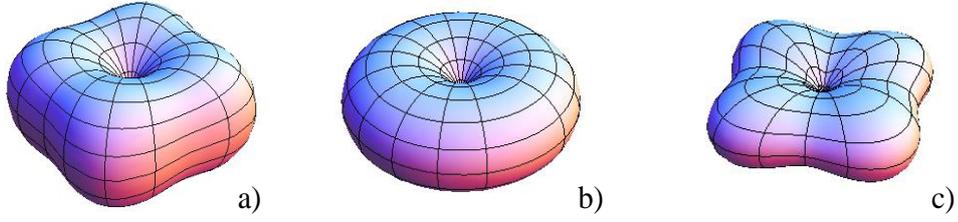

FIG. 18. 3D plots of the mixed anisotropy potential for $\zeta = 1$ (a: flat saddles), 0 (b: uniaxial anisotropy), and −1 (c: flat wells).

$$V = \sigma \left[ \sin^2 \vartheta + \frac{\zeta}{4} (\sin^4 \vartheta \sin^2 2\varphi + \sin^2 2\vartheta) \right], \quad (160)$$

where $\sigma = vK_1/(kT)$, $K_1$ denotes the uniaxial anisotropy constant and $\zeta$ is the cubic-to-uniaxial anisotropy ratio, which may be either positive or negative. For $|\zeta|/4 \gg 1$, Eq. (160) represents *cubic* anisotropy. For $|\zeta| \leq 1$, on the other hand, Eq. (160) represents a double-well potential with two equivalent wells and four equivalent saddle points; these saddle points disappear at $\zeta = 0$ corresponding to a uniaxial anisotropy (see Fig. 18). For $\zeta = -1$, the well frequency $\omega_i = 0$, while for $\zeta = 1$ the damped saddle frequency $\Omega_0 = 0$. These values of $\zeta$ are of particular interest, since the existing escape rate formulas described in Sections IV.C-E cannot be used to estimate the reversal time $\tau$ without modification.

The mixed anisotropy equation, Eq. (160), appears in various applications. For example, it appears in the "effective macrospin" model,[148] in which a many-spin cluster is mapped onto a macrospin representing the net moment of the cluster, with the corresponding energy comprising mixed uniaxial and cubic anisotropies. Here the effective anisotropy energy landscape depends on the size and shape of the cluster, the crystalline structure of the medium, and other physical parameters such as the exchange coupling and local anisotropy constants. The model provides a compromise between the macrospin approach, based on the Stoner–Wohlfarth concept of *coherent rotation* of all the atomic spins, and the *many-spin approach*. The effective constants of the model, e.g., the parameter $\zeta$, etc., must, however, be computed from microscopic considerations in order to account for the crystallographic structure, the shape of the particle, the number of spins, etc., so that the model can be directly compared with experimental data or numerical simulations. Furthermore, the results may also be applied[149] to the temperature dependence of the switching field curves of isolated Co nanoclusters characterized by mixed anisotropy (see Section V). The mixed anisotropy energy, Eq. (160), also occurs in paleomagnetism and rock magnetism.[150]. Hence, thermal relaxation is important for both thermo-remanent magnetization and related measurements determining the blocking temperature(s) characteristic of a given material. Moreover, the anisotropy energy, Eq. (160), permits many remanent states to coexist for $|\zeta|>1$, thereby leading to multiple blocking temperatures and a transition towards a single blocking temperature. The magnetization relaxation rate problem for particles with mixed anisotropy, Eq. (160), was implicitly identified by Smith and de Rozario,[82] Brown,[9] and Dormann et al.,[48] while Newell[150] explicitly evaluated the IHD escape rate from the IHD Eq. (84), noting the absurd prediction of a vanishing escape rate for $|\zeta|=1$. Now, insofar as the rate calculation in the vicinity of nonparabolic stationary points is concerned, a method for point Brownian particles or rigid rotators with separable and additive Hamiltonians has been suggested by Talkner and Ryter[151] and Hänggi et al.[53] This has been applied to a Brownian single-axis rotator in a periodic potential with *nonparabolic* barriers[152] and also has been recently extended to a single-domain particle with mixed anisotropy,[95] yielding a corresponding turnover formula for the reversal time of the magnetization $\tau$. Following the exposition of Ref. 95, we shall, for the purposes of illustration, confine our treatment to anisotropy ratios $|\zeta|\leq 1$. The calculations can be extended to $|\zeta|>1$ without major difficulties.

For $0<|\zeta|\leq 1$, i.e., for a double-well potential with two equivalent wells and four equivalent saddle points, $\tau$ is given in terms of the depopulation factor $A(\Delta)$ and the escape rate $\Gamma_1^{IHD}$ from a single well as[95]

$$\tau = \frac{A(2\alpha S)}{8\Gamma_1^{IHD} A^2(\alpha S)}. \tag{161}$$

Because $S_1 = S_2 = S$, $\Gamma_1^{IHD} = \Gamma_2^{IHD}$, and the IHD $\Gamma_1^{IHD}$ refers to *one* saddle point only. The factor 8 occurs because (i) there are *four* magnetization escape routes from the well over the saddle points, and (ii) *two* equivalent wells are involved in the relaxation process.

In order to evaluate the action $S$ in Eq. (161) from the contour integral, Eq. (91), as usual we first need an explicit equation for the separatrix. For the distinct cases of *positive* and *negative* cubic anisotropy, i.e., $0<\zeta\leq 1$ and $-1\leq\zeta<0$, the respective separatrixes are determined by the trigonometric equations

$$\sin^2\vartheta + \zeta(\sin^4\vartheta\sin^2 2\varphi + \sin^2 2\vartheta)/4 = 1$$

and

$$\sin^2\vartheta - |\zeta|(\sin^4\vartheta\sin^2 2\varphi + \sin^2 2\vartheta)/4 = 1-|\zeta|/4.$$

The physically meaningful solutions are[95]

$$\cos\vartheta(\varphi)\Big|_{V=V_0} = \sqrt{\frac{\zeta(3+\cos 4\varphi)-4+4\sqrt{(\zeta-1)^2+\zeta\sin^2 2\varphi}}{\zeta(7+\cos 4\varphi)}}, \tag{162}$$

$(0<\zeta\leq 1)$ and

$$\cos\vartheta(\varphi)\big|_{V=V_0} = \sqrt{\frac{|\zeta|(3+\cos 4\varphi)+4-2\sqrt{4+8|\zeta|+|\zeta|(|\zeta|-4)\sin^2 2\varphi}}{|\zeta|(7+\cos 4\varphi)}} \quad (163)$$

($-1 \leq \zeta < 0$). By substituting Eqs. (162) and (163) into Eq. (91), we can evaluate the actions $S$ analytically as Taylor series expansions up to any desired order of $\zeta$ in two distinct regions, viz.,[95]

$$S = \sigma\sqrt{\zeta}\left(1 - \frac{\zeta}{6} + \frac{\zeta^2}{8} + \frac{17\zeta^3}{240} - \frac{\zeta^4}{128} - \cdots\right) \quad (164)$$

for *positive* cubic anisotropy and

$$S = \sigma\sqrt{|\zeta|}\left(1 + \frac{7}{12}|\zeta| - \frac{15}{32}\zeta^2 + \frac{363}{640}|\zeta|^3 - \frac{1569}{2048}\zeta^4 + \cdots\right) \quad (165)$$

for *negative* cubic anisotropy, where $\sigma = vK_1/(kT)$. Furthermore, $S$ can also be computed numerically from Eq. (91).

Thus, the only remaining quantity remaining in Eq. (161) is $\Gamma_i^{\text{IHD}}$ itself, which cannot be evaluated by naively applying the IHD equation, Eq. (84), and requires separate analysis for $0 < \zeta \leq 1$ and $-1 \leq \zeta < 0$. In the first instance, the *hyperbolic* paraboloid approximation at the saddle point breaks down, so that the *Kramers method of determining the crossover function between the wells needs to be modified*. In the second instance, the *elliptic* paraboloid approximation at the bottom of the wells breaks down, so that the *calculation of well populations via steepest descents* needs to be modified. We treat both cases separately.

*IHD escape rate for $0 < \zeta \leq 1$*

Despite the breakdown of the hyperbolic paraboloid approximation near the saddle point, the IHD magnetization escape rate can still be expressed as a flux-over-population. In order to see this, we first recall that, in general, in IHD the picture is that inside the well the distribution function of $\mathbf{M}(t)$ is almost the equilibrium Boltzmann distribution. However, very near the saddle, the distribution *deviates* from that equilibrium distribution as a result of the quasi-stationary reversal of $\mathbf{M}(t)$ over the saddle point. Now the saddle point (separatrix) region where nonequilibrium prevails is very small, and the saddle point itself is a stationary point; therefore, near that point the Fokker–Planck equation may be written in terms of the direction cosines of $\mathbf{M}(t)$ as[95]

$$2\tau_N \dot{W} \approx \frac{1}{\alpha}\left(\frac{\partial V}{\partial u_1}\frac{\partial W}{\partial u_2} - \frac{\partial W}{\partial u_1}\frac{\partial V}{\partial u_2}\right) + \frac{\partial}{\partial u_1}\left(W\frac{\partial V}{\partial u_1} + \frac{\partial W}{\partial u_1}\right) + \frac{\partial}{\partial u_2}\left(W\frac{\partial V}{\partial u_2} + \frac{\partial W}{\partial u_2}\right). \quad (166)$$

Here, by approximating $V$ near a saddle point to *fourth* order in the direction cosines $(u_1, u_2)$, we have $V$ as

$$V(u_1, u_2) \approx \sigma - \sigma(1-\zeta)u_1^2 + \zeta\sigma u_2^2 - \zeta\sigma\left(u_1^4 + u_1^2 u_2^2 + u_2^4\right). \quad (167)$$

Since the barrier-crossing process is exponentially slow, we may now assume a *quasi-stationary* solution of Eq. (166) in the separatrix region, of the form

$$W(u_1, u_2, t) = w(u_1, u_2) e^{-\Gamma_i^{\text{IHD}} t}, \tag{168}$$

leading to the partial differential equation

$$-2\Gamma_i^{\text{IHD}} \tau_N w \approx \frac{1}{\alpha}\left(\frac{\partial V}{\partial u_1}\frac{\partial w}{\partial u_2} - \frac{\partial w}{\partial u_1}\frac{\partial V}{\partial u_2}\right) + \frac{\partial}{\partial u_1}\left(w\frac{\partial V}{\partial u_1} + \frac{\partial w}{\partial u_1}\right) + \frac{\partial}{\partial u_2}\left(w\frac{\partial V}{\partial u_2} + \frac{\partial w}{\partial u_2}\right). \tag{169}$$

Thus by integrating this equation with respect to the direction cosines $u_1$ and $u_2$, limiting the integration to a *single well*, then using Green's theorem in the $(u_1, u_2)$ plane, we may *formally* obtain $\Gamma_i^{\text{IHD}}$ as the closed-line integral along the saddle-point contour

$$\Gamma_i^{\text{IHD}} \approx -\frac{1}{2\alpha \tau_N Z_i} \oint_{\text{well } i} e^{-V}\left\{\left(g\frac{\partial V}{\partial u_2} - \alpha\frac{\partial g}{\partial u_1}\right)du_2 + \left(g\frac{\partial V}{\partial u_1} + \alpha\frac{\partial g}{\partial u_2}\right)du_1\right\}. \tag{170}$$

Here $g = e^V w$ is the crossover function; this was originally used by Kramers to obtain the IHD solution of the Klein–Kramers equation pertaining to point particles, by converting that equation in the vicinity of the saddle into an ordinary differential equation (see Appendix B). The partition function $Z_i$ represents the well population, where the elliptic paraboloid approximation for the energy near the bottom of the well still holds, so that $Z_i$ may be evaluated (as usual) by steepest descents. For $V$ given by Eq. (160), we obtain

$$Z_i = \int_0^{2\pi}\int_0^{\pi/2} e^{-V(\vartheta,\varphi)} \sin\vartheta\, d\vartheta\, d\varphi \approx \frac{\pi}{\sigma(1+\zeta)}. \tag{171}$$

In order to evaluate the escape rate, we require an expression for the relevant Kramers crossover function $g = e^V w$ describing the saddle dynamics and its first derivatives at the well boundary, along with a suitable parameterization of the well boundary itself. Since the distribution function $w$ must always be finite, and because a high barrier is assumed, the left-hand side of Eq. (169) almost vanishes by quasi-stationarity. Hence, in terms of the Kramers crossover function $g$, we have

$$\left[\frac{\partial V}{\partial u_2} + \alpha\frac{\partial V}{\partial u_1}\right]\frac{\partial g}{\partial u_1} - \left[\frac{\partial V}{\partial u_1} - \alpha\frac{\partial V}{\partial u_2}\right]\frac{\partial g}{\partial u_2} \approx \alpha\left[\frac{\partial^2 g}{\partial u_1^2} + \frac{\partial^2 g}{\partial u_2^2}\right]. \tag{172}$$

Following Hänggi *et al.*,[53] we now seek $g$ as[95]

$$g(u_1, u_2) = C^{-1}\int_{-\infty}^{\rho(u_1,u_2)} e^{-A_1 z^2 - A_2 z^4}dz,\quad C = \int_{-\infty}^{\infty} e^{-A_1 z^2 - A_2 z^4}dz, \tag{173}$$

where $A_1$ and $A_2$ are unknown coefficients that account for both the shape of the saddle region *and the energy loss at the saddle*, and where $\rho(u_1, u_2)$ is a function to be determined. We note that the Kramers IHD calculation, as adapted to magnetization reversal by Brown, corresponds to setting $A_2 = 0$ in Eq.(173), and dropping altogether the fourth-order terms in the Taylor series expansion of the free energy density in Eq. (167). Here in contrast, we must have $A_2 \neq 0$ *in order to account for the nonparaboloidal shape of the saddle*, and all terms of the fourth-order Taylor expansion of the free energy density in Eq. (167) are retained in Eqs. (172) *et seq.* In succinct terms, because of the *nonparaboloidal* shape of the saddle region, the crossover function *deviates* from the error function originally used by Kramers for *parabolic* barriers.

Next, one must transform the partial differential equation Eq. (172) in the two variables $(u_1, u_2)$, into an ordinary differential equation in the single variable $\rho$. We may do this, following Kramers, by implicitly seeking $\rho = qu_1 + u_2$ as a linear combination of $u_1$ and $u_2$ in the saddle region. Thus we obtain (the details are in Ref. 95).

$$q = -(2\zeta)^{-1}[\alpha + \sqrt{\alpha^2 + 4\zeta(1-\zeta)}], \tag{174}$$

$$A_1 = -\frac{\zeta\sigma(1+q/\alpha)}{1+q^2}, \quad A_2 = -\frac{(q-1/\alpha)\zeta\sigma}{q^3(1+q^2)}. \tag{175}$$

These expressions for $q$, $A_1$, and $A_2$ determine the crossover function $g$.

According to Brown[9] and Geoghegan *et al.*[59] the well boundary is parameterized by $u_1 = 0$. Therefore, setting $du_1 = 0$ in the contour integral, Eq. (170), and retaining the parabolic approximation *only* in the factor $e^{-V(0,u_2)}$, Eq. (170) finally becomes[95]

$$\begin{aligned}
\Gamma_i^{IHD} &\approx \frac{\sigma(1+\zeta)(1-\alpha q)}{2\pi\tau_0(1+\alpha^2)C} \int_{-\infty}^{\infty} e^{-(\zeta\sigma+A_1)u_2^2 - A_2 u_2^4} du_2 \\
&= \sqrt{1+\frac{\zeta\sigma}{A_1}} \frac{\sigma(1+\zeta)(1-\alpha q) K_{1/4}\left[(A_1+\zeta\sigma)^2/(8A_2)\right]}{\tau_0 2\pi(1+\alpha^2) K_{1/4}\left[A_1^2/(8A_2)\right]} e^{\frac{\zeta\sigma(\zeta\sigma+2A_1)}{8A_2} - \sigma},
\end{aligned} \tag{176}$$

where $0 < \zeta \leq 1$, and $K_{1/4}(z)$ is a modified Bessel function of the third kind.[153] For $\zeta = 1$, when $A_1 \to 0$, Eq. (176) becomes[95]

$$\tau_0 \Gamma_i^{IHD} \sim \frac{\alpha\sigma^{5/4}}{\pi\Gamma(1/4)} K_{1/4}(\alpha^4\sigma/8) e^{-\sigma(1-\alpha^4/8)} \tag{177}$$

which for $\alpha^4\sigma/8 \gg 1$ reduces to

$$\Gamma_i^{IHD} \sim \frac{2\sigma^{3/4} e^{-\sigma}}{\tau_0 \alpha\pi^{1/2}\Gamma(1/4)}. \tag{178}$$

where $\Gamma(z)$ is the gamma function.[153] This completes the solution of the flat-saddle problem.

*IHD escape rate for $-1 \leq \zeta < 0$*

We now consider negative anisotropy ratio, $-1 \leq \zeta < 0$, where the free-energy potential, Eq. (160), near the bottom of a well may not be approximated by an elliptic paraboloid, which will obviously affect the well partition function. Nevertheless $\Gamma_i^{IHD}$ can still be estimated from Langer's equation, Eq. (89), viz.,

$$\Gamma_i^{IHD} = \frac{\Omega_0 Z_0}{2\pi Z_i}, \tag{179}$$

where $Z_i$ and $Z_0$ are the well and saddle partition functions, respectively. First, we recall that near the saddle for $-1 \leq \zeta < 0$, the hyperbolic paraboloid approximation still holds, so that $Z_0$ and the damped saddle frequency $\Omega_0$ can be evaluated as usual. Hence[95]

$$Z_0 = 2\pi \left[ |\zeta|(2+|\zeta|) \right]^{-1/2} e^{-\sigma(1-|\zeta|/4)} \tag{180}$$

and

$$\Omega_0 = \frac{\sigma}{4\tau_0(\alpha+\alpha^{-1})} \left[ 2-|\zeta|+\sqrt{(3|\zeta|+2)^2 + 8\alpha^{-2}|\zeta|(2+|\zeta|)} \right]. \tag{181}$$

The calculation of the well partition function $Z_i$ in Eq. (179), where the elliptic paraboloid approximation fails, can be accomplished using Eq. (167). Hence, we have an accurate approximation for $Z_i$, viz.,[95]

$$Z_i \approx \frac{(1-|\zeta|)}{4|\zeta|} \exp\left[ \sigma(1-|\zeta|)^2 / 4|\zeta| \right] K_{1/4}^2 \left[ \sigma(1-|\zeta|)^2 / (8|\zeta|) \right], \tag{182}$$

which approximates the well partition function $Z_i$ with an error of the order of 5% for $|\zeta| \leq 1$. Thus, by substituting Eqs. (180), (181), and (182) into Langer's equation, Eq. (179), we finally obtain

$$\tau_0 \Gamma_i^{IHD} \sim \sqrt{\frac{|\zeta|}{2(2+|\zeta|)}} \frac{2-|\zeta|+\sqrt{(2+3|\zeta|)^2 + 8\alpha^{-2}|\zeta|(2+|\zeta|)}}{(\alpha+\alpha^{-1})(1-|\zeta|) K_{1/4}^2 [(1-|\zeta|)^2 \sigma / (8|\zeta|)]} e^{-\sigma(1+2|\zeta|)/(4|\zeta|)}. \tag{183}$$

For $\zeta = -1$, Eq.(183) becomes[95]

$$\tau_0 \Gamma_i^{IHD} = \frac{\sqrt{\sigma/6}(\alpha+\sqrt{25\alpha^2+24})}{(1+\alpha^2)\Gamma^2(1/4)} e^{-3\sigma/4}. \tag{184}$$

This equation, combined with the turnover, Eq. (161), completes the *flat minimum* magnetization escape rate.

The results of the numerical[95] and asymptotic (from Eqs. (161) (turnover), (176), (177), (183), and (184): all pertaining to IHD) calculations of the normalized reversal time $\tau/\tau_0$ as functions of the anisotropy ratio parameter $\zeta$ and $\alpha$ are shown in Figs. 19. Figure 19 shows that the universal equation, Eq. (161), describes the behavior of the reversal time in the *entire* dissipation range for both $\zeta = -1$ and $\zeta = 1$. Moreover, Eq. (161) is valid in the range $0.2 \leq |\zeta| \leq 1$ (see Fig. 19). Here, the *usual* Brown–Kramers IHD formula, Eq. (84), based on the paraboloidal approximation, does not describe the relaxation rate at all (as is apparent from Fig. 19). In fact, for mixed anisotropy, Eq. (84) yields

$$\Gamma_i^{IHD} = \frac{\sigma(1+\zeta)[1-2\zeta+\sqrt{1+4\alpha^{-2}\zeta(1-\zeta)}]e^{-\sigma}}{4\pi\tau_0(\alpha+\alpha^{-1})\sqrt{\zeta(1-\zeta)}} \qquad (185)$$

for $0 < \zeta \leq 1$, and

$$\Gamma_i^{IHD} = \frac{\sigma(1-|\zeta|)[2-|\zeta|+\sqrt{(2+3|\zeta|)^2+8\alpha^{-2}|\zeta|(2+|\zeta|)}]e^{-\sigma(1-|\zeta|/4)}}{4\pi\tau_0(\alpha+\alpha^{-1})\sqrt{2|\zeta|(2+|\zeta|)}} \qquad (186)$$

For $-1 \leq \zeta < 0$. By inspection, Eq. (186) predicts *zero* escape rate for $|\zeta| = 1$, while yielding an *infinite* escape rate as $\zeta \to 0$. Fig. 19 also indicates that Eqs. (161), (176), (177), (183), and (184) correctly reproduce the behavior of the relaxation time for $0.2 \leq |\zeta| \leq 1$, in contrast to the IHD equation, Eq. (84). Notice that the correct escape-rate equation for the uniaxial case, $\zeta = 0$, is in fact provided by Brown's *uniaxial* anisotropy formula, Eq. (79).

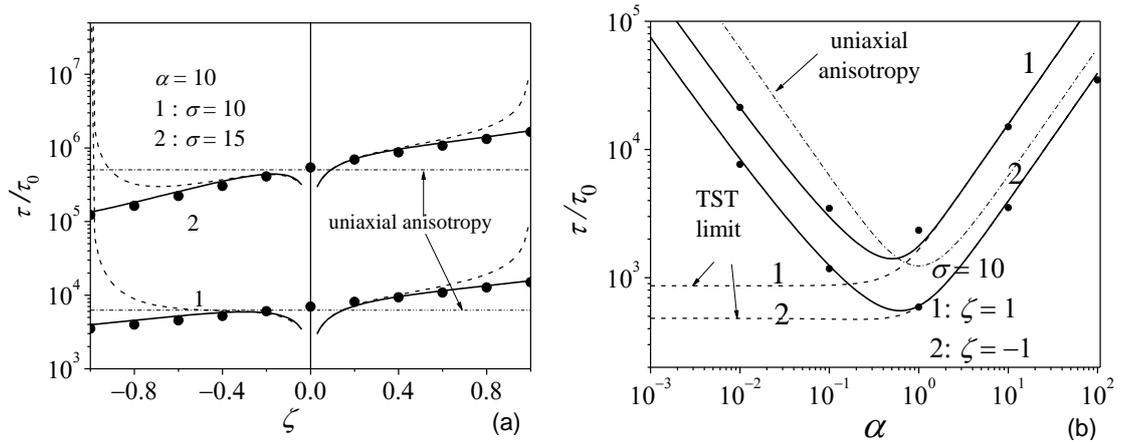

FIG. 19. (a) Reversal time $\tau/\tau_0$ vs. $\zeta$ for $\alpha = 10$ and $\sigma = 10, 15$. Filled circles: matrix continued fraction solution;[95] solid lines: turnover Eq. (161); dashed lines: the IHD Eqs. (185) and (186); dashed-dotted lines: Brown's axially symmetric asymptote, Eq. (79). (b) $\tau/\tau_0$ vs. $\alpha$ for $\zeta = -1$ and $\zeta = 1$. Filled circles: matrix continued fraction solution.[95] Solid lines: Eq.(161). Dashed lines: the IHD Eqs. (184) and (177). Dotted line: Brown's equation, Eq. (79). Reprinted figure with permission from W. T. Coffey, P. M. Déjardin, and Yu. P. Kalmykov, [Phys. Rev. B **79**, 054401 (2009)](#).[95] Copyright (2009) by the American Physical Society.

The turnover formula, Eq. (161), is valid for *all* $\alpha$. We remark, however, that as $\zeta \to 0$, axial symmetry is regained and the azimuthal dependence of the distribution function disappears. The saddle region now becomes *very wide*, so that the method of Garanin *et al.*[56,131] should be used; as the Mel'nikov method fails since the action *S* is zero in this case, once again yielding zero escape rate. By comparing the uniaxial asymptotes in Eq. (79) with Eq. (161), as shown in Fig. 19, we see that $\tau$ for pure uniaxial anisotropy can differ by as much as an order of magnitude from $\tau$ for mixed anisotropy as rendered by Eq. (161). This may be attributed to the difference in the prefactors between the uniaxial and nonaxially symmetric results.

## V. SWITCHING FIELD CURVES AND SURFACES

### A. Geometrical method

If one knows the reversal time of the magnetization **M** as a function of the direction of an external magnetic field, one can include thermal effects in the calculation of switching field curves and/or surfaces. We recall that the first calculation of the magnetization reversal of single-domain ferromagnetic particles with uniaxial anisotropy subjected to an applied field was made by Stoner and Wohlfarth.[6] They made the hypothesis of *coherent rotation* of the magnetization and *zero* temperature, so that *thermally induced switching between the potential minima is ignored*. In the simplest uniaxial anisotropy, as considered by them, the magnetization reversal occurs at that particular value of the applied field (switching field) which destroys the bistable nature of the potential. The 2D parametric plot, of the parallel vs. the perpendicular component of the switching field, then yields the famous *critical (*or *limit of metastability) curves,* or *astroids.*[41] The calculation of the switching field curves or surfaces at *zero* temperature has been given by Thiaville[42] for *arbitrary* anisotropy, and may be summarized as follows.

The starting point of this calculation is the normalized free-energy potential

$$\bar{V}(\mathbf{u}) = G(\mathbf{u}) - 2(\mathbf{u} \cdot \mathbf{h}), \qquad (187)$$

where **h** is the normalized external field $\mathbf{H}/H_K$ ($H_K$ is the anisotropy field) and *G* is the normalized anisotropy in the absence of the field. The unit vector $\mathbf{u} = \mathbf{M}/M_S$ is described by the polar angle $\vartheta$ relative to some axis *OZ*, and the azimuthal angle $\varphi$, while the vector **h** is described by the polar angle $\psi$ and the azimuthal angle $\phi$ (see Fig. 20).

The switching field is characterized by the fact that both the first and second derivatives of $\bar{V}$ with respect to $\vartheta$ and $\varphi$ vanish. In fact, this condition indicates that one metastable minimum and one saddle point in the potential $\bar{V}$ merge, giving rise to a point of inflexion. These conditions correspond to a *switching field surface* in 3D space. This surface, as it generalizes the critical curves of the 2D problem of Stoner and Wohlfarth,[6] is called the *limit of metastability surface*. At any point of that surface, $\bar{V}$ must satisfy the stationary conditions

$$\frac{\partial \bar{V}}{\partial \vartheta} = \frac{\partial G}{\partial \vartheta} - 2(\mathbf{h} \cdot \mathbf{e}_\vartheta) = 0, \quad \frac{\partial \bar{V}}{\partial \varphi} = \frac{\partial G}{\partial \varphi} - 2(\mathbf{h} \cdot \mathbf{e}_\varphi)\sin\vartheta = 0,$$

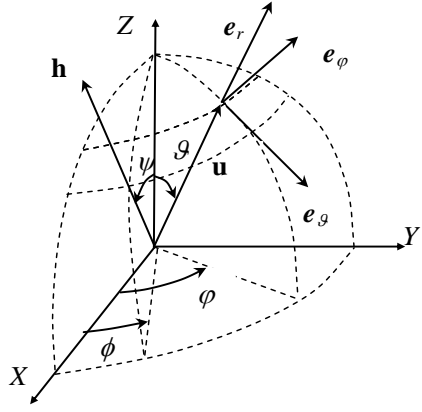

FIG. 20. Spherical polar coordinate system.

so that the field vector **h** can be described by a parameter $\lambda$, viz.,

$$\mathbf{h} = \lambda \mathbf{e}_r + \frac{1}{2}\frac{\partial G}{\partial \vartheta}\mathbf{e}_\vartheta + \frac{1}{2\sin\vartheta}\frac{\partial G}{\partial \varphi}\mathbf{e}_\varphi, \tag{188}$$

where the unit vectors $\mathbf{e}_r$, $\mathbf{e}_\vartheta$, and $\mathbf{e}_\varphi$ forming the orthonormal direct basis are defined as

$$\mathbf{e}_r = \begin{pmatrix} \sin\vartheta\cos\varphi \\ \sin\vartheta\sin\varphi \\ \cos\vartheta \end{pmatrix},\ \mathbf{e}_\vartheta = \begin{pmatrix} \cos\vartheta\cos\varphi \\ \cos\vartheta\sin\varphi \\ -\sin\vartheta \end{pmatrix},\ \mathbf{e}_\varphi = \begin{pmatrix} -\sin\varphi \\ \cos\varphi \\ 0 \end{pmatrix}. \tag{189}$$

The switching conditions are now determined by the equation

$$\frac{\partial^2 \overline{V}}{\partial \vartheta^2}\frac{\partial^2 \overline{V}}{\partial \varphi^2} - \left[\frac{\partial^2 \overline{V}}{\partial \vartheta \partial \varphi}\right]^2 = 0. \tag{190}$$

Because the second derivatives of $\overline{V}$ are given by

$$\frac{\partial^2 \overline{V}}{\partial \vartheta^2} = \frac{\partial^2 G}{\partial \vartheta^2} + 2\lambda,\quad \frac{\partial^2 \overline{V}}{\partial \varphi^2} = \frac{\partial^2 G}{\partial \varphi^2} + \left(\cot\vartheta\frac{\partial G}{\partial \vartheta} + 2\lambda\right)\sin^2\vartheta,\quad \frac{\partial^2 \overline{V}}{\partial \vartheta \partial \varphi} = \frac{\partial^2 \overline{V}}{\partial \varphi \partial \vartheta} = \sin\vartheta\frac{\partial}{\partial \vartheta}\left(\frac{1}{\sin\vartheta}\frac{\partial G}{\partial \varphi}\right),$$

Eq. (190) reduces to a quadratic equation in $\lambda$, viz.,

$$4\lambda^2 + 2\lambda\left[\frac{1}{\sin^2\vartheta}\frac{\partial^2 G}{\partial \varphi^2} + \cot\vartheta\frac{\partial G}{\partial \vartheta} + \frac{\partial^2 G}{\partial \vartheta^2}\right] + \left[\frac{1}{\sin^2\vartheta}\frac{\partial^2 G}{\partial \varphi^2} + \cot\vartheta\frac{\partial G}{\partial \vartheta}\right]\frac{\partial^2 G}{\partial \vartheta^2} - \left[\frac{\partial}{\partial \vartheta}\left(\frac{1}{\sin\vartheta}\frac{\partial G}{\partial \vartheta}\right)\right]^2 = 0,$$

which has two roots $\lambda^+(\vartheta,\varphi)$ and $\lambda^-(\vartheta,\varphi)$ given by

$$\lambda^\pm = -\frac{1}{4}\left\{\frac{1}{\sin^2\vartheta}\frac{\partial^2 G}{\partial \varphi^2} + \cot\vartheta\frac{\partial G}{\partial \vartheta} + \frac{\partial^2 G}{\partial \vartheta^2} \mp \sqrt{\left[\frac{1}{\sin^2\vartheta}\frac{\partial^2 G}{\partial \varphi^2} + \cot\vartheta\frac{\partial G}{\partial \vartheta} - \frac{\partial^2 G}{\partial \vartheta^2}\right]^2 + 4\left[\frac{\partial}{\partial \vartheta}\left(\frac{1}{\sin\vartheta}\frac{\partial G}{\partial \varphi}\right)\right]^2}\right\}.$$

Now the half line, Eq. (188), described by $\lambda > \lambda^+$ is the locus of the fields for which the magnetization is stable. At $\lambda = \lambda^+$ the metastable minimum in the potential $\overline{V}$ disappears so that the magnetization vector **M** can then escape from the potential well. Thus the Switching field surface may be obtained from the vector $\mathbf{h}_S$ defined as[42]

$$\mathbf{h}_S = \lambda^+ \mathbf{e}_r + \frac{1}{2}\frac{\partial G}{\partial \vartheta}\mathbf{e}_\vartheta + \frac{1}{2\sin\vartheta}\frac{\partial G}{\partial \varphi}\mathbf{e}_\varphi. \tag{191}$$

## B. Limit of metastability curves

In order to illustrate the Thiaville geometrical method,[42] we summarize in Fig. 21 the switching field surfaces and curves for uniaxial anisotropy:

$$G_{\text{un}}(\vartheta,\varphi) = \sin^2\vartheta, \tag{192}$$

biaxial anisotropy:

$$G_b(\vartheta,\varphi) = \sin^2\vartheta + \delta\sin^2\vartheta\cos^2\varphi, \tag{193}$$

and positive and negative cubic anisotropies:

$$G_{c\pm}(\vartheta,\varphi) = \pm(\sin^4\vartheta\sin^2 2\varphi + \sin^2 2\vartheta). \tag{194}$$

and mixed anisotropy

$$G_m(\vartheta,\varphi) = \sin^2\vartheta + \zeta(\sin^4\vartheta\sin^2 2\varphi + \sin^2 2\vartheta)/4. \tag{195}$$

The transverse ($h_\perp$) and longitudinal ($h_\parallel$) components of the normalized 2D switching field for uniaxial, biaxial, and positive and negative cubic anisotropy, respectively, are given by

$$h_\perp^{\text{un}} = \sin^3\vartheta, \quad h_\parallel^{\text{un}} = -\cos^3\vartheta, \tag{196}$$

$$\begin{aligned} h_\perp^b &= \frac{1}{2}\Big[\delta + (1+\delta)\sin^2\vartheta + \big|1-(1+\delta)\cos^2\vartheta\big|\Big]\sin\vartheta, \\ h_\parallel^b &= -\frac{1}{2}\Big[1+(1+\delta)\cos^2\vartheta - \big|1-(1+\delta)\cos^2\vartheta\big|\Big]\cos\vartheta, \end{aligned} \tag{197}$$

$$h_\perp^{c+} = (2+3\cos 2\vartheta)\sin^3\vartheta, \quad h_\parallel^{c+} = (2-3\cos 2\vartheta)\cos^3\vartheta, \tag{198}$$

$$\begin{aligned} h_\perp^{c-} &= \frac{1}{8\sqrt{2}}\sin^3\vartheta(11+9\cos 2\vartheta - 3|1+3\cos 2\vartheta|), \\ h_\parallel^{c-} &= \frac{1}{32}(5-28\cos 2\vartheta - 9\cos 4\vartheta - 12|1+3\cos 2\vartheta|\sin^2\vartheta)\cos\vartheta. \end{aligned} \tag{199}$$

$$\begin{aligned} h_\perp^m &= \frac{1}{2}\sin^3\vartheta\big(1+\zeta+3\zeta\cos 2\vartheta + |1+3\zeta+3\zeta\cos 2\vartheta|\big), \\ h_\parallel^m &= -\frac{1}{4}\cos\vartheta\big(3-\zeta+(1+2\zeta)\cos 2\vartheta + 3\zeta\cos^2 2\vartheta - 2|1+3\zeta+3\zeta\cos 2\vartheta|\sin^2\vartheta\big). \end{aligned} \tag{200}$$

The concept of the limit-of-metastability surfaces and curves plays a fundamental role, because it leads to an elegant graphical representation of the stability properties of nanoparticles.[41, 42] Switching field curves and surfaces have been measured yielding, therefore, estimates of magnetic anisotropy in a single nanoparticle.[11, 43, 44] In particular, the distinct anisotropy contributions (uniaxial, biaxial, cubic) can be separated.[43] Switching field surface measurements[11, 43, 44] have also verified that the angular dependence of the switching field is in agreement with Brown's model.

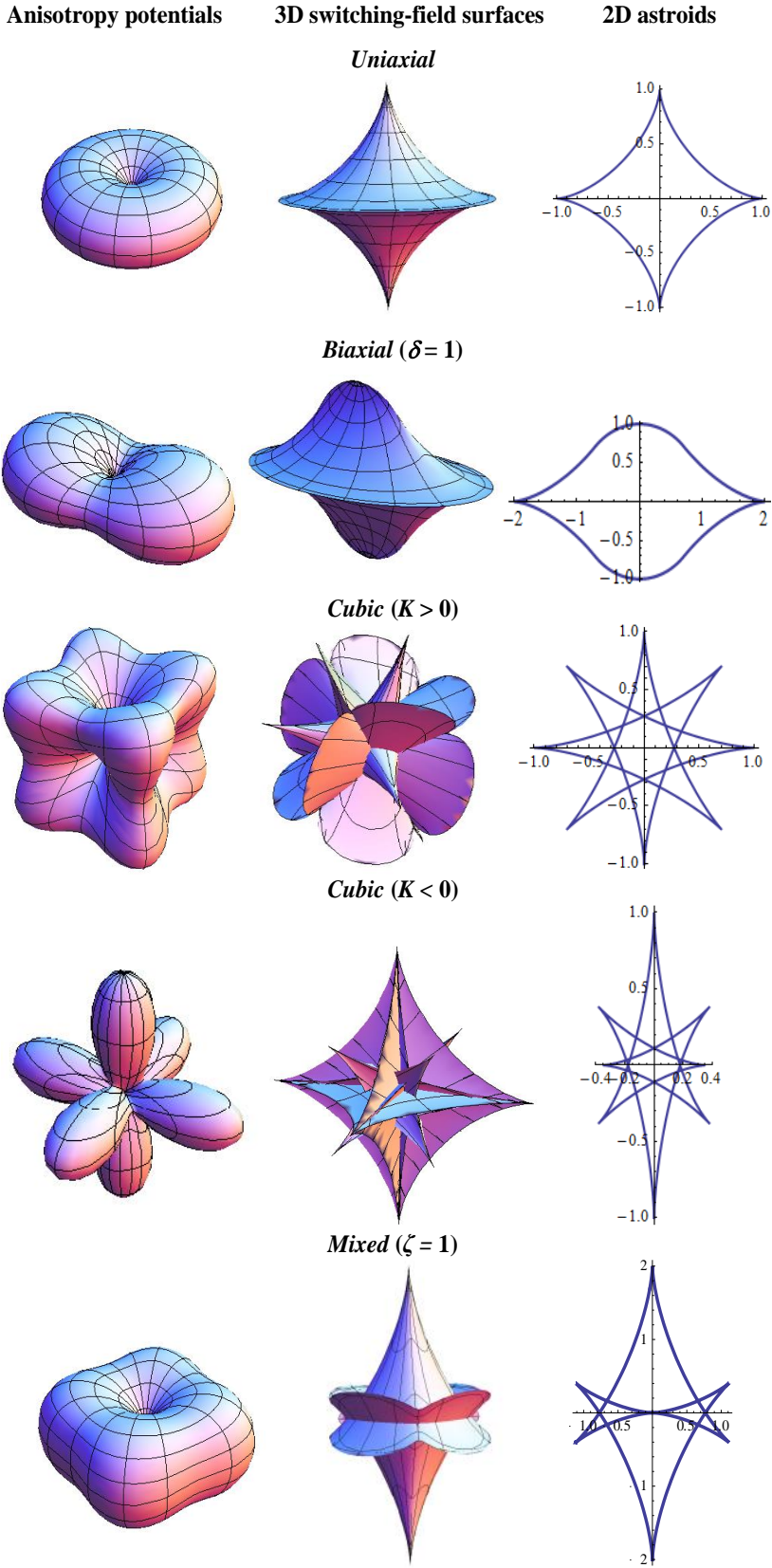

FIG. 21. 3D Switching field surfaces and 2D curves (astroids) for uniaxial, biaxial, positive and negative cubic, and mixed anisotropies.

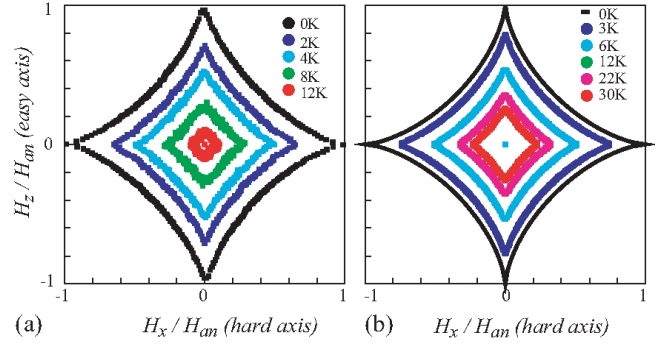

FIG. 22. Temperature dependence of the switching field of a 3 nm Co particle. (a) Experimental data [43] and (b) numerical calculations for a uniaxial anisotropy with $\alpha = 0.1$. Reprinted from C. Vouille, A. Thiaville, and J. Miltat, Thermally activated switching of nanoparticles: a numerical study, J. Magn. Magn. Mater., Vol. **272–276**, e1237-e1238,[45] Copyright (2004), with permission from Elsevier

## C. Finite temperatures

At zero temperature, magnetization reversal is only possible if the energy barrier is *fully suppressed*. At finite temperatures, the switching field becomes (intuitively) smaller, therefore relaxation of the magnetization must be accounted for. Moreover, *experimental* observation of the magnetization reversal depends on the relaxation time of the cluster *and* on the measuring time $\tau_m$ of the experimental setup. Therefore the magnetization reversal can be *experimentally* observed only if the relaxation time lies *in the time window of the experiment, or* equivalently, if the relaxation rate is equal to the measuring frequency $f_m = 1/(2\pi\tau_m)$. Hence, for experimental observation of the magnetization reversal at finite temperatures, Néel's criterion for the observation of the magnetization reversal must hold, namely,[45,149]

$$\tau_m = \tau(\psi, \phi, h_S, \alpha, \text{anisotropy parameters}), \qquad (201)$$

where $\tau$ is the reversal time of the magnetization and $h_S$ is the normalized switching field. This equation can be solved numerically for $h_S$ and $\tau_m$ for given values of $\psi, \phi, \alpha$, and anisotropy parameters.

The temperature dependences of the Switching field curves and surfaces of magnetic nanoparticles have been measured experimentally[11,44] using the micro-SQUID technique. This technique allows one to measure the 3D Switching field surfaces as well as the temperature dependence of individual grains, simultaneously yielding their magnetic anisotropy and allowing one to probe the magnetization dynamics. Temperature effects on the Switching field curves can also be treated theoretically. For example, Vouille *et al.*[45] have calculated temperature-dependent switching curves by numerically solving the stochastic Landau–Lifshitz–Gilbert equation for uniaxial anisotropy. As in the experimental results, they have shown by numerical solution that Brown's diffusion model[9] can reproduce the main features of the measured astroids, which continuously diminish with increasing temperature and ultimately vanish at the blocking temperature, thus corroborating the experimental results (see Fig. 22). They also found that an Arrhenius law for the relaxation time $\tau$ corresponds closely to the calculations, thus allowing a determination of the attempt frequency. The results are also found to compare favorably with the escape-rate formulas for $\tau$.

## VI. CONCLUSIONS

We have comprehensively reviewed Brown's diffusion model of the magnetization relaxation of superparamagnetic nanoparticles, treating the reversal time both via the Kramers escape-rate theory adapted to classical spin systems and numerical solution of the Fokker-Planck equation for the evolution of the surface density of magnetic moment orientations. The numerical solution of that equation yields the smallest nonvanishing eigenvalue so providing a benchmark solution for the reversal time on which all the asymptotic and computer simulation solutions must be judged. Using the Kramers method, as ingeniously adapted to magnetic relaxation by Brown,[8] we have given simple approximate analytical solutions for the reversal time over wide ranges of temperature and damping. The good agreement with the numerical solutions also amply demonstrates that escape-rate theory provides us with an ideal tool for treating the relaxation processes in superparamagnets.

Concerning the range of applicability of particular results of escape-rate theory, we have demonstrated that

(1) the TST escape rate $\Gamma_i^{\text{TST}}$ defined as

$$\Gamma_i^{\text{TST}} = \frac{\omega_i}{2\pi} e^{-\Delta V_i}, \tag{202}$$

comprises the *intermediate damping (ID) limit* ($\alpha S_i \sim 1$), where it yields a quantitative estimate of the escape rate. In general, $\Gamma_i^{\text{TST}}$ yields an *upper bound* for the escape rate in a superparamagnet.

(2) In the *true* VLD limit, ($\alpha S_i \ll 1$), or *energy controlled diffusion*, where the energy loss per cycle of the almost periodic motion of the magnetization on the saddle point energy (escape) trajectory is much less than the thermal energy, the VLD escape rate $\Gamma_i^{\text{VLD}}$ is just as that for point particles, viz.,

$$\Gamma_i^{\text{VLD}} = \alpha S_i \Gamma_i^{\text{TST}}, \tag{203}$$

where $S_i$ is the dimensionless action at the saddle point energy defined by Eq. (91). Here $\Gamma_i^{\text{VLD}}$ is *directly proportional to the damping constant*,

(3) In IHD or *spatially controlled diffusion* ($\alpha S_i \gtrsim 1$), the IHD escape rate $\Gamma_i^{\text{IHD}}$ of classical spins for $\alpha \gg 1$ [the so called very high damping (VHD) limit] is given by Eq. (84) and is essentially *proportional to the inverse of the dissipation constant,* i.e.,

$$\Gamma_i^{\text{IHD}} = \frac{\Omega_0}{\omega_0}\Gamma_i^{\text{TST}}, \quad \Omega_0/\omega_0 \underset{\alpha \gg 1}{\sim} \alpha^{-1}, \quad (204)$$

where $\Omega_0/\omega_0$ is a damping-dependent prefactor given by Eqs. (86) and (87). While, if the dissipation constant tends to zero, $\alpha \to 0$, $\Omega_0/\omega_0 \approx 1$, $\Gamma_i^{\text{IHD}}$ yields Néel's TST formula Eq. (202).

(4) In general, the correction (depopulation) factor $A(\alpha S_i)$ must be incorporated in the escape rate equation, viz.,

$$\Gamma_i = A(\alpha S_i)\frac{\Omega_0(\alpha)}{\omega_0}\Gamma_i^{\text{TST}} \quad (205)$$

An explicit equation for the depopulation factor $A(\alpha S_i)$ is given by Eqs. (14) and (91), providing a reliable estimate of the escape rate for all damping.

The quantitative agreement of the escape rate formulas with the results of numerical solution of the Fokker-Planck equation (18) in damping behavior may be explained as follows. The behavior of the escape rate as a function of the barrier height parameter $\sigma$ for large $\sigma$ is approximately Arrhenius-like and arises from an *equilibrium* property of the system (namely the Boltzmann distribution at the bottom of the well). On the other hand, the damping dependence of the escape rate is entirely due to *nonequilibrium* (dynamical) properties of the system and so is contained in the prefactor of the exponential only, the detailed nature of which depends on the behavior of the energy distribution function at the saddle point. The Mel'nikov approach[55] yields the distribution function at the saddle point for all values of the damping allowing one to evaluate the damping dependence of this prefactor. We remark that as emphasized by Kramers, it is hardly ever of any practical importance to improve on the accuracy of the IHD or VLD formulas themselves because in experimental situations where relaxation is studied, one has only estimates of the prefactor within a certain degree of accuracy which is difficult to evaluate. For example, little detailed information about the value of $\alpha$ exists. Nevertheless, it is important to determine the relaxation times as an intrinsic function of $\alpha$ using asymptotic methods since they reveal the nature of the physical mechanisms governing the relaxation process which cannot be gleaned from the purely numerical methods. A prominent example is Garanin's physical explanation of the depletion effect of a bias field on the relaxation in a double-well potential[125] which was first discovered via numerical solution of Brown's Fokker-Planck equation.[123]

Regarding the most important problem in superparamagnetism, namely the magnetization reversal owing to thermal agitation, we have obtained simple analytical formulas for the superparamagnetic relaxation time $\tau$ for magnetic nanoparticles. These are valid in the high-barrier limit and for *various kinds of anisotropy;* for uniaxial, biaxial, and cubic anisotropies, these formulas are summarized in Table 2. Equations for $\tau$ for other types of anisotropy, despite their relative complexity, are also presented in the review in a form suitable for comparison with experiments. These include uniaxial anisotropy plus the Zeeman term with an arbitrary orientation of the d.c. magnetic field (Section IV.C), biaxial anisotropy plus the Zeeman term (Section IV.E), and mixed (uniaxial and cubic) anisotropy (Section IV.F).

**Table 2.** Reversal time of magnetization for uniaxial, biaxial, and cubic anisotropy.

| Section | Dimensionless anisotropy $V(\vartheta,\varphi)$ | Reversal time $\tau$ and conditions of validity |
|---|---|---|
| III.C | Uniaxial $\sigma\sin^2\vartheta$ | $\tau_0(\alpha+\alpha^{-1})\dfrac{\sqrt{\pi}}{2\sigma^{3/2}}e^{\sigma}\left(1+\dfrac{1}{\sigma}+\dfrac{7}{4\sigma^2}+...\right)$ <br> $\sigma \gg 1$ |
| III.C | Uniaxial+longitudinal d.c. field $\sigma\left(\sin^2\vartheta - 2h\cos\vartheta\right)$ | $\dfrac{\tau_0(\alpha+\alpha^{-1})\sqrt{\pi}e^{\sigma(1-h)^2}}{\sigma^{3/2}(1-h^2)\left[1-h+(1+h)e^{-4\sigma h}\right]}$ <br> $\sigma(1-h)^2 \gg 1,\ 0\le h<1$ |
| IV.C | Uniaxial+transverse d.c. field $\sigma\left(\sin^2\vartheta - 2h\sin\vartheta\cos\varphi\right)$ | $\dfrac{2\tau_0(\alpha+\alpha^{-1})\pi\sqrt{h/(1+h)}e^{\sigma(1-h)^2}A(2\alpha S)}{\sigma[1-2h+\sqrt{1+4h(1-h)\alpha^{-2}}]A^2(\alpha S)}$ <br> $S=\sigma\sqrt{h}\left(16-\dfrac{104}{3}h+22h^2-3h^3+\cdots\right)$ <br> $\sigma(1-h)^2 \gg 1,\ 0.03\le h<1$ |
| IV.D | Cubic ($K>0$) $\sigma\left(\sin^4\vartheta\sin^2 2\varphi + \sin^2 2\vartheta\right)$ | $\dfrac{\tau_0(\alpha+\alpha^{-1})\pi e^{\sigma}}{2\sqrt{2}\sigma\left(\sqrt{9+8/\alpha^2}+1\right)A(8\sqrt{2}\sigma\alpha/9)}$ <br> $\sigma \gg 1$ |
| IV.D | Cubic ($K<0$) $-|\sigma|\left(\sin^4\vartheta\sin^2 2\varphi + \sin^2 2\vartheta\right)$ | $\dfrac{3\tau_0(\alpha+\alpha^{-1})\pi e^{|\sigma|/3}}{2\sqrt{2}|\sigma|\left(\sqrt{9+8/\alpha^2}-1\right)A(\alpha|\sigma|8\sqrt{2}/9)}$ <br> $|\sigma|/3 \gg 1$ |
| IV.E | Biaxial $\sigma\sin^2\vartheta\left(1+\delta\cos^2\varphi\right)$ | $\dfrac{\tau_0(\alpha+\alpha^{-1})\pi e^{\sigma}A(8\alpha\sigma\sqrt{\delta})A^{-2}(4\alpha\sigma\sqrt{\delta})}{\sigma\sqrt{1+1/\delta}\left[1-\delta+\sqrt{(1+\delta)^2+4\delta/\alpha^2}\right]}$ <br> $\sigma \gg 1,\ \delta \gtrsim 0.03$ |
| | $\tau_0=\dfrac{vM_S}{2kT\gamma},\quad A(\Delta)=\exp\left[\dfrac{1}{\pi}\displaystyle\int_0^{\infty}\dfrac{\ln\left\{1-\exp\left[-\Delta(\lambda^2+1/4)\right]\right\}}{\lambda^2+1/4}d\lambda\right]$ | |

Thus the problem of calculation of the reversal time of the magnetization of magnetic nanoparticles for the simplest nonaxially symmetric cases in all damping ranges may be considered as solved and more complicated nonaxially symmetric cases involving noninteracting particles can in principle be solved by the same techniques. However, other relatively more complex problems, such as the reversal time of systems of two interacting spins and of assemblies of interacting spins,[154-156] the reversal time of noninteracting single domain particles with nonaxially symmetric anisotropy driven by an external a.c. magnetic field,[157-159] and reversal driven by a colored noise, i.e., a random field with a finite correlation time,[84] remain to be investigated. Finally, the methods of Brown can also be applied, with small modifications, to thermal agitation in current-induced magnetization dynamics in nanomagnets[160, 161] whereby a current of spin polarized electrons is capable of applying nonconservative torques to a nanoscale ferromagnet. In the spin-torque case, the dynamical equation for the magnetization $\mathbf{M}$ augmented by a random field $\mathbf{h}(t)$ with Gaussian white-noise properties (assuming uniform magnetization in the free layer) is (in our notation)

$$\dot{\mathbf{M}} = \gamma \left[ \mathbf{M} \times \left( -\frac{\partial V}{\partial \mathbf{M}} - M_S^{-1} \mathbf{M} \times \frac{\partial \Phi}{\partial \mathbf{M}} + \mathbf{h} - \frac{\alpha}{\gamma M_S} \dot{\mathbf{M}} \right) \right], \quad (206)$$

where

$$\Phi(\mathbf{M}) = \frac{M_S^2 b_p J_e}{c_p J_p} \ln\left(1 + c_p M_S^{-1} \mathbf{M} \cdot \mathbf{e}_p\right),$$

the unit vector $\mathbf{e}_p$ identifies the magnetization direction in the fixed layer, $J_e$ is the current density, taken as positive when the electrons flow from the free into the fixed layer, while $J_p = M_S^2 |e| d / \hbar$ ($e$ is the electron charge, $\hbar$ is the reduced Planck constant, and $d$ is the free layer thickness). The coefficients $b_p$ and $c_p$ are model-dependent. In the treatment originally proposed in Ref. 160 these coefficients are determined by

$$b_p = \frac{4P^{3/2}}{3(1+P)^3 - 16P^{3/2}}, \quad c_p = \frac{(1+P)^3}{3(1+P)^3 - 16P^{3/2}}.$$

One finds that $0 < b_p < 1/2$ and $1/3 < c_p < 1$ when $P$ is increased from 0 to 1. The typical value of $J_p$ for a few nanometers thick layer is $J_p = 10^9$ A/cm$^2$. The Langevin equation (206) and the corresponding Fokker-Planck equation can be analyzed by the methods outlined in this review, i.e., we can estimate the reversal time of the magnetization via the matrix continued fraction approach and escape rate theory, etc. (see, e.g., refs. 167 and 168). The overall situation (albeit more complicated) is in some way reminiscent of that occurring in the resistively shunted junction (RSJ) model[58,108] of a Josephson junction, which is an electric analog of the motion of a Brownian particle in a tilted periodic potential. This is so because like the bias current in the junction (which constitutes a nonconservative electrical source giving rise to the motion in a tilted cosine periodic potential) ensuing *inter alia* that the stationary distribution is no longer the Boltzmann distribution, the spin-torque term in Eq. (206) also constitutes a nonconservative source. Thus once again, the stationary distribution is no longer the Boltzmann distribution as it depends both on the spin-polarized current and damping analogous to the dependence of the stationary distribution in the RSJ model on the bias current or tilt parameter. Some of the consequences being that the switching time is systematically smaller than Brown's intrinsic thermally activated time in the low damping regime and that the damping parameter now governs the barrier heights. Moreover, the effect of the spin polarized current may be as much as orders of magnitude. The spin-torque effect is very important in applications to current controlled memory cells or microwave sources and resonators.[161] Yet another application is in fast and reliable nanosecond level writing for spin-torque induced switching in memory and recording technologies,[162] where the overall applied field is greater than the critical field at which the double-well nature of the potential disappears. In the context of spin-torque effects, which always involve a current it should be reiterated that the purely mathematical method of approximate minimization for the calculation of the smallest nonvanishing eigenvalue of the Sturm-Liouville equation based on the calculation of variations is also extremely useful as it automatically avoids the concept of zero divergence of the current which always enters into the escape rate theory.

**ACKNOWLEDGEMENTS**

We would like to thank a number of individuals who have greatly helped us, directly or indirectly, in the preparation of the present review. In particular, we thank B. Barbara, R. Chantrell, O. Chubykalo-Fesenko, P.J. Cregg, D.S.F. Crothers, the late J.-L. Dormann, P. Fannin, D. Garanin, L. Geoghegan, E. Kennedy, D. J. McCarthy, H. El Mrabti, B. Mulligan, B. Ouari, Y.L. Raikher, B. Scaife, V.I. Stepanov, A. Thiaville, J. T. Waldron, J.-E. Wegrowe, and W. Wernsdorfer. Special thanks are due to S. V. Titov and P.-M. Déjardin, who have carefully read the entire manuscript and proposed a number of corrections and improvements in the presentation. The partial support of the work by the European Community (Programme FP7, project DMH, No. 295196) is gratefully acknowledged.

## APPENDIX A: DIFFERENTIAL-RECURRENCE EQUATION FOR THE STATISTICAL MOMENTS

We shall now demonstrate how the hierarchy of differential-recurrence relations (36) for the averages governing the relaxation dynamics of single-domain ferromagnetic particles can be obtained from the stochastic Gilbert and Fokker-Planck equations. Following Ref. 110, we first transform Gilbert's equation, Eq. (15), to this hierarchy. As we have seen, in magnetic applications, the relevant observables are averages involving the spherical harmonics. Thus, we have from Eqs. (98) and (99) the stochastic differential equation of motion for the spherical harmonic $Y_{l,m}[\vartheta(t), \varphi(t)]$, viz.

$$\dot{Y}_{l,m} = \dot{\varphi}(t)\frac{\partial Y_{l,m}}{\partial \varphi} + \dot{\vartheta}(t)\frac{\partial Y_{l,m}}{\partial \vartheta} = -\frac{b}{\sin \vartheta(t)}\left(\frac{1}{\sin \vartheta(t)}\frac{\partial V}{\partial \varphi} + \frac{1}{\alpha}\frac{\partial V}{\partial \vartheta}\right)\frac{\partial Y_{l,m}}{\partial \varphi}$$

$$-b\left(\frac{\partial V}{\partial \vartheta} - \frac{1}{\alpha \sin \vartheta(t)}\frac{\partial V}{\partial \varphi}\right)\frac{\partial Y_{l,m}}{\partial \vartheta} + bM_S\left[h_\vartheta(t) - \frac{h_\varphi(t)}{\alpha}\right]\frac{\partial Y_{l,m}}{\partial \vartheta} + \frac{bM_S}{\sin \vartheta(t)}\left[h_\varphi(t) + \frac{h_\vartheta(t)}{\alpha}\right]\frac{\partial Y_{l,m}}{\partial \varphi}$$

or, equivalently, in vector notation,

$$\dot{Y}_{l,m} = -bM_S\mathbf{h}\cdot\left\{[\mathbf{u}\times\nabla Y_{l,m}] + \alpha^{-1}\nabla Y_{l,m}\right\} - b\nabla Y_{l,m}\cdot\left\{\nabla V + \alpha^{-1}[\mathbf{u}\times\nabla V]\right\}. \quad (A.1)$$

Here $\nabla$ is the orientation space gradient operator defined as

$$\nabla = \mathbf{u}\times\frac{\partial}{\partial \mathbf{u}} = -\mathbf{e}_\vartheta\frac{1}{\sin \vartheta}\frac{\partial}{\partial \varphi} + \mathbf{e}_\varphi\frac{\partial}{\partial \vartheta}.$$

On averaging the stochastic equation, Eq.(A.1), according to the Stratonovich rule as described in detail in Refs. 58 and 110, we have

$$\frac{d}{dt}\langle Y_{l,m}\rangle = \frac{1}{2\tau_N}\left\langle \Delta Y_{l,m} + \frac{v}{2kT}\left\{V\Delta Y_{l,m} + Y_{l,m}\Delta V - \Delta(VY_{l,m}) + \frac{2}{\alpha \sin \vartheta}\left[\frac{\partial V}{\partial \varphi}\frac{\partial Y_{l,m}}{\partial \vartheta} - \frac{\partial V}{\partial \vartheta}\frac{\partial Y_{l,m}}{\partial \varphi}\right]\right\}\right\rangle. \quad (A.2)$$

Next, we can express the right-hand side of Eq. (A.2) in terms of the angular momentum operator $\hat{L} = -i\nabla$.[109] We recall first that the operators $\hat{L}_Z$, $\hat{L}_\pm$, $\hat{L}^2$ are defined as[109]

$$\hat{L}^2 = -\Delta, \ \hat{L}_Z = -i\frac{\partial}{\partial \varphi}, \ \hat{L}_\pm = e^{\pm i\varphi}\left(\pm\frac{\partial}{\partial \vartheta} + i\cot\vartheta\frac{\partial}{\partial \varphi}\right). \quad (A.3)$$

The right hand side of Eq. (A.2) may ultimately be written as a linear combination of averages of spherical harmonics by using the theory of angular momentum,[109] because the action of the operators $\hat{L}_Z$, $\hat{L}_\pm$, $\hat{L}^2$ on $Y_{l,m}$ is[109]

$$\hat{L}_Z Y_{l,m} = mY_{l,m},$$

$$\hat{L}^2 Y_{l,m} = l(l+1)Y_{l,m},$$

$$\hat{L}_\pm Y_{l,m} = \sqrt{l(l+1) - m(m\pm 1)}Y_{l,m\pm 1}.$$

Thus for an arbitrary magnetocrystalline anisotropy, which can be expressed in terms of spherical harmonics as

$$\frac{vV}{kT} = \sum_{R=1}^{\infty}\sum_{S=-R}^{R} A_{R,S}Y_{R,S} \quad (A.4)$$

we can transform Eq. (A.2) into a differential-recurrence equation (36), namely, (details in Ref. 110)

$$\tau_N \frac{d}{dt}\langle Y_{l,m}\rangle(t) = \sum_{s,r} e_{l,m,l+r,m+s}\langle Y_{l+r,m+s}\rangle(t), \quad (A.5)$$

where $e_{l,m,l',m'}$ is defined by Eq. (38).

Moreover, we can also derive the same results from the Fokker–Planck equation (18) by seeking a solution of the form

$$W(\vartheta,\varphi,t) = \Psi(\vartheta,\varphi,t)\Psi^*(\vartheta,\varphi,t), \tag{A.6}$$

where $\Psi(\vartheta,\varphi,t)$ is given by

$$\Psi(\vartheta,\varphi,t) = \sum_{l,m} f_{l,m}(t) Y_{l,m}(\vartheta,\varphi). \tag{A.7}$$

The normalization condition for $W(\vartheta,\varphi,t)$ is

$$\int_0^{2\pi}\int_0^{\pi} W(\vartheta,\varphi,t) d\Omega = \sum_{l,m} |f_{l,m}(t)|^2 = 1. \tag{A.8}$$

The representation in Eq. (A.6) has the advantage that it is unnecessary to apply additional conditions to the distribution function in order that it should be physically meaningful (e.g., $W$ should be positive and real). Moreover, the *direct quantum-mechanical analogy* is obvious, because $W$ is now similar to the quantum probability density $|\Psi|^2$ ($\Psi$ is the wave function), which obeys the continuity equation (Ref. 163, p.75) $\partial_t |\Psi|^2 + \text{div}\,\mathbf{j} = 0,$ where $\mathbf{j}$ is the probability current density.

We then have from Eqs. (18), (A.6), and (A.7) the moment system for the averaged spherical harmonics, via the transformation

$$\frac{d}{dt}\langle Y_{l,m}\rangle(t) = \int_0^{2\pi}\int_0^{\pi} Y_{l,m} \dot{W} d\Omega = \sum_{l',l'',m',m''} f_{l',m'}(t) f^*_{l'',m''}(t) \int_0^{2\pi}\int_0^{\pi} Y_{l,m} L_{\text{FP}}\left(Y_{l',m'}Y^*_{l'',m''}\right) d\Omega$$

$$= \sum_{l',l'',l''',m',m'',m'''} \sqrt{\frac{(2l'+1)(2l''+1)}{4\pi(2l'''+1)}} C^{l''',0}_{l',0,l'',0} C^{l''',m'''}_{l',m',l'',m''} f_{l',m'}(t) f^*_{l'',m''}(t) \int_0^{2\pi}\int_0^{\pi} Y_{l,m} L_{\text{FP}} Y^*_{l''',m'''} d\Omega$$

$$= \sum_{l',m'} d_{l',m',l,m} \langle Y_{l',m'}\rangle(t), \tag{A.9}$$

where

$$d_{l',m',l,m} = \int_0^{2\pi}\int_0^{\pi} Y_{l,m} L_{\text{FP}} Y^*_{l',m'} d\Omega \tag{A.10}$$

are the matrix elements of the Fokker–Planck operator $L_{\text{FP}}$,

$$\langle Y_{l,m}\rangle(t) = \sum_{l',l'',m',m''} \sqrt{\frac{(2l+1)(2l'+1)}{4\pi(2l''+1)}} C^{l'',0}_{l,0,l',0} C^{l'',m''}_{l,m,l',m'} f_{l',m'}(t) f^*_{l'',m''}(t), \tag{A.11}$$

and $C^{c,\gamma}_{a,\alpha,b,\beta}$ are the Clebsch–Gordan coefficients.[109] Moreover, we can express the operator $L_{\text{FP}}$ in Eq. (A.10) as before in terms of the angular momentum operators $\hat{L}_z, \hat{L}_{\pm}, \hat{L}^2$ and we have (details in Ref. 110)

$$d_{l',m',l,m} = \frac{1}{\tau_N} e_{l,m,l',m'}. \tag{A.12}$$

Equation (A.12) demonstrates the equivalence of the approaches based on either the Langevin or the Fokker–Planck equation.

# APPENDIX B. LANGER'S GENERALIZATION OF KRAMERS' THEORY TO MANY DIMENSIONS IN THE IHD LIMIT

We have seen that the original IHD treatment of Kramers pertained to a mechanical system of one degree of freedom specified by the coordinate $x$ with additive Hamiltonian $H = p^2/2m + V(x)$. Thus, the motion is separable and described by a 2D phase space with state variables $(x, p)$. However, this is not always so. For example, the motion of the magnetic moment in a single-domain ferromagnetic particle is governed by a nonadditive Hamiltonian, which is simply the magnetocrystalline anisotropy energy of the particle, so that the system is nonseparable.

The phase-space trajectories in the Kramers problem of the underdamped motion are approximately ellipses. The corresponding trajectories in the magnetic problem are much more complicated because of the nonseparable form of the energy. Similar considerations hold in the extension of the Debye theory of dielectric relaxation to include inertia, as in this case one would usually have a six-dimensional phase space corresponding to the orientations and angular momenta of the rotator. These, and other considerations, suggest that the Kramers theory should be extended to a multi-dimensional phase space.

Such generalizations, having been instigated by Brinkman,[164] were further developed by Landauer and Swanson.[165] However, the most complete treatment is due to Langer in 1969,[86] who considered the IHD limit. As a specific example of the application of the theory, we have used it to calculate the IHD magnetic relaxation time for a single-domain ferromagnetic particle for an arbitrary nonaxially symmetric potential of the magnetocrystalline anisotropy in that limit (see Section IV.A *et seq.* and Appendix C).

Before proceeding, we remark that a number of other interesting applications of the theory, which, as the reader will appreciate, is generally concerned with the nature of metastable states and the rates at which these states decay, have been mentioned by Langer[86] and we briefly summarize these. Examples are:

(1) A supersaturated vapor[87] which can be maintained in a metastable state for a very long time but which will eventually undergo condensation into the more stable liquid phase.
(2) A ferromagnet, which can persist with its magnetization pointing in a direction opposite to that of an applied magnetic field.
(3) In metallurgy, an almost identical problem occurs in the study of alloys whose components tend to separate on ageing or annealing.
(4) The final examples quoted by Langer are the theories of superfluidity and superconductivity, where states of nonzero superflow are metastable and so may undergo spontaneous transitions to states of lower current and greater stability.

According to Langer,[86] all the phase transitions above take place via the nucleation and growth of some characteristic disturbance within the metastable system. Condensation of the supersaturated vapor is initiated by the formation of a sufficiently large droplet of the liquid. If this droplet is big enough, it will be more likely to grow than to dissipate, and so will bring about condensation of the entire sample. If the nucleating disturbance appears spontaneously as a thermodynamic fluctuation it is said to be *homogeneous*. This is an intrinsic thermodynamic property of the system and is the type of disturbance described by Langer,[86] which we shall summarize here. The other type of nucleation is *inhomogeneous nucleation*, which occurs when the disturbance leading to the phase transition is caused by a foreign object, for example an irregularity in the walls of the container or some agent that is not part of the system of direct interest.

The above examples have been chosen in order to illustrate the breadth of applicability of the theory. However, Langer's method, since it can in effect be applied to a system of multiple degrees of freedom, is likely to be of much use in calculating relaxation times for fine particle magnetic systems in which other types of interaction, such as exchange and dipole–dipole coupling, also appear. We also emphasize that Langer's treatment of the homogeneous nucleation problem contains within it the magnetic case of the Kramers' IHD calculation. The multi-dimensional Kramers problem was first solved in the VHD limit by Brinkman[164] and Landauer and Swanson.[165] A general discussion of this problem is given in Chapter 7 of Frenkel[166] on the kinetics of phase transitions.

For easy comparison with previous work, we shall adopt the notation of Ref. 56. Thus, we shall consider the Fokker–Planck equation for a multi-dimensional random process governed by a state vector $\{\eta\}$ which is[53,86]

$$\frac{\partial}{\partial t}\rho(\{\eta\},t) = \sum_{i=1}^{2N}\sum_{n=1}^{2N}\frac{\partial}{\partial \eta_i}\left[M_{in}\left(\frac{\partial E}{\partial \eta_n} + kT\frac{\partial}{\partial \eta_n}\right)\right]\rho(\{\eta\},t). \tag{B.1}$$

In Eq. (B.1), $E(\{\eta\})$ is a Hamiltonian (energy) function having two minima at points *A* and *B*, separated by a saddle point *C* surrounded by two wells. One well, say the one at *B*, is at a much lower energy than the other. The particles have to transverse the saddle point, which acts as a barrier at *C*. We again assume that the barrier height $\Delta V = E_C - E_A$ is very high (at least of the order of $5kT$), so that the diffusion over the barrier is slow enough to ensure that a Maxwell–Boltzmann distribution is established and maintained near *A* at all times. The high barrier also assures that the contribution to the flux over the saddle point will come mainly from a small region around *C* and that quasi-stationary conditions will prevail. The 2*N* state variables $\{\eta\} = \{\eta_1, \eta_2, \ldots, \eta_{2N}\}$ are parameters, which could equally well be the coordinates and momenta of a point in phase space, or angular coordinates describing the orientation of the magnetization vector of a single-domain ferromagnetic particle. Generally, however, the first *N* of the $\eta_i$'s will be functions of the *N* coordinates of position[53]

$$\eta_i = \eta(x_i), \quad i = 1, 2, \ldots, N. \tag{B.2}$$

The second $N$ of the $\eta_i$'s will be the conjugate momenta $\pi_i$, namely

$$\eta_{i+N} = \pi_i \quad i=1,2,\ldots,N. \tag{B.3}$$

In fact, the $\eta_i$'s will often (although not necessarily) be the coordinates themselves, in which case (obviously) $\eta_i = x_i, i = 1,2,\ldots,N$. Here, when the noise term in the Langevin equation is ignored, the system evolves in accordance with the deterministic equation

$$\dot{\eta}_i = -\sum_n M_{in} \frac{\partial E}{\partial \eta_n}, \tag{B.4}$$

where $M_{ij}$ are the matrix elements of the transport matrix $\mathbf{M}$, which, for simplicity, we shall assume to be constant.

We may define the matrices $\mathbf{D}$ and $\mathbf{A}$ by the equations

$$\mathbf{D} = \frac{1}{2}(\mathbf{M} + \mathbf{M}^T) \text{ and } \mathbf{A} = \frac{1}{2}(\mathbf{M} - \mathbf{M}^T), \tag{B.5}$$

where $\mathbf{M} = (M_{ij})$ is the *transport* matrix resulting from Eq. (B.4), and the symbol "T" means matrix transposition. Matrix $\mathbf{D}$ is called the *diffusion* matrix, which characterizes the thermal fluctuations due to the heat bath, while matrix $\mathbf{A}$ describes the motion in *the absence of the bath*, i.e., the inertial term in the case of mechanical particles, and if $\mathbf{D}$ is not identically zero, then the dissipation of energy satisfies[53]

$$\dot{E} = -\sum_{i,n} \frac{\partial E}{\partial \eta_i} D_{in} \frac{\partial E}{\partial \eta_n} \leq 0. \tag{B.6}$$

We consider, as before, a single well and suppose that, at finite temperatures, a Maxwell–Boltzmann distribution is set up and the density at equilibrium is

$$\rho_{eq}(\{\boldsymbol{\eta}\}) = \frac{1}{Z} e^{-E(\{\boldsymbol{\eta}\})/(kT)}, \tag{B.7}$$

where

$$Z \equiv \int_{-\infty}^{\infty} \cdots \int_{-\infty}^{\infty} e^{-E/(kT)} d\eta_1 \cdots d\eta_{2N} \tag{B.8}$$

is the partition function. The IHD escape rate for this multivariable problem may again be calculated by the flux-over-population method.

We make the following assumptions about $\rho(\{\boldsymbol{\eta}\})$:

(1) It obeys the quasi-stationary Fokker–Planck equation (i.e., $\dot{\rho} = 0$), which is (on linearization about the saddle point):

$$\sum_{i,n} \frac{\partial}{\partial \eta_i} M_{in} \left[ \sum_k e_{nk}(\eta_k - \eta_k^s) + kT \frac{\partial}{\partial \eta_n} \right] \rho = 0, \tag{B.9}$$

where the $e_{jk}$ are the coefficients in the Taylor expansion of the energy about the saddle point truncated at the second term, namely the quadratic (form) approximation

$$E(\{\boldsymbol{\eta}\}) = E_C - \frac{1}{2}\sum_{i,n} e_{in}(\eta_i - \eta_i^C)(\eta_n - \eta_n^C), \tag{B.10}$$

$\{\boldsymbol{\eta}\} \approx \{\boldsymbol{\eta}^C\}$, and $E_C$ is the value of the energy function at the saddle point (compare Kramers' method above: there the saddle point is a one-dimensional maximum). Equation (B.10) constitutes the paraboloidal approximation to the potential in the vicinity of the saddle point. For example, in magnetic relaxation in a uniform field with uniaxial anisotropy, the energy surface in the vicinity of the saddle point will be a hyperbolic paraboloid.[59] Equation (B.9) is the multi-dimensional Fokker–Planck equation linearized in the region of the saddle point.

(2) Owing to the high barrier, just as in the Kramers high-damping problem, a Maxwell–Boltzmann distribution is set up in the vicinity of the bottom of the well, i.e., at $A$, so:

$$\rho(\{\boldsymbol{\eta}\}) \approx \rho_{eq}(\{\boldsymbol{\eta}\}), \quad \{\boldsymbol{\eta}\} \approx \{\boldsymbol{\eta}^A\}. \tag{B.11}$$

(3) Practically speaking, no particles have arrived at the far side of the saddle point, so we have the sink boundary condition

$$\rho(\{\boldsymbol{\eta}\}) = 0, \quad \{\boldsymbol{\eta}\} \text{ beyond } \{\boldsymbol{\eta}^C\}. \tag{B.12}$$

This is Kramers' condition that only rare particles of the assembly ever cross the barrier. Just as in the Klein–Kramers problem for one degree of freedom, we make the substitution

$$\rho(\{\boldsymbol{\eta}\}) = g(\{\boldsymbol{\eta}\})\rho_{eq}(\{\boldsymbol{\eta}\}), \tag{B.13}$$

where the function $g$ is known as the crossover function. Thus, we obtain from Eqs. (B.7) and (B.9), as before, an equation for $g$, namely

$$\sum_{i,n} M_{ni}\left[-\sum_k e_{nk}\left(\eta_k - \eta_k^C\right) - kT\frac{\partial}{\partial \eta_n}\right]\frac{\partial}{\partial \eta_i} g(\{\boldsymbol{\eta}\}) = 0, \tag{B.14}$$

where $\{\boldsymbol{\eta}\} \approx \{\boldsymbol{\eta}^C\}$. We postulate that $g$ may be written in terms of a single variable $u$, viz.,

$$g(u) = (2\pi kT)^{-1}\int_u^\infty e^{-z^2/(2kT)} dz, \tag{B.15}$$

where $u$ has the form of the linear combination

$$u = \sum_i U_i(\eta_i - \eta_i^C). \tag{B.16}$$

This is simply Kramers' method of forcing the multi-dimensional Fokker–Planck equation into an equation in a single variable $u$ (in his original case, a linear combination of the two variables, position and velocity, so that $u = p - ax'$). We must now determine the coefficients $U_i$ of the linear combination $u$ of the $\eta_j$. This is accomplished as follows. We define the matrix $\tilde{\mathbf{M}} = -\mathbf{M}^T$. Then we shall have the coefficients $U_i$ of the linear combination as a solution of the *eigenvalue problem*

$$-\sum_{i,n} U_i \tilde{M}_{in} e_{nk} = \lambda_+ U_k. \tag{B.17}$$

The eigenvalue $\lambda_+$ is the *deterministic growth rate of a small deviation from the saddle point*, and is the positive eigenvalue of the system matrix of the noiseless Langevin equations, linearized about the saddle point. It characterizes the unstable barrier-crossing mode. Thus, in order to calculate $\lambda_+$, all that is required is a knowledge of the energy landscape; Eq. (B.17) need not, in practice, be involved. Equation (B.17) is obtained essentially by substituting the linear combination $u$, i.e., Eq. (B.16), into Eq. (B.14) for the crossover function, and requiring the resulting equation to be a proper ordinary differential equation in the single variable $u$ with solution given by Eq. (B.15) (the details of this are given in Ref. 56). Equation (B.17) may also be written in the matrix form

$$-\mathbf{U}^T \tilde{\mathbf{M}} \mathbf{E}^C = \lambda_+ \mathbf{U}^T. \tag{B.18}$$

(Hänggi *et al.* [53] describe this equation by stating that $\mathbf{U}^T$ is a "left eigenvector" of the matrix $-\tilde{\mathbf{M}} \mathbf{E}^C$. The usual eigenvalue equation of an arbitrary matrix $\mathbf{A}$ is $\mathbf{A}\mathbf{X} = \lambda \mathbf{X}$. In the above terminology, $\mathbf{X}$ would be a "right eigenvector" of $\mathbf{A}$ ). In Eq. (B.18), $\mathbf{E}^C \equiv (e_{ij})$ is the matrix of the second derivatives of the potential evaluated at the saddle point, which is used in the Taylor expansion of the energy near the saddle point. The determinant of this (Hessian) matrix is the Hessian itself. The normalization of $U_i$ is fixed, so that

$$\lambda_+ = \sum_{i,n} U_i M_{in} U_n, \tag{B.19}$$

which is equivalent to

$$\sum_{i,n} U_i e_{in}^{-1} U_n = -1. \tag{B.20}$$

This condition ensures that the crossover function, Eq. (B.15), retains the form of an error function and so may describe diffusion over a barrier. Alternatively, one may say that the foregoing conditions require that the entry in the diffusion matrix in the direction of flow (that is, the unstable direction) is nonzero; that is, *we have current over the barrier and so particles escape the well*.

Now the Fokker–Planck equation, Eq. (B.1), is in essence a continuity equation for the representative points so that

$$\dot{\rho} + \nabla \cdot \mathbf{J} = 0. \tag{B.21}$$

Thus by inspection, we find that the current density becomes

$$j_i = -\sum_n M_{in} \left( \frac{\partial E}{\partial \eta_n} + kT \frac{\partial}{\partial \eta_n} \right) \rho \tag{B.22}$$

and we obtain, using Eqs. (B.7), (B.14), and (B.15) for the *stationary* current density, i.e., $\dot{\rho} = 0$,

$$j_i(\{\eta\}) = \frac{1}{\sqrt{2\pi}} \sum_n M_{in} U_n \rho_{eq}(\{\boldsymbol{\eta}\}) e^{-\frac{u^2}{2kT}}. \tag{B.23}$$

We now take advantage of the condition stated above, namely that the flux over the barrier emanates from a small region around the saddle point *C*. We integrate the current density over a plane containing the saddle point but not parallel to the flow of particles. The plane $u = 0$ will suffice here. Thus the total current is

$$J = \sum_i \int_{u=0} j_i(\{\boldsymbol{\eta}\}) dS_i. \tag{B.24}$$

Using Eq. (B.24) with the quadratic approximation of Eq. (B.10) for the energy near the saddle point, the integration for the total flux (current) now yields, after a long calculation,[56]

$$J \approx \frac{1}{2\pi Z} \sum_{i,j} U_i M_{ij} U_j \left| \sum_{i,j} U_i e_{ij}^{-1} U_j \det\left((2\pi kT)^{-1} \mathbf{E}^C\right) \right|^{-1/2} e^{-\frac{E_C}{kT}}. \tag{B.25}$$

From Eqs. (B.19) and (B.20), we immediately obtain

$$J = \frac{\lambda_+}{2\pi Z} \left| \det\left((2\pi kT)^{-1} \mathbf{E}^C\right) \right|^{-1/2} e^{-\frac{E_C}{kT}}. \tag{B.26}$$

Now, we assume that the energy function near the bottom of the well *A* may again be written in the quadratic approximation

$$E = E_A + \frac{1}{2} \sum_{i,j} a_{ij} \left(\eta_i - \eta_i^A\right)\left(\eta_j - \eta_j^A\right), \tag{B.27}$$

and we write $\mathbf{E}^A = (a_{ij})$ so that the number of particles in the well is[56]

$$n_A = \left\{\det[(2\pi kT)^{-1} \mathbf{E}^A]\right\}^{-1/2} Z^{-1}. \tag{B.28}$$

Now the escape rate $\Gamma$, by the usual flux-over-population method, is defined to be $\Gamma = J/n_A$, and so from Eqs. (B.26) and (B.28), in terms of the unique positive eigenvalue $\lambda_+$ of the set of *noiseless* Langevin equations linearized about the saddle point, we have

$$\Gamma = \frac{\lambda_+}{2\pi} \sqrt{\frac{\det\{\mathbf{E}^A\}}{|\det\{\mathbf{E}^C\}|}} e^{-\frac{(E_C - E_A)}{kT}}, \tag{B.29}$$

which is Langer's[86] expression in terms of the Hessians of the saddle and well energies for the escape rate for a multi-dimensional process in the IHD limit. The result again pertains to this limit because of our postulate that the potential in the vicinity of the saddle point may be approximated by the first two terms of its Taylor series. Thus Eq. (B.29) fails for very small damping corresponding to *energy controlled diffusion*, because the region of deviation from the Maxwell–Boltzmann distribution prevailing in the depths of the well extends far beyond the narrow region at the top of the barrier in which the potential may be replaced by its quadratic approximation.

## APPENDIX C: ESCAPE RATE FORMULAS FOR SUPERPARAMAGNETS: LANGER'S METHOD

In this Appendix, we show in detail how Langer's method may be used to solve the problem of superparamagnetic relaxation in the IHD limit. Again, we deal with an energy (or Hamiltonian) function, $E = V(\vartheta, \varphi)$, with minima at points *A* and *B* separated by a barrier (saddle point) at *C*. We use spherical polar coordinates $(\vartheta, \varphi)$, where $\vartheta$ is the polar angle and $\varphi$ is the azimuthal angle as usual. The noiseless Gilbert equation, Eq. (27), takes the form in the coordinates $(p = \cos\vartheta, \varphi)$ [9]

$$\dot{p} = -h'(1-p^2)\partial_p V - h'\alpha^{-1}\partial_\varphi V, \tag{C.1}$$

$$\dot{\varphi} = h'\alpha^{-1}\partial_p V - h'(1-p^2)^{-1}\partial_\varphi V, \tag{C.2}$$

where subscripts denote the partial derivatives. We linearize these equations about the saddle point and determine $\lambda_+$ from the transition matrix. Thus, expanding the Hamiltonian $V(p,\varphi)$ as a Taylor series about the saddle point $(p_C = \cos\vartheta_C, \varphi_C)$, we obtain

$$V = V_0 + \frac{1}{2}\left[V_{pp}^{(0)}(p-p_C)^2 + 2V_{p\varphi}^{(0)}(p-p_C)(\varphi-\varphi_C) + V_{\varphi\varphi}^{(0)}(\varphi-\varphi_C)^2\right], \tag{C.3}$$

with the superscript (0) denoting evaluation at the saddle point. We remark, following Klik and Gunther,[83, 84] that the Hamiltonian is defined on a phase space which is a closed manifold (the space $(\vartheta, \varphi)$ is the surface of a unit sphere) and that a local energy minimum is thus surrounded by two or more saddle points, depending on the symmetry of the problem. The total probability flux away from the metastable minimum equals the sum of the fluxes through all the saddle points. In asymmetric cases, e.g., when an external field is applied, some of these fluxes become exponentially small and may safely be neglected. The total flux away from the metastable minimum is then dominated by the energetically most favorable path. Now, if the coordinates of the saddle point are $(\varphi_C, p_C)$, we have

$$\frac{\partial V}{\partial p} = (p-p_C)V_{pp}^{(0)} + (\varphi-\varphi_C)V_{p\varphi}^{(0)}, \tag{C.4}$$

$$\frac{\partial V}{\partial \varphi} = (p-p_C)V_{p\varphi}^{(0)} + (\varphi-\varphi_C)V_{p\varphi}^{(0)}. \tag{C.5}$$

Now, let the saddle point *C* of interest lie on the equator $p = 0$ and make the transformation $\varphi - \varphi_C \to \varphi$. Equations (C.1) and (C.2) then become in matrix notation

$$\begin{pmatrix} \dot{\varphi} \\ \dot{p} \end{pmatrix} = h' \begin{pmatrix} -1 & \alpha^{-1} \\ -\alpha^{-1} & -1 \end{pmatrix} \begin{pmatrix} \dfrac{\partial V}{\partial \varphi} \\ \dfrac{\partial V}{\partial p} \end{pmatrix}. \tag{C.6}$$

Thus, the linearized Eq.(C.6) has the form of the canonical Eqs. (B.4), and so Langer's equation, Eq. (B.29), may be used to calculate the IHD escape rate. In particular, the transport matrix $\mathbf{M}$ and the matrix $\tilde{\mathbf{M}}$ are given by

$$\mathbf{M} = h' \begin{pmatrix} 1 & -\alpha^{-1} \\ \alpha^{-1} & 1 \end{pmatrix}, \quad \tilde{\mathbf{M}} = h' \begin{pmatrix} -1 & -\alpha^{-1} \\ \alpha^{-1} & -1 \end{pmatrix}.$$

The equations of motion (C.6) linearized about the saddle point become [83]

$$\begin{pmatrix} \dot{\varphi} \\ \dot{p} \end{pmatrix} = h' \begin{pmatrix} \alpha^{-1} V_{p\varphi}^{(0)} - V_{\varphi\varphi}^{(0)} & \alpha^{-1} V_{pp}^{(0)} - V_{p\varphi}^{(0)} \\ -V_{p\varphi}^{(0)} - \alpha^{-1} V_{\varphi\varphi}^{(0)} & -V_{pp}^{(0)} - \alpha^{-1} V_{p\varphi}^{(0)} \end{pmatrix} \begin{pmatrix} \varphi \\ p \end{pmatrix}. \tag{C.7}$$

Equations (C.7) are the noiseless Langevin equations given by Klik and Gunther; see Ref. 83, Eq. (3.2). The secular equation of Eq. (C.7) then yields

$$\lambda_{\pm} = h' \left\{ -\frac{V_{pp}^{(0)} + V_{\varphi\varphi}^{(0)}}{2} \pm \sqrt{\left( \frac{V_{pp}^{(0)} + V_{\varphi\varphi}^{(0)}}{2} \right)^2 - \frac{1+\alpha^2}{\alpha^2} \left[ V_{pp}^{(0)} V_{\varphi\varphi}^{(0)} - \left( V_{p\varphi}^{(0)} \right)^2 \right]} \right\}. \tag{C.8}$$

The Hessian matrix of the system is

$$\begin{pmatrix} V_{\varphi\varphi} & V_{p\varphi} \\ V_{p\varphi} & V_{pp} \end{pmatrix}, \tag{C.9}$$

and the Hessian itself is *negative* at the saddle point. Thus, to ensure a growing disturbance at the saddle point, we must again take the positive sign in Eq. (C.8). Now the well angular frequency is defined as

$$\omega_i = \frac{\gamma}{M_S} \sqrt{V_{pp}^{(i)} V_{\varphi\varphi}^{(i)} - (V_{p\varphi}^{(i)})^2}, \tag{C.10}$$

the superscript (*i*) denoting evaluation at the minimum of well *i*, while the saddle angular frequency is

$$\omega_0 = \frac{\gamma}{M_S} \sqrt{\left| V_{pp}^{(0)} V_{\varphi\varphi}^{(0)} - (V_{p\varphi}^{(0)})^2 \right|}, \tag{C.11}$$

which, with Eq.(B.29), leads to the Klik and Gunther result[83]

$$\Gamma = \frac{\lambda_+ \omega_i}{2\pi \omega_0} e^{-v(V_0 - V_i)/(kT)}. \tag{C.12}$$

This formula shows clearly how, once the potential landscape is known, the IHD escape rate may be calculated. If we now choose a local coordinate system $(\varphi, p)$ at the saddle point, where $V_{p\varphi} = 0$, then we obtain a more compact expression for $\omega_i$, $\omega_0$, and $\lambda_+$, namely

$$\omega_i = \frac{\gamma}{M_S}\sqrt{V_{pp}^{(i)}V_{\varphi\varphi}^{(i)}}, \quad \omega_0 = \frac{\gamma}{M_S}\sqrt{\left|V_{pp}^{(0)}V_{\varphi\varphi}^{(0)}\right|},$$

$$\lambda_+ = \frac{h'}{2}\left\{-\left[V_{pp}^{(0)} + V_{\varphi\varphi}^{(0)}\right] + \sqrt{[V_{pp}^{(0)} - V_{\varphi\varphi}^{(0)}]^2 - 4\alpha^{-2}V_{pp}^{(0)}V_{\varphi\varphi}^{(0)}}\right\},$$

where we observe that the $\alpha^{-2}$ term represents the effect of the precessional term in the Gilbert equation on the longitudinal relaxation. This longitudinal and transverse *mode coupling effect* is always present in a nonaxially symmetric potential, as the smallest eigenvalue of the Fokker–Planck equation will always *intrinsically* depend on the damping. *This is quite unlike axial symmetry, where the damping only enters via the free diffusion time.*

The IHD Eq. (C.12) was also derived from first principles directly using Kramers' escape-rate theory, without recourse to Langer's work, by Smith and de Rozario[82] and Brown,[9] and has been reviewed by Geoghegan *et al.*[59] In Brown's calculation,[9] the free energy density is diagonalized so that, in the vicinity of the saddle point and minimum, respectively, we have

$$V = V_0 + \frac{1}{2}\left(c_1^{(0)}\varphi^2 + c_2^{(0)}p^2\right) \text{ and } V = V_i + \frac{1}{2}\left(c_1^{(i)}\varphi^2 + c_2^{(i)}p^2\right),$$

where $c_1^{(0)}$ and $c_2^{(0)}$ are the coefficients of the second-order term of the Taylor series of the expansion of $V$ at the saddle point, and $c_1^{(i)}$ and $c_2^{(i)}$ are the coefficients of the second-order term in the Taylor series expansion of the energy in the well. Thus Brown's IHD result for the escape rate is given by Eq. (84). Obviously Brown's equation, Eq. (84), coincides with Eq. (C.12).

We emphasize that Langer's formula for classical spins, Eq. (C.12), which is valid for $\alpha S_i \gtrsim 1$ only, is never applicable at low damping, $\alpha S_i < 1$ (the most interesting damping range from experimental point of view), where it coincides with TST. The range $\alpha S_i < 1$ for classical spins constitutes the so-called *turnover-energy controlled diffusion range*, where the energy loss per cycle of the motion of the magnetization on the saddle point energy (escape) trajectory is less (and even much less at $\alpha S_i \ll 1$) than the thermal energy so that that escape rate theory methods using the *spatially controlled diffusion concept (as does Langer's method)* are no longer applicable here. To ensure consistent results in that damping range, the *energy controlled diffusion* approach of Klik and Gunther[83] in conjunction with the Landau-Lifshitz or Gilbert's equation must be used rather than that based on Langer's IHD or spatially controlled diffusion method (see Section IV.A).

# APPENDIX D. DISCRETE ORIENTATION MODEL

When the energy barriers are large in comparison with $kT$, but not so large as to prevent changes in the magnetization orientation occurring altogether, we may assume[9,52,58,59] that the magnetization **M** is restricted to stable orientations along the local minima of the free energy. The time behavior of **M** is then treated as a discrete Markov process, and the continuous distribution of orientations $W$ is replaced by $n_i$, the number of particles in the $i^{th}$ orientation. For a large number $n$ of noninteracting particles, $n_i$ changes with time in accordance with the master equation

$$\dot{n}_i = \sum_{j \neq i}(\nu_{ji} n_j - \nu_{ij} n_i), \quad n = \sum_i n_i, \tag{D.1}$$

where $\nu_{ij}$ is the transition probability (i.e., the relative frequency or escape rate) from orientation $i$ to orientation $j$, i.e., the probability of the magnetization in orientation $i$ undergoing a transition to orientation $j$. If there are $m$ directions of easy magnetization, there are $m$ equations in Eq.(D.1) with $i = 1, 2, ..., m$. Because $\dot{n} = 0$, so that the total number of particles remains at its initial value (i.e., conserved), one may drop one of the equations in Eq. (D.1).

For two orientations (as in a uniaxial or biaxial superparamagnet; see Sections IV.C and IV.E), we let 1 refer to the positive orientation and 2 to the negative, so that Eq. (D.1) reduces to

$$\dot{n}_1 = \nu_{21} n_2 - \nu_{12} n_1 \text{ and } \dot{n}_2 = \nu_{12} n_1 - \nu_{21} n_2. \tag{D.2}$$

Setting $n_2 = n - n_1$ gives

$$\dot{n}_1 = \nu_{21} n - (\nu_{21} + \nu_{12}) n_1, \tag{D.3}$$

so that the $n_1$, and hence $n_2$, and the relative magnetization $M/M_S = n_1 - n_2$ approach their final value according to the factor $e^{-(\nu_{21} + \nu_{12})t}$, i.e., with time constant or reversal time

$$\tau = (\nu_{21} + \nu_{12})^{-1}. \tag{D.4}$$

For two equivalent wells with one saddle point, where $\nu_{21} = \nu_{12} = \Gamma$, e.g., for uniaxial anisotropy with a transverse magnetic field (see Fig. 10), we have $\tau = (2\Gamma)^{-1}$. However, for two equivalent wells with two saddle points, e.g., for biaxial anisotropy (see Fig. 16), we have $\tau = (4\Gamma)^{-1}$. Here the factor 4 occurs because (i) there are *two* escape routes from the well over the saddle points, and (ii) *two* equivalent wells are involved in the relaxation process.

For more than two orientations, the case of greatest interest is when the free energy per unit volume arises from cubic anisotropy, namely

$$V = K(u_1^2 u_2^2 + u_2^2 u_3^2 + u_3^2 u_1^2), \tag{D.5}$$

where the $u_i$ denote the direction cosines with respect to the cubic axes. For $K > 0$ (Fe-type crystals), $V$ has minima at six orientations of type {100} (i.e., $u_1 = 1$, $u_2 = u_3 = 0$, **M** along a cube edge of the lattice). It has maxima at eight orientations of type {111} (**M** along a body diagonal) and saddle points at twelve orientations of type {110} (**M** along a face diagonal); see Fig. 13. For $K < 0$ (Ni-type crystals), the maxima and minima are interchanged. The values of $V$ at the orientations {100}, {110}, and {111} are 0, $K/4$, and $K/3$, respectively.[9,59]

*Positive cubic anisotropy, $K > 0$*

Let $n_1$, $n_2$, and $n_3$ denote the numbers of particles at the {100}, {010}, and {001} orientations, respectively, and $n_{-1}$, $n_{-2}$, and $n_{-3}$ denote the corresponding numbers in the opposite orientations[59] (see Fig. 23). To get from orientation 1 to orientation 2, a particle must surmount an energy barrier whose lowest point is the saddle point at {110}. To get from orientation 1 to orientation −1, it must surmount two successive energy barriers. If the energy barriers are high, it will be unlikely to do this in just a single event, and thus we may take

$$v_{i-i} = 0, \quad v_{ij} = \Gamma \quad \text{for} \quad j \neq \pm i. \tag{D.6}$$

Equation (D.1) then becomes

$$\dot{n}_{\pm i} = \Gamma(n_j + n_k + n_{-j} + n_{-k} - 4n_{\pm i}), \quad i \neq j \neq k, \quad i, j, k \in \{1, 2, 3\}. \tag{D.7}$$

Setting $x_i = n_i + n_{-i}$ and $y_i = n_i - n_{-i}$, we obtain by addition and subtraction

$$\dot{x}_i = 2\Gamma(x_j + x_k - 2x_i), \quad i \neq j \neq k, \quad i, j, k \in \{1, 2, 3\}, \tag{D.8}$$

$$\dot{y}_i = -4\Gamma y_i, \quad i = 1, 2, 3. \tag{D.9}$$

Hence each component of the magnetization, $M_i = M_s y_i / n$, decays with time constant $1/(4\Gamma)$. By symmetry the equilibrium values attained at $t \to \infty$ are $n_i = n/6$ so that $x_i = 1/3$ and $y_i = 0$, and the solutions of Eq. (D.9) are

$$y_i(t) = y_i(0) e^{-4\Gamma t}, \quad i = 1, 2, 3. \tag{D.10}$$

To solve Eq. (D.8), we set $x_j + x_k = n - x_i$ so that

$$\dot{x}_i = 2\Gamma(n - 3x_i), \quad i = 1, 2, 3. \tag{D.11}$$

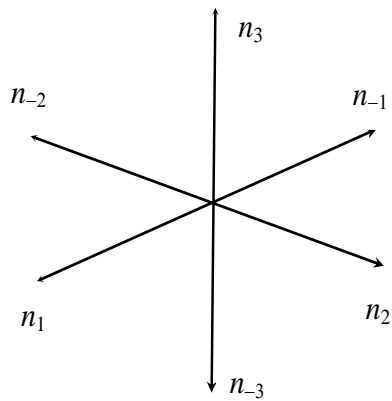

FIG. 23. Populations at the stable magnetization orientations for cubic anisotropy for $K > 0$.

The solutions are then

$$x_i(t) = n/3 + [x_i(0) - n/3]e^{-6\Gamma t}, \quad i = 1, 2, 3. \tag{D.12}$$

Thus the deviations of the $x_i(t)$ from their equilibrium values $n/3$ decay with the time constant $1/(6\Gamma)$, while the time behavior of the $n_i(t)$ is governed by *two* time constants, namely, $1/(4\Gamma)$ and $1/(6\Gamma)$.

*Negative cubic anisotropy, $K < 0$*

The stable magnetization orientations correspond to the eight corners of a cube. Let $n_1$, $n_2$, $n_3$, and $n_4$ denote the numbers of particles with $\{111\}$, $\{11\bar{1}\}$, $\{1\bar{1}1\}$, and $\{\bar{1}11\}]$ orientations respectively and $n_{-1}$, $n_{-2}$, $n_{-3}$, and $n_{-4}$ denote the corresponding numbers in the opposite orientations;[59] see Fig. 24. Just as $K > 0$, we suppose that only one barrier at a time can be surmounted. If we let $i.\text{ADJ}.j$ mean that the subscripts $i$ and $j$ correspond to adjacent minima and $i.\text{NA}.j$ mean the opposite, then[59]

$$v_{ij} = 0, \quad i.\text{NA}.j, \quad v_{ij} = \Gamma, \quad i.\text{ADJ}.j. \tag{D.13}$$

Equation (D.1) becomes[59]

$$\dot{n}_{\pm i} = \Gamma\left(n_j + n_k + n_l - 3n_i\right), \quad i.\text{ADJ}.j, k, l \quad i \neq j \neq k \neq l. \tag{D.14}$$

By setting $x_i = n_i + n_{-i}$ and $y_i = n_i - n_{-i}$, we obtain by addition and subtraction

$$\dot{x}_i = \Gamma(x_j + x_k + x_l - 3x_i), \quad i \neq j \neq k \neq l, \tag{D.15}$$

$$\dot{y}_1 = \Gamma(-3y_1 + y_2 + y_3 + y_4), \quad \dot{y}_2 = \Gamma(y_1 - 3y_2 - y_3 - y_4),$$

$$\dot{y}_3 = \Gamma(y_1 - y_2 - 3y_3 - y_4), \quad \dot{y}_4 = \Gamma(y_1 - y_2 - y_3 - 3y_4). \tag{D.16}$$

Setting $x_j + x_k + x_l = n - x_i$ in Eq. (D.15) yields

$$\dot{x}_i = \Gamma(n - 4x_i). \tag{D.17}$$

Hence the $x_i$ approach their equilibrium value $n/4$ with time constant $1/(4\Gamma)$. Eq. (D.16) can be expressed in matrix form as

$$\frac{d}{dt}\begin{pmatrix} y_1 \\ y_2 \\ y_3 \\ y_4 \end{pmatrix} = \Gamma \begin{pmatrix} -3 & 1 & 1 & 1 \\ 1 & -3 & -1 & -1 \\ 1 & -1 & -3 & -1 \\ 1 & -1 & -1 & -3 \end{pmatrix} \begin{pmatrix} y_1 \\ y_2 \\ y_3 \\ y_4 \end{pmatrix}. \tag{D.18}$$

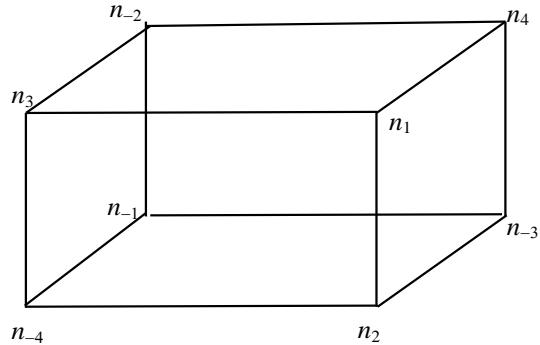

FIG. 24. Populations at the stable magnetization orientations for cubic anisotropy for $K < 0$.

Assuming a solution of the form

$$y_i(t) = y_i(0) e^{-\lambda_i \Gamma t}, \quad i = 1, 2, 3, 4,$$

leads to the requirement that $\lambda_i$ be an eigenvalue of the system matrix in Eq. (D.18). The eigenvalues are the solutions of[59]

$$\begin{vmatrix} -3+\lambda & 1 & 1 & 1 \\ 1 & -3+\lambda & -1 & -1 \\ 1 & -1 & -3+\lambda & -1 \\ 1 & -1 & -1 & -3+\lambda \end{vmatrix} = (\lambda - 2)^3 (\lambda - 6) = 0. \tag{D.19}$$

The eigenvalues are $\lambda = 2$ and $\lambda = 6$. Thus, in general, the decay of the $n_i$ is governed by *three* time constants, namely $1/(2\Gamma)$, $1/(4\Gamma)$, and $1/(6\Gamma)$. The $x$ component of the magnetization $M_X$ is proportional to $y_1 + y_2 + y_3 - y_4$. From Eqs. (D.16), we find that[59]

$$\frac{d}{dt}(y_1 + y_2 + y_3 - y_4) = -2\Gamma(y_1 + y_2 + y_3 - y_4), \tag{D.20}$$

so that the rate of change of $y_1 + y_2 + y_3 - y_4$ is given by $-2\Gamma$ times the quantity itself; thus each component of the magnetization, $M_X$ (or $M_Y$ or $M_Z$) decays with time constant $1/(2\Gamma)$.

The comment of Déjardin *et al.* amply demonstrates both the absolute need for and the timeliness of our review article. Their comment hinges on the statement made in Sec. IV.B of the review, where both the stochastic Gilbert and the Landau-Lifshitz equations are discussed, namely,

"*..... Unfortunately, some authors (see, e.g., Ref. 137 and 138) have ignored this property of the Landau-Lifshitz equation and, in consequence, have used this intrinsically under-damped equation in conjunction with the intrinsically IHD [Intermediate-to-high damping] Langer formula for the calculation of the escape rate in all damping ranges. Thus the ensuing escape rate formulas [Refs. 137, 138] are misleading and not valid for experimental comparison both at low damping, where they coincide with the TST rate, and also in the IHD range, $\alpha \gtrsim 1$, where they predict nonphysical behavior of the rate, namely, a rate in excess of the TST one.*"

We first explain in more detail why using the Landau-Lifshitz equation in conjunction with the IHD Langer formula for the escape rate leads to unphysical behavior of the rate (taking as a particular example, the results of Refs. 137 and 138). In order to draw this conclusion, comparison with experiment is unnecessary, instead one can merely plot (i) the escape rate from *any* of the equations obtained in Refs. 137 and 138 via Langer's theory and (ii) the escape rate $\Gamma^{TST}$ estimated via transition state theory (TST). For example, take Eq. (36) of Ref. 138 for the IHD escape rate, which is

$$\Gamma^{Eq.36}(\alpha) \sim \alpha \left[ (1-\xi/2) + \sqrt{(1+3\xi/2)^2 + 2\xi\alpha^{-2}(2+\xi)} \right] (3\xi+2) \sqrt{\frac{\sigma(2+\xi)}{\pi\xi(3\xi-2)}} e^{-\sigma(2+\xi)}, \qquad (1)$$

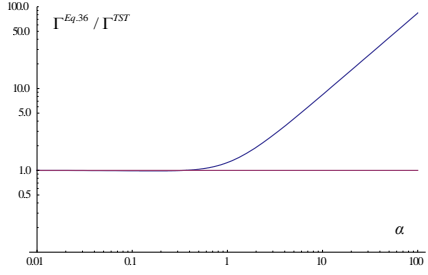

FIG. 25. $\Gamma^{Eq.36}(\alpha)/\Gamma^{TST}$ (blue line) vs. the damping parameter $\alpha$ for $\xi = 5$ and $\sigma = 10$ [$\Gamma^{TST} = \Gamma^{Eq.36}(0)$ (magenta line)].

where $\sigma$ and $\xi$ are dimensionless model parameters characterizing the free energy, then let us plot $\Gamma^{Eq.36}(\alpha)/\Gamma^{TST}$ as a function of *the damping parameter $\alpha$* (see Fig 25). Clearly, by inspection of Fig. 25,

$$\Gamma^{Eq.36}(\alpha) \geq \Gamma^{TST} \text{ for all damping, } \Gamma^{Eq.36}(\alpha \to 0) = \Gamma^{TST}, \text{ and } \Gamma^{Eq.36}(\alpha \to \infty) = \infty \ ^3 \quad (2)$$

Thus, one may conclude that *the ensuing escape rate formula, Eq.(1), for low damping coincides with the TST rate, and in the IHD range, $\alpha \gtrsim 1$, predicts a monotonically increasing rate in excess of the TST one* (Do the Authors of the comment now agree with that statement?). Then, two rhetorical questions may be posed:

1) *Is the damping dependence of the escape rate shown in Fig. 1 physically acceptable*? and
2) *Can $\Gamma^{Eq.36}(\alpha)$ be used as it is for comparison with experiment (numerical or real)?*

The answers (No, it is not and No, it cannot) have already been given in our paragraph quoted above. Apparently, the Authors of the comment disagree with these answers. However, our conclusions may be succinctly restated as follows: *an equation contradicting known physical principles is misleading and cannot be used for comparison with experiment.* Is that not so? It is highly likely that to many, including the Authors of the comment, who have specialized, e.g., in electro-optics, supergravity, or fluid mechanics, Eqs. (1) and (2) do not at first glance appear wrong as they may not necessarily be familiar with escape rate theory. Therefore, they may never have read carefully a textbook or a review article on the subject, whereupon they could easily see that $\Gamma^{TST}$ must be the upper bound of the escape rate $\Gamma(\alpha)$ with limiting values $\Gamma(\alpha \to 0) = 0$ and $\Gamma(\alpha \to \infty) = 0$ [Ref. 56, Fig. 18; see also Figs. 2, 3, and 9 of the review and Fig. 4 of Ref. 56.). Thus, if they can now bring themselves to agree that the correct escape rate $\Gamma(\alpha)$ must satisfy the inequality $\Gamma(\alpha) \leq \Gamma^{TST}$ and that the correct undamped and overdamped limits are, respectively, $\Gamma(\alpha \to 0) = 0$ and $\Gamma(\alpha \to \infty) = 0$, we can together logically conclude that the predictions of Eq. (1) [Eq. (36) of Ref. 138] shown in Fig. 25 are *unphysical and misleading*. Hence, $\Gamma^{Eq.36}(\alpha)$ from Eq. (36) of Ref. 138 is *not valid for comparison with experiment*.

Now, the reason for such absurd behavior of $\Gamma^{Eq.36}(\alpha)$ is that in Ref. 138, the Landau-Lifshitz equation, *which strictly applies only in the underdamped limit, where energy controlled diffusion dominates*, was used in conjunction with *the spatially controlled diffusion theory of Langer, which strictly applies only in the IHD limit*. Recall that Langer's calculation is merely an extension of the Kramers escape rate theory to multidimensional systems and nonseparable Hamiltonians in the so called IHD limit alias the spatially controlled diffusion range. Therefore, in order to avoid such unphysical behavior of the escape rate for classical spins in the IHD range, $\alpha \gtrsim 1$, Langer's theory must be used in conjunction with Gilbert's equation as suggested by Brown (the arguments of Brown are reproduced in Sec. IV.B). Now, *for low damping*, $\Gamma^{Eq.36}(\alpha)$ is approximately equal to $\Gamma^{TST}$ (see Fig. 25), which may deviate substantially from the actual escape rate $\Gamma(\alpha)$ (recalling that $\Gamma(\alpha \to 0) = 0$) as repeatedly shown in the review. In a nutshell, in order to ensure physically acceptable results in the low damping range (*the most interesting damping range from an experimental point of view*), *the energy controlled diffusion* approach must be used rather than that based on Langer's IHD or *spatially controlled diffusion* method (as also repeatedly shown in the review; see Sec. IV.A).

Consider now some other ("last but not least") points raised in the comment of Déjardin *et al.*:

• *"It is curious how the authors' select their ... a kind of phase diagram was obtained for uniaxial anisotropy with precise crossovers between various damping regimes."*

We have cited Ref. 137 and 138 as *typical* examples chosen, firstly, to illustrate the inconsistencies in the application of the IHD Langer escape rate theory for spin systems in conjunction with the Landau-Lifshitz equation, and, secondly, chosen so as to warn the reader of possible errors due to *uncritical* application of escape rate theory to classical spin systems in various damping ranges (the crucial point which is sometimes misunderstood). We also emphasise that the crossovers between the various damping regimes for classical spin systems pertain only to the *symmetry breaking* properties of the Fokker-Planck operator and do not arise from the normalizing free diffusion time. However, these crossovers due to breakdown of axial symmetry are masked by the factor $\alpha + \alpha^{-1}$.[4] Therefore, to observe the crossovers, one can simply plot $\tau/(\alpha+\alpha^{-1})$ given in Table II of our review as a function of $\alpha$ for any nonaxially symmetric potential (cubic, biaxial, etc.).

---

[3] Hence Eq. (1) predicts (i) escape from a metastable state in the absence of coupling to the bath, $\alpha \to 0$, and (ii) instantaneous escape in the overdamped limit, $\alpha \to \infty$!

Fortunately, most calculations of the IHD escape rate for classical spin systems require neither correction nor clarification because they were made using the Gilbert equation [8, 9, 56, 82, 85, etc.][5]. We reiterate, however, that *all* IHD formulas, since they are based on spatially controlled diffusion, are inapplicable at low damping, where the energy loss per period < *kT*, because they *always* predict, just as with the TST formula, *escape in the absence of coupling to the bath*. This is so irrespective of whether or not the Landau-Lifshitz or the Gilbert equation is used since for low damping both are the same anyway.[6] Instead the energy controlled diffusion approach must be used. This is of course the very (and so often misunderstood) point raised by Kramers in 1940 in his treatment of the escape rate of point particles over potential barriers in the low damping case, which represents the core result of his historic paper.

• *"The authors claim that the analytical expressions published in Refs. 137, 138 are misleading and "not valid for experimental comparison"...."*

See above discussion and the review.

• *"The 2nd article in Ref. 137 ... summarized the main steps of Langer's calculation of the relaxation rate and clearly started the validity of the approach with respect to damping."*

Indeed, in the latter paper, it is correctly stated that Langer's theory is applied in the IHD range. Unfortunately, here one of the Authors applied that theory in conjunction with the *intrinsically underdamped* Landau-Lifshitz equation, which has led to absurd results like those shown in Fig. 25.

To conclude, we have demonstrated in our reply that *the escape rate formulas derived in Refs. 137 and 138 predict a rate in excess of the TST one in the IHD range, $\alpha \gtrsim 1$. Such a damping dependence of the escape rate is physically unacceptable and cannot be used for comparison with experiment (real or numerical)*. Indeed, the results of Refs. 137 and 138 with their bizarre damping dependence Eq. (2) would never have appeared (i) had the Authors used in the first place the form of escape rate theory (viz., energy controlled diffusion) appropriate to the range of applicability of the Landau-Lifshitz equation (low damping) or (ii) had they used the Gilbert equation appropriate to the IHD range $\alpha \gtrsim 1$ (as has been accomplished, e.g., by Brown *et al*. [8,9,56,82,85, etc.]). We now hope that with the help of our reply, the Authors of the comment will prepare appropriate *corrigenda* to their previous results in order to avoid further confusion in the literature. Furthermore, we also hope that in future they will *themselves* always compare (prior to publication) the results of their calculations of the escape rate $\Gamma$ with the inequality $\Gamma \leq \Gamma^{TST}$ as well as with the undamped and overdamped limits, viz., $\Gamma(\alpha \to 0) = 0$ and $\Gamma(\alpha \to \infty) = 0$.[7]

---

[4] NB: this is the only factor which determines the damping dependence of the escape rate for axially symmetric potentials.

[5] We suggest to the Authors of the comment simply to answer the question: *why is this so*? Even "the well-known specialist in this area, Dmitry Garanin," "who [more or less] exclusively used the Landau-Lifshitz equation" to estimate the escape rate for a small damping case, has occasionally used the Gilbert equation to treat the IHD escape rate (see [56]). However, we underline once again the fact that the Landau-Lifshitz equation is perfectly suited to the escape rate theory for classical spins in the range of its applicability (low damping), where it was successfully used by many authors (including D. Garanin).

[6] For low damping (the most interesting damping range from an experimental point of view), the Landau-Lifshitz and Gilbert equations are equivalent and yield identical results for the escape rate and for the reversal time of the magnetization.

[7] In reality, the *main problem associated with the escape rate equations obtained in Refs. 137 and 138 is that they have almost never been compared either with actual experiments or with computer simulations*. Actually, *the only* attempt to compare them with independent calculations of the reversal time of the magnetization via the smallest nonvanishing eigenvalue of the Fokker-Planck operator was made in Ref. 156c constituting a mutual cross-check of two independent, viz., numerically exact and asymptotic methods. However, during this cross-checking, it was discovered, in particular, that in the IHD damping range, the escape rate equations of Ref. 137 predict an escape rate in excess of $\Gamma^{TST}$, and, moreover, strong deviations from the *numerically exact* solution of the corresponding Fokker-Planck equation for two interacting spins. Thus they had to be corrected. The corrected equations are listed in Appendix B of Ref. 156c of the comment. Unfortunately, one of the coauthors of Ref. 156c (H. Kachkachi) did not prepare any errata or corrigenda of his previous results [137]. Minor comment: *the Authors themselves are always responsible for all errors in their publications* and not the Editors of EPL, JML, or PRB.

Finally, our principal objective in writing our review (as formulated in the Introduction) was "*to present an overview of the various theoretical approaches for the estimation of the magnetization relaxation time of superparamagnetic nanoparticles fifty years after Brown's seminal paper"* so that many aspects of the dynamics of magnetic nanoclusters remain outside its scope. There is no doubt that the Authors of the comment will be able to prepare a more complete and useful review "for the benefit of a newcomer to the field."